\documentclass[traditabstract]{aa} 
\usepackage{txfonts,graphics,lscape,longtable,epsfig,color}

\begin{document}

\title{Parkes H\,I observations of galaxies behind the southern Milky Way.}


\subtitle{II. The Crux and Great Attractor regions ($l\approx$ 289$\degr$ to 338$\degr$)}

\author{Anja C. Schr\"oder\inst{1,2}
\and Ren\'ee C. Kraan-Korteweg\inst{3} 
\and Patricia A. Henning\inst{4}}

\offprints{A. Schr\"oder}

\institute{
Hartebeesthoek Radio Astronomy Observatory, PO Box 443, Krugersdorp 1740,
South Africa
\and
Dept. of Physics and Astronomy, University of Leicester, University Road,
Leicester LE1 7RH, U.K.
\and 
Department of Astronomy, University of Cape Town, Private Bag X3,
Rondebosch 7701, South Africa
\and
Institute for Astrophysics, University of New Mexico, MSC07 4220,
800 Yale Blvd., NE, 
Albuquerque, NM, 87131, USA
}

\date{Received date;accepted date}

  
\abstract{As part of our programme to map the large-scale distribution of
  galaxies behind the southern Milky Way, we observed 314
  optically-selected, partially-obscured galaxies in the Zone of Avoidance
  (ZOA) in the Crux and Great Attractor (GA) regions. An additional 29
  galaxies were observed in the Vela ZOA survey region (because of the
  small numbers they are not discussed any further). The observations were
  conducted with the Parkes 64\,m (210\,ft) radio telescope, in a
  single-pixel pointed mode, reaching an rms noise level of typically $2 -
  6$\,mJy over the velocity search range of $400 < v <
  10\,500$\,km\,s$^{-1}$. A total of 162 galaxies were detected (plus 14
  galaxies in the Vela region). The detection rate is slightly higher than
  for the Hydra/Antlia region (52\% versus 45\%) observed in the same
  way. This can be explained by the prominence of the GA overdensity in the
  survey regions, which leads to a relatively higher fraction of nearby
  galaxies. It is also evident from the quite narrow velocity distribution
  (largely confined to $3000-6000$\,km\,s$^{-1}$) and deviates
  significantly from the expectation of a uniform galaxy distribution for
  the given sensitivity and velocity range. No systematic differences were
  found between detections and non-detections, in terms of latitude,
  foreground extinction, or environment, except for the very central part
  of the rich Norma cluster, where hardly any galaxies were detected. A
  detailed investigation of the \ion{H}{i} content of the galaxies reveals
  strong \ion{H}{i} deficiency at the core of the Norma cluster (within
  about a 0.4 Abell radius), similar to what has been found in the Coma
  cluster. The redshifts obtained by this observing technique result in a
  substantial reduction of the so-called redshift ZOA. This is obvious when
  analysing the large-scale structure of the new \ion{H}{i} data in
  combination with data from other (optical) ZOA redshift surveys. The
  lower latitude detections provide further evidence of the extension of
  the Norma Wall, across the ZOA, in particular its bending towards the
  Cen-Crux clusters above the Galactic plane at slightly higher redshift,
  rather than a straight continuation towards the Centaurus clusters.

}

\maketitle


\section{Introduction}

Revealing the three-dimensional distribution of galaxies over the entire
sky including the regions behind the dust and stars of our Milky Way is
important for understanding the motion of the Local Group with respect to
the microwave background as well as the peculiar flow fields in the nearby
Universe (e.g., review by Kraan-Korteweg \& Lahav 2000; Kraan-Korteweg
2005; and contributions in ``Mapping the Hidden Universe'', 2000, ASP CS
218, eds. Kraan-Korteweg et al. 2000; ``Nearby Large-Scale Structures and
the Zone of Avoidance'', 2005, ASP CS 329, eds Fairall \& Woudt
2005). Except for blind \ion{H}{i} surveys where both the angular
coordinates and redshifts of galaxies are simultaneously detected, this is
a two-step process: first the galaxies have to be identified (in the
optical, near or far-infrared), then redshifts have to be determined in
follow-up studies. This has been done in the optical, either as
single-channel or multi-fibre spectroscopy depending on surface brightness
of the galaxies and the galaxy density on the sky or in the radio using the
21\,cm spectral line of neutral hydrogen. The latter is most effective for
the most obscured and/or low surface brightness spirals and irregular
galaxies.

Based on the deep optical galaxy catalogues in the southern Zone of
Avoidance (ZOA; Kraan-Korteweg 2000; Woudt \& Kraan-Korteweg 2001), we have
obtained pointed \ion{H}{i} observations of a sample of obscured spiral
galaxies with the 64\,m Parkes radio telescope in Australia. Previous
results of the Hydra/Antlia region ($266\degr \la \ell \la 296\degr$) were
presented in Kraan-Korteweg et al. (2002; hereafter Paper I). The second
part, presented here, covers the observations of spiral galaxies in the
Crux and Great Attractor regions (hereafter GA; $289\degr \la \ell \la
338\degr, -10\degr \la b \la +10\degr$; Woudt \& Kraan-Korteweg 2001). The
optical search detected galaxies above a diameter limit of $D \ga 0\farcm2$
on IIIaJ film copies of the ESO/SRC sky survey. For a detailed description
of the optical search, see Paper I and the optical catalogue papers
(Kraan-Korteweg 2000; Woudt \& Kraan-Korteweg 2001). In summary, our target
list consisted of spiral galaxies without redshift information at the time
of observation, which have extinction-corrected diameters larger than $D^0
> 60\arcsec$ (based on the Schlegel at al. 1998 extinction maps, and the
Cameron 1990 correction laws). This corresponds to the completeness limit
of the deep optical ZOA galaxy catalogues to foreground extinction levels
of $A_B \le 3\fm0$ (Kraan-Korteweg 2000). It will therefore complement
existing optical whole-sky catalogues such as Nilson (1973) for the
northern sky and Lauberts (1982) for the southern sky. The ESO/SRC IIIaJ
film-copies that were used for the searches have such fine-grained emulsion
and sensitivity that the spiral morphology could always be discerned with
our $50\times$ magnifying viewer -- though not always the spiral
sub-type. At the highest extinction levels, we also targeted smaller
galaxies since optical spectroscopy is unlikely to succeed in yielding
redshifts for these heavily extincted galaxies. We also sampled deeper in
some of the overdensities like, e.g., the Norma cluster.

Our sample has a sensitivity of $2 - 6$\,mJy. This study is therefore
complementary to the systematic blind \ion{H}{i} survey of the southern ZOA
conducted with the Parkes multibeam receiver (Staveley-Smith et al. 2000;
for preliminary results see Kraan-Korteweg et al. 2005; Henning et
al. 2005), which covers the optically opaque part of the southern ZOA
($212\degr \le \ell \le 36\degr$, $|b| \leq 5\degr$) for the velocity range
$-1200$ to 12\,700\,km\,s$^{-1}$, with a sensitivity similar to the
observations presented here. A subsample of this work consists of the
shallow \ion{H}{i} ZOA survey (hereafter HIZSS; Henning et al. 2000), based
on 8\% of the integration time of the full survey with a sensitivity of
15\,mJy\,beam$^{-1}$ after Hanning smoothing). In another subsample which
is based on 16\% of the integration time of the full survey, Juraszek et
al. (2000; hereafter JS00) focused on the area of the GA in the ZOA
($308\degr \la \ell \la 332\degr$).

In the following section, a short description of the observations is given
(see Paper I for further details). Section~\ref{det} presents the
\ion{H}{i} data and line profiles of the detected galaxies. In
Sect.~\ref{ndet}, the non-detected galaxies are listed with their
respective velocity search range. An analysis of the properties of the
detected galaxies, the velocity distribution as well as the detection rate
(Sect.~\ref{results}) is followed by a detailed discussion of
\ion{H}{i}-deficiency in the Norma cluster (Sect.~\ref{hidef}). We then
present a description of the three-dimensional galaxy distribution in and
around the investigated area (Sect.~\ref{lss}). A summary is given in
Sect.~\ref{summary}. Throughout the paper we assume a Hubble constant of
$H_0 = 70$\,km\,s$^{-1}$\,Mpc$^{-1}$.

In Appendix A, we discuss cross-identifications of optical galaxies for
which the detected \ion{H}{i} signal might not necessarily be associated
with the optical counterpart, or where more than one galaxy was detected in
one pointing. In Appendix B, we present detections in the Vela region
($245\degr \la \ell \la 275\degr$) which have also been observed during the
course of these observing runs.

\section{Observations} \label{obs}

The Parkes 64\,m radio telescope\footnote{The Parkes telescope is part of
the Australia Telescope which is funded by the Commonwealth of Australia
for operation as a National Facility managed by CSIRO.} was used over
four observing periods of about 10 -- 14 days each (June 1993, April 1994,
July 1995 and September 1996). Here, we report on the observations that
cover the Crux and GA ZOA regions. A detailed description of the
observational set-up is given in Paper~I. A summary of the main
characteristics of the observations is given below.

At 21\,cm, the telescope has a half-power beam-width (HPBW) of 15$\arcmin$.
The system temperature was typically 39\,K at the time of these
observations. Typical integration times were a total of 30 minutes each on
the source (ON) and on a reference position (OFF). Strong sources had
shorter integration times (10 or 20 minutes) while weaker possible
detections were reobserved until the reality of the signal was clearly
determined. Based on calibrating observations the internal consistency of
the flux scale is about $\pm15$\%.

In 1993, we used the 1024-channel auto correlator with a bandwidth of
32\,MHz, covering, in most cases, the radial velocity range $300 -
5500$\,km\,s$^{-1}$, with some additional observations centred at
7500\,km\,s$^{-1}$; the channel spacing was 6.6\,km\,s$^{-1}$ and the
velocity resolution after Hanning smoothing was 13.2\,km\,s$^{-1}$. From
1994 on, we covered the range $300 - 10\,500$\,km\,s$^{-1}$ with a channel
spacing of 13.2\,km\,s$^{-1}$ and a velocity resolution after Hanning
smoothing of 27.0\,km\,s$^{-1}$.

\section{Detections}	\label{det}

In the following, we present the parameters of the 162 detected galaxies,
from the sample of 314 target galaxies. The data were reduced using the
Spectral Line Analysis Program (Staveley-Smith 1985). The two orthogonal
polarisations were averaged during reduction, and a low-order polynomial
baseline subtracted from each spectrum. The reduced \ion{H}{i} spectra are
shown in Fig.~1 which is available online at A\&A. The optical properties
as well as the \ion{H}{i} parameters are given in Table~\ref{cxgadet}. The
columns in the table are described below. A colon after an entry indicates
an uncertain value.


\onlfig{1}{
\begin{figure*}[p]
\vspace{-1cm}
\resizebox{\hsize}{!}{\includegraphics{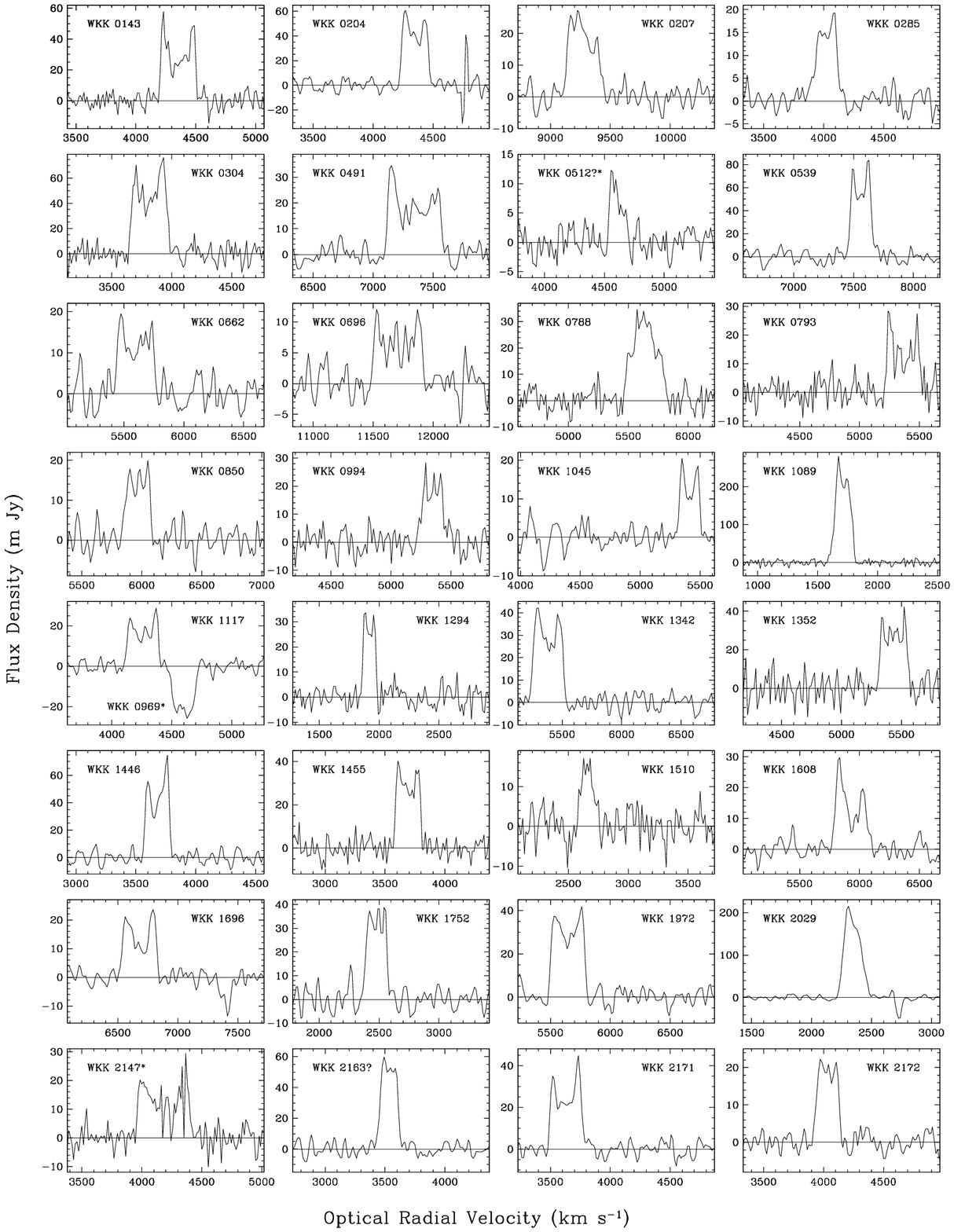}}
\vspace{-1cm}
\caption[]{Baseline-subtracted \ion{H}{i} profiles of the 162 detections in
  the Crux/GA region. The vertical axis gives the flux density in mJy, the
  horizontal axis the velocity range (optical convention), generally
  centred on the velocity of the galaxy and displaying a width of
  1600\,km\,s$^{-1}$. All spectra are baseline-subtracted and generally
  Hanning-smoothed. The identifications are given within the
  panels. Question marks indicate uncertain identifications, stars denote a
  detection not at the centre of the beam.
  }
\label{hiprofile}
\end{figure*}
\addtocounter{figure}{-1} 
\begin{figure*}
\resizebox{\hsize}{!}{\includegraphics{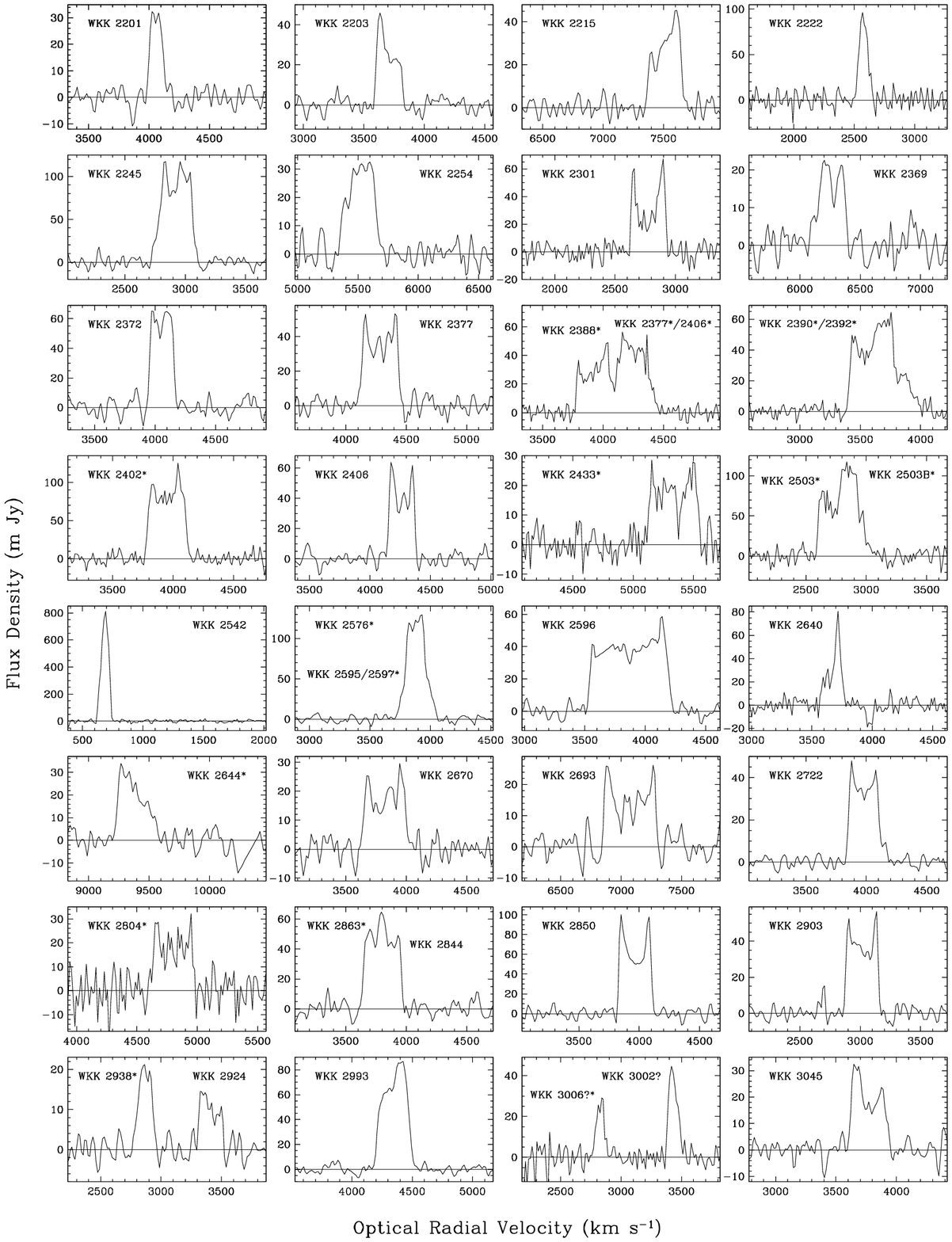}}
\vspace{-1cm}
\caption[]{continued.}
\end{figure*}
\addtocounter{figure}{-1} 
\begin{figure*}
\resizebox{\hsize}{!}{\includegraphics{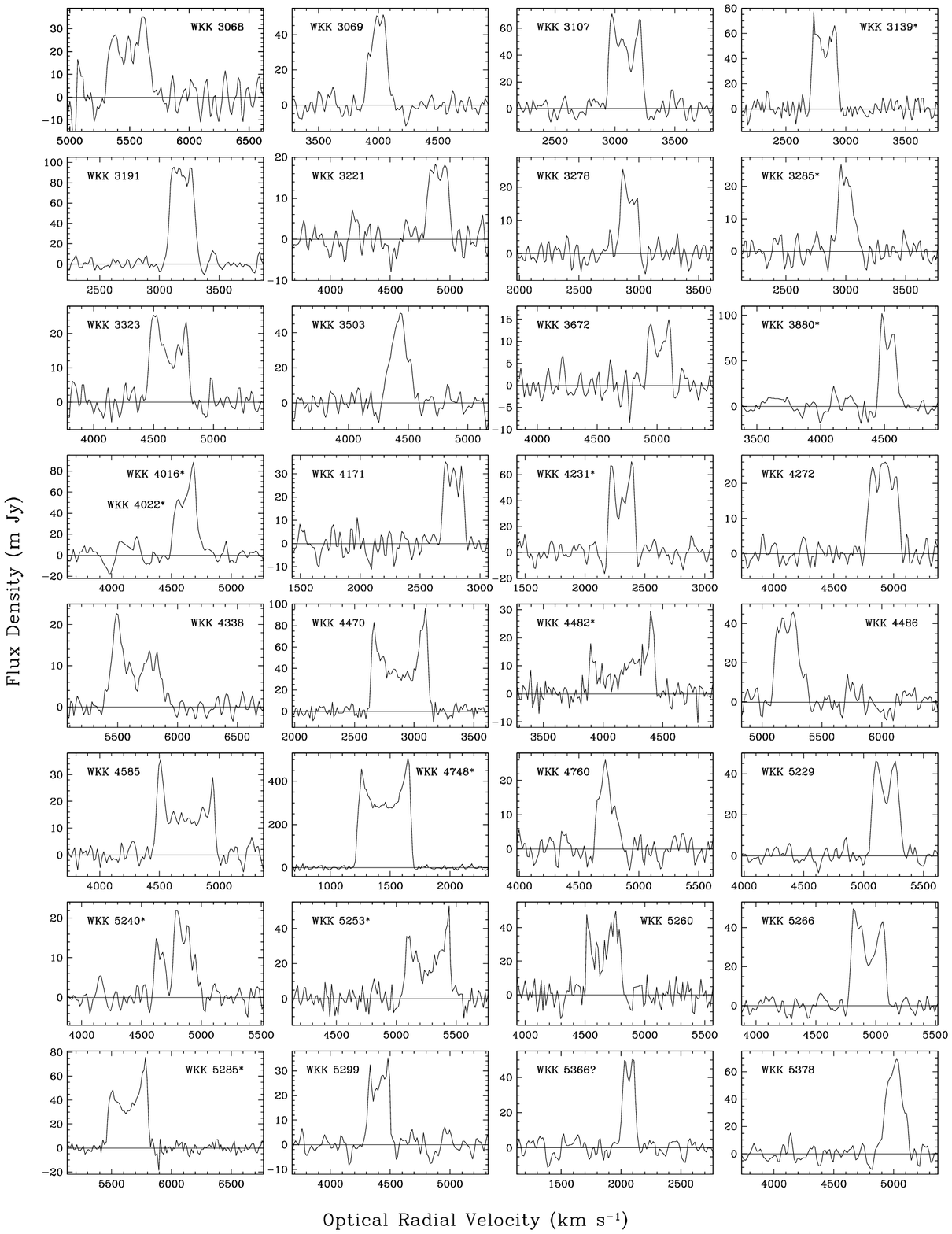}}
\vspace{-1cm}
\caption[]{continued.}
\end{figure*}
\addtocounter{figure}{-1} 
\begin{figure*}
\resizebox{\hsize}{!}{\includegraphics{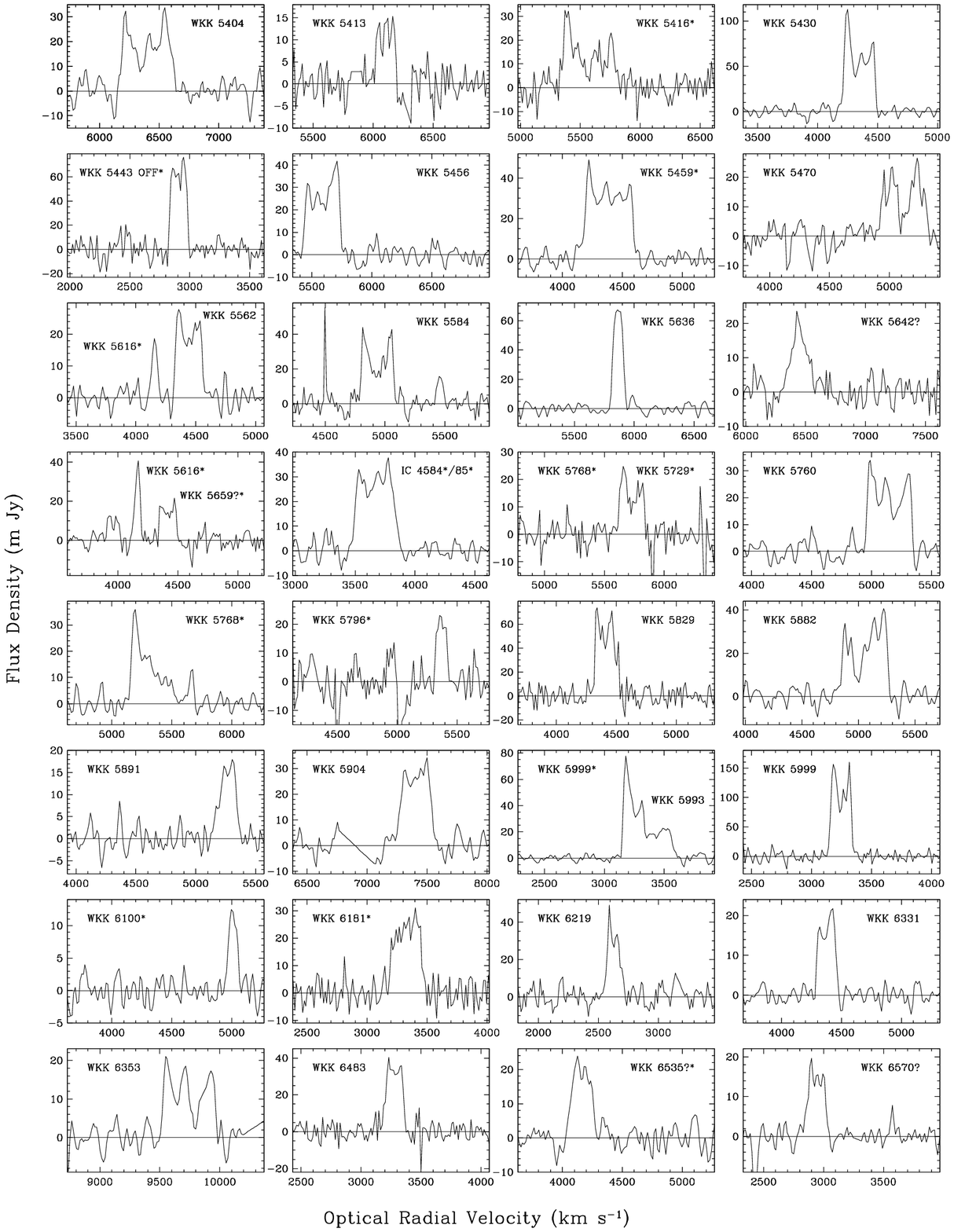}}
\vspace{-1cm}
\caption[]{continued.}
\end{figure*}
\addtocounter{figure}{-1} 
\begin{figure*}
\vspace{-6cm}
\resizebox{\hsize}{!}{\includegraphics{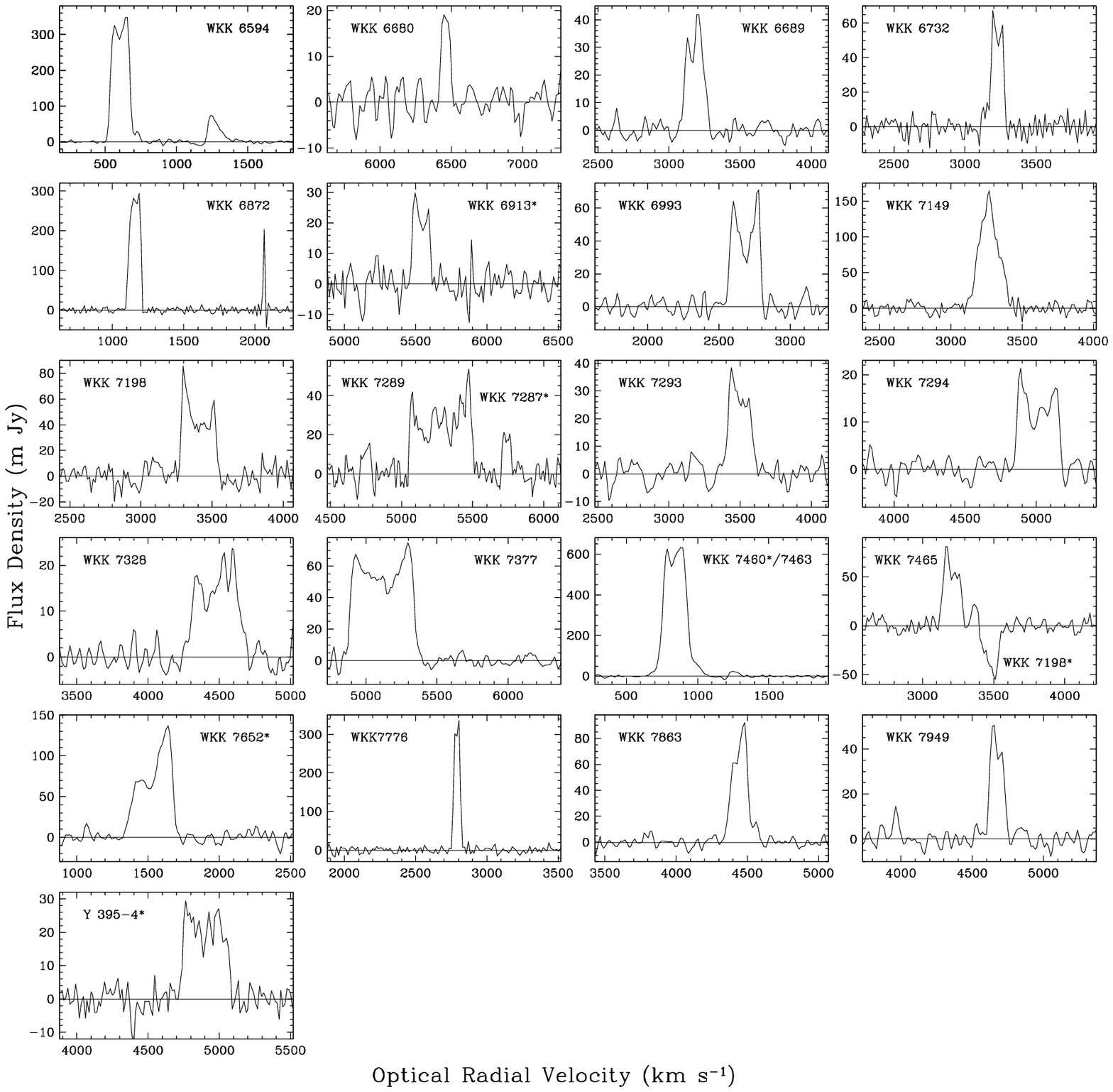}}
\vspace{-1cm}
\caption[]{continued.}
\end{figure*}
}

\addtocounter{figure}{1} 

{\it Column 1:} Identification as given in the optical Crux/GA ZOA galaxy
catalogue (Woudt \& Kraan-Korteweg 2001). A few galaxies from Yamada et
al. (1993), which are all IRAS-selected galaxies, had been added to the
observing programme. Their names start with `Y' and have the ESO-galaxy
number or the IRAS number attached to it. A question mark after the name
denotes an uncertain identification of the \ion{H}{i} signal, and a plus
indicates that more than one signal was found in the pointing or in the
associated OFF-observation.

{\it Column 2:} Second name as found by NED\footnote{the NASA/IPAC
Extragalactic Database} in the order of NGC, IC, ESO, and other catalogue
names. Most of the second identifications originate from the ESO/Uppsala
Survey of the ESO(B) Atlas (Lauberts 1982). Other names come from the
following catalogues: NGC stands for the New General Catalogue (Dreyer
1888), IC stands for the Index Catalogue (Dreyer 1908), FGCE stands for the
Flat Galaxy Catalog (Karachentsev et al. 1993), AM for Arp \& Madore
(1987), CSRG for Catalog of Southern Ringed Galaxies (Buta 1995), HIZSS for
the \ion{H}{i} Parkes ZOA Shallow Survey (Henning et al. 2000), and HIZOA
for Juraszek et al. (2000).

{\it Column 3:} Identification in the infrared (IR) and near-infrared
(NIR): `I' indicates an entry in the IRAS Point Source Catalog according to
the precepts explained in Woudt \& Kraan-Korteweg (2001); `M' or `D'
indicates an entry in the 2MASS Extended Objects Catalogue (2MASS, 2003)
and the DENIS catalogue by Vauglin et al. (2002), respectively.

{\it Column 4 and 5:} Right Ascension RA and Declination Dec (J2000.0).

{\it Column 6 and 7:} Galactic longitude and latitude $\ell$ and $b$.

{\it Column 8:} Morphological type. The morphological types are coded
similarly to the precepts of the RC2 (de Vaucouleurs et al. 1976) with the
addition of the subtypes E, M and L, which stand for early spiral
(S0/a\,--\,Sab), middle spiral (Sb\,--\,Sd) and late spiral or irregular
(Sdm\,--\,Im), respectively. (Note that the uncertainty of the bar presence
is indicated by 'Y' as in Woudt \& Kraan-Korteweg (2001), rather than the
standard 'X' as in the RC2. We have kept their denotation.)

{\it Column 9:} Major axis diameter $D$ and minor axis diameter $d$ in
arcseconds.

\begin{landscape}  
\begin{table}[tb]
 \normalsize
 \renewcommand{\baselinestretch}{0.65}
\caption{\ion{H}{i}-detections in the Crux and Great Attractor region}
\label{cxgadet}
\scriptsize  
\begin{tabular*}{21.5cm}{
  l  @{\extracolsep{3mm}} l @{\extracolsep{5mm}} l@{\extracolsep{-1mm}} l@{\extracolsep{5mm}}      
  l@{\extracolsep{3mm}} l @{\extracolsep{3mm}}  r @{\extracolsep{2mm}} r @{\extracolsep{3mm}} 
  r @{\extracolsep{0mm}}  l @{\extracolsep{0mm}}  c @{\extracolsep{0mm}}   l@{\extracolsep{1mm}}
  r @{\extracolsep{0mm}} l @{\extracolsep{2mm}}
  l @{\extracolsep{2mm}} c @{\extracolsep{0mm}}   
 r @{\extracolsep{0.mm}} c @{\extracolsep{2mm}} r @{\extracolsep{0.mm}} c @{\extracolsep{2mm}}
 r @{\extracolsep{0.mm}} c @{\extracolsep{1mm}} r @{\extracolsep{0.mm}} c @{\extracolsep{1mm}}
 r @{\extracolsep{1mm}} c @{\extracolsep{1mm}} 
 c @{\extracolsep{1mm}} r @{\extracolsep{1mm}} r @{\extracolsep{3mm}} l @{\extracolsep{0mm}}
}
\noalign{\smallskip}
\hline
\noalign{\smallskip}
 \multicolumn{1}{c}{Ident.} & \multicolumn{1}{c}{Other} & IR & & \multicolumn{1}{c}{R.A.} & \multicolumn{1}{c}{Dec.}
& gal $\ell$ \ & gal $b$ & \multicolumn{4}{c}{Type} & \multicolumn{2}{c}{$D$ x $d$} & 
 $B_{J}$ &  $E_{(B-V)}$ & 
 {$V_{hel}$} & & {$\Delta V_{50}$} & & {$\Delta V_{20}$} & & {$I \ \ $} & &
 {rms} & hann & 
 N & dist & {$I_c \ $} &  excised RFI \\
& &  &  & (h\,\, m\,\, s) & \ ($\deg$\,\, $\arcmin$\,\, $\arcsec$) & ($\deg$) \ \ & ($\deg$) \ &
& & & & \multicolumn{2}{c}{($\arcsec$)} & ($^{\rm m}$) & ($^{\rm m}$) & 
km/s & & km/s & & km/s & & {Jy\,km/s} & & m\,Jy & & & {($\arcmin$)}& {Jy\,km/s} & km/s \\
\vspace{-1mm} \\
\multicolumn{1}{c}{(1)} & \multicolumn{1}{c}{(2)} & (3) && \multicolumn{1}{c}{(4)} & \multicolumn{1}{c}{(5)} 
& \multicolumn{1}{c}{(6)} & (7) \ & \multicolumn{4}{c}{(8)} \ & \multicolumn{2}{c}{(9)} & (10) &
\multicolumn{1}{c}{(11)} & (12) & & (13) & & (14) & & (15) & & (16)  & (17) & (18) & (19) & (20) & (21) \ \ \\
\noalign{\smallskip}
\hline
\noalign{\smallskip}
WKK0143   & FGCE0850      &M& & 10 57 03.6 & -70 10 34 & 293.46 & -9.44 & S& & &5    &112x&\phantom{0}11 & 16.4 & 0.23 & 4353& & 296& & 316& &  9.64& &  4.6 & & &        &      &   \\
WKK0204   & ESO063-G006   &M& & 11 09 03.2 & -68 34 42 & 293.77 & -7.55 & S& & & ?   & 51x&\phantom{0}27 & 16.0 & 0.24 & 4349& & 227& & 249& & 10.46& &  4.5 & & &        &      &   \\
WKK0207   &               &M&I& 11 09 15.5 & -70 15 17 & 294.45 & -9.08 & S&Y& &4    & 48x&\phantom{0}32 & 15.8 & 0.22 & 9290& & 266& & 317& &  5.57& &  3.0 & & &        &      &   \\
WKK0285   & ESO063-G014   & &I& 11 30 21.8 & -70 10 59 & 296.11 & -8.39 & S& & &7    & 54x&\phantom{0}28 & 16.2 & 0.41 & 4017& & 183& & 227& &  2.90& &  2.0 & & &        &      &   \\
WKK0304   & ESO063-G017   &M&I& 11 38 55.2 & -69 57 43 & 296.75 & -7.96 & S& & &5    & 81x&\phantom{0}23 & 15.7 & 0.45 & 3813& & 294& & 331& & 15.51& &  6.8 & & &        &      &   \\[0.1cm]
WKK0491+  & ESO171-G003   &M& & 11 58 48.4 & -55 19 01 & 295.44 &  6.79 & S& & &5    & 73x&\phantom{0}15 & 16.3 & 0.24 & 7353& & 451& & 475& &  8.86& &  3.3 & & &        &      &   \\
WKK0512?+ &               &M& & 11 59 54.9 & -55 11 03 & 295.57 &  6.95 & S& & &L    & 28x&\phantom{00}8 & 18.0 & 0.25 & 4617& & 152&:& 169&:&  1.12&:&  2.1 & & &12.4    &  7.4 &   \\
WKK0539   &               &M& & 12 01 20.0 & -53 47 56 & 295.49 &  8.35 & S& & &5    & 73x&\phantom{0}62 & 15.0 & 0.21 & 7562& & 175& & 203& & 11.30& &  5.3 & & &        &      &   \\
WKK0662   & ESO171-G011   &M& & 12 08 35.3 & -54 05 59 & 296.61 &  8.25 & S& & &4    & 55x&\phantom{00}8 & 17.0 & 0.18 & 5603& & 311& & 333& &  3.97& &  3.7 & & &        &      &   \\
WKK0696   &               &M& & 12 10 13.3 & -70 02 06 & 299.40 & -7.45 & S& & &5    & 60x&\phantom{0}31 & 15.9 & 0.26 &11729& & 410& & 433& &  2.81& &  2.4 & & &        &      &   \\[0.1cm]
WKK0788   &               & &I& 12 14 25.4 & -58 26 53 & 298.09 &  4.08 & S& & &5    & 59x&\phantom{0}11 & 17.0 & 0.60 & 5643& & 292& & 337& &  7.15& &  3.9 & & &        &      & \phantom{0}5900  \\
WKK0793   &               & & & 12 14 45.5 & -69 41 15 & 299.74 & -7.04 & S& & &5    & 78x&\phantom{00}8 & 17.1 & 0.29 & 5366& & 269& & 296& &  4.29& &  4.1 & & &        &      & \phantom{0}4750  \\
WKK0850   &               & & & 12 17 29.3 & -53 21 25 & 297.81 &  9.18 & S& &P&     & 48x&\phantom{0}24 & 15.9 & 0.17 & 5964& & 214& & 245& &  3.21& &  2.9 & & &        &      &   \\
WKK0969+  &               &M& & 12 23 06.5 & -58 06 48 & 299.19 &  4.55 &? & & &     & 16x&\phantom{0}12 & 18.5 & 0.66 & 4598& & 195& & 228& &  4.25&:&  2.5 & &*& 8.5    & 10.4 &   \\
WKK0994   &               &M& & 12 24 43.7 & -70 16 09 & 300.66 & -7.52 & S& & &6    & 40x&\phantom{0}34 & 16.2 & 0.36 & 5343& & 172& & 232& &  3.74& &  4.0 & & &        &      &   \\[0.1cm]
WKK1045   &               &M& & 12 27 47.6 & -53 14 45 & 299.35 &  9.46 & S& & &3    & 50x&\phantom{0}42 & 15.4 & 0.17 & 5411& & 172& & 190& &  2.47& &  2.8 & & &        &      &   \\
WKK1089   & ESO172-G004   &M& & 12 31 02.2 & -55 08 14 & 299.99 &  7.62 & I& & &     & 85x&\phantom{0}54 & 14.5 & 0.44 & 1712& & 146& & 183& & 33.61& &  6.8 & & &        &      &   \\
WKK1117+  & ESO131-G010   &M& & 12 32 50.9 & -58 01 30 & 300.46 &  4.76 & S& & &5:   & 43x&\phantom{0}19 & 16.5 & 0.71 & 4256& & 264& & 287& &  4.88& &  2.5 & & &        &      &   \\
WKK1294   & ESO172-G006   & & & 12 43 12.7 & -52 47 47 & 301.67 & 10.06 & S& & &3    & 74x&\phantom{0}31 & 15.6 & 0.17 & 1919& & 110& & 128& &  3.16& &  3.8 & & &        &      &   \\
WKK1342   &               & & & 12 45 46.6 & -55 40 42 & 302.13 &  7.18 & S& & &5    & 65x&\phantom{0}17 & 16.5 & 0.55 & 5386& & 256& & 286& &  8.18& &  3.1 & & &        &      &   \\[0.1cm]
WKK1352   & ESO172-G008   &M& & 12 46 12.9 & -56 44 22 & 302.21 &  6.12 & S&Y& &4:   & 67x&\phantom{0}42 & 15.2 & 0.42 & 5437& & 227& & 257& &  6.71& &  6.8 & & &        &      &   \\
WKK1446   &               & & & 12 51 41.0 & -52 43 10 & 302.97 & 10.15 & S& & &3    & 47x&\phantom{0}27 & 16.2 & 0.23 & 3677& & 207& & 226& &  9.98& &  4.3 & & &        &      &   \\
WKK1455   & ESO172-G011   &M& & 12 52 12.2 & -53 29 37 & 303.05 &  9.38 & S& & &7    & 81x&\phantom{0}34 & 15.5 & 0.32 & 3692& & 211& & 227& &  6.43& &  4.0 & & &        &      &   \\
WKK1510   &               &M& & 12 54 51.6 & -55 53 54 & 303.42 &  6.97 & S& & &4    & 66x&\phantom{0}11 & 16.5 & 0.43 & 2668& & 123& & 176& &  1.76& &  3.8 & & &        &      &   \\
WKK1608   &               &M& & 12 58 43.4 & -56 48 54 & 303.93 &  6.04 & S&B& &4    & 63x&\phantom{0}40 & 15.4 & 0.52 & 5938& & 262& & 319& &  4.39& &  2.6 & & &        &      &   \\[0.1cm]
WKK1696   &               & &I& 13 02 26.8 & -56 08 55 & 304.48 &  6.69 & S& & &5    & 78x&\phantom{0}15 & 16.7 & 0.36 & 6684& & 288& & 312& &  4.31& &  2.3 & & &        &      &   \\
WKK1752   &               & & & 13 05 06.1 & -70 16 38 & 304.09 & -7.44 & S& & &7    & 95x&\phantom{0}12 & 16.6 & 0.38 & 2475& & 179& & 204& &  6.17& &  4.5 & & &        &      &   \\
WKK1972   & ESO173-G008   &M& & 13 12 38.2 & -57 03 45 & 305.82 &  5.69 & S& & &6    & 85x&\phantom{0}31 & 15.6 & 0.48 & 5651& & 287& & 309& &  8.96& &  4.0 & & &        &      &   \\
WKK2029   & HIZSS077      &M& & 13 15 09.9 & -58 56 11 & 306.00 &  3.79 & S& & &M    & 47x&\phantom{0}31 & 15.8 & 0.90 & 2349& & 157& & 221& & 31.67& &  4.2 & & &        &      &   \\
WKK2147   & ESO173-G011   &M&I& 13 21 59.7 & -54 36 45 & 307.39 &  7.99 & S& & &5    &105x&\phantom{0}28 & 15.0 & 0.41 & 4180& & 431& & 454& &  5.87& &  3.7 & & & 0.1    &  5.9 &   \\[0.1cm]
WKK2163?  & ESO220-G006   &M& & 13 22 52.0 & -52 44 36 & 307.75 &  9.83 & S& & &6    & 74x&\phantom{0}56 & 15.3 & 0.31 & 3533& & 155& & 185& &  8.27& &  4.0 & & &        &      &   \\
WKK2171   &               &M&I& 13 23 30.3 & -56 18 31 & 307.40 &  6.28 & S& & &L    & 43x&\phantom{0}38 & 16.5 & 0.75 & 3630& & 262& & 289& &  7.31& &  3.2 & & &        &      &   \\
WKK2172   &               & & & 13 23 27.9 & -53 30 05 & 307.74 &  9.07 & S& & &5    & 60x&\phantom{0}36 & 15.9 & 0.39 & 4029& & 191& & 223& &  3.73& &  2.3 & & &        &      &   \\
WKK2201   & ESO173-G013   & & & 13 25 19.8 & -53 32 13 & 308.02 &  9.00 & S& & &7    & 56x&\phantom{0}46 & 15.9 & 0.33 & 4061& & 121& & 146& &  3.58& &  3.6 & & &        &      &   \\
WKK2203   &               & & & 13 25 24.8 & -56 09 47 & 307.68 &  6.40 & S& & &5    & 59x&\phantom{00}9 & 17.6 & 0.58 & 3709& & 218& & 238& &  6.13& &  3.3 & & &        &      &   \\[0.1cm]
WKK2215   &               & &I& 13 25 46.5 & -55 09 36 & 307.86 &  7.38 & S& & &5    & 70x&\phantom{0}70 & 15.1 & 0.49 & 7520& & 274& & 299& &  8.36& &  3.4 & & &        &      &   \\
WKK2222   & ESO173-G014   &M& & 13 26 16.3 & -55 29 07 & 307.89 &  7.05 & S& & &3:   & 70x&\phantom{0}51 & 15.3 & 0.48 & 2582& &  81& & 123& &  7.72& &  8.8 &n& &        &      &   \\
WKK2245   & HIZSS078      & & & 13 27 43.0 & -57 29 05 & 307.81 &  5.04 & S&Y& &     & 70x&\phantom{0}50 & 15.6 & 0.80 & 2912&:& 272&:& 335&:& 28.17&:&  5.5 & & &        &      &   \\
WKK2254   &               & &I& 13 27 59.0 & -54 58 09 & 308.21 &  7.53 & S& & &5:   & 74x&\phantom{0}23 & 15.9 & 0.43 & 5518& & 279& & 319& &  7.67& &  3.7 & & &        &      &   \\
WKK2301   &               &M& & 13 31 48.7 & -65 16 37 & 307.14 & -2.74 &  & & &     & 20x&\phantom{0}12 & 17.7 & 1.26 & 2775& & 286& & 303& & 10.05& &  5.9 & & &        &      &   \\[0.1cm]
WKK2369   &               & & & 13 35 38.8 & -52 41 34 & 309.70 &  9.60 & S& & &6    & 54x&\phantom{00}7 & 17.8 & 0.34 & 6264& & 224& & 270& &  4.06& &  3.5 & & &        &      &   \\
WKK2372   &               & &I& 13 36 12.3 & -56 32 22 & 309.11 &  5.79 & S& & &L    & 71x&\phantom{0}54 & 15.3 & 0.70 & 4058& & 201& & 223& & 11.88& &  5.3 & & &        &      &   \\
WKK2377   &               &M& & 13 36 35.9 & -54 55 55 & 309.44 &  7.37 & S& & &7:   & 70x&\phantom{0}23 & 16.1 & 0.56 & 4282& & 301& & 329& & 11.93& &  4.5 & & &        &      &   \\
WKK2388++ &               &M&I& 13 37 17.0 & -55 04 34 & 309.52 &  7.21 & S& & &5    & 36x&\phantom{0}16 & 17.0 & 0.57 & 3938& & 281& & 331& &  9.29& &  3.8 & & & 0.7    &  9.3 &   \\
WKK2390+  &               &M&I& 13 37 24.5 & -58 52 21 & 308.85 &  3.47 & S& & &E    & 54x&\phantom{0}12 & 16.6 & 1.09 & 3659&:& 364&:& 516&:& 20.78&:&  3.7 & & & 1.2    & 21.2 &   \\[0.1cm]
WKK2392+  &               &M&I& 13 37 32.6 & -58 50 06 & 308.88 &  3.50 & I& & & ?   & 24x&\phantom{0}13 & 17.2 & 1.09 & 3659&:& 364&:& 516&:& 20.78&:&  3.7 & & & 3.2    & 23.6 &   \\
WKK2402   &               & &I& 13 38 10.5 & -56 28 39 & 309.39 &  5.81 & S&B& &5    &120x&\phantom{0}85 & 14.1 & 0.58 & 3947& & 311& & 343& & 27.32& &  6.7 & & & 0.5    & 27.4 &   \\
WKK2406   &               &M& & 13 38 18.2 & -55 04 16 & 309.66 &  7.19 & S& & &L    & 55x&\phantom{0}38 & 15.9 & 0.62 & 4264& & 216& & 235& &  9.76& &  4.1 & & &        &      &   \\
WKK2433   & ESO066-G001   &M&I& 13 40 21.7 & -71 07 45 & 306.91 & -8.64 & S& & &5    & 78x&\phantom{0}44 & 14.8 & 0.42 & 5335& & 403& & 431& &  7.38& &  4.4 & & & 0.1    &  7.4 &   \\
WKK2503+  & HIZSS083      &M&I& 13 44 55.8 & -65 40 52 & 308.41 & -3.38 & S& & &     & 85x&\phantom{0}43 & 15.2 & 0.89 & 2794&:& 329&:& 396&:& 29.93&:&  7.8 & & & 0.8    & 30.2 &   \\[0.1cm]
WKK2503B+ &               & & &            &           &        &       &\multicolumn{4}{c}{\hspace{-5mm}(LSB)}&&&&0.89& 2794&:& 329&:& 396&:& 29.93&:&  7.8 & & & ...    &  ... &   \\
WKK2542   & ESO174-G001   & & & 13 47 58.3 & -53 20 55 & 311.41 &  8.59 & I& & & ?   &208x&\phantom{0}56 & 14.0 & 0.50 &  684& &  75& & 109& & 57.91& &  8.8 &n& &        &      &   \\
WKK2576++ &               &M&I& 13 50 29.2 & -52 56 47 & 311.87 &  8.90 & S& & &5:   & 86x&\phantom{0}75 & 14.6 & 0.41 & 3883&:& 171&:& 216&:&  ... & &  3.7 & &*& 8.4    &  ... &   \\
WKK2595++ & AM1348-524    &M& & 13 51 23.3 & -52 54 49 & 312.01 &  8.90 & S& & &6    &102x&\phantom{0}40 & 14.9 & 0.39 & 3886&:& 171&:& 268&:& 23.30&:&  3.7 & & &        &      &   \\
WKK2596   & HIZSS084      &M&I& 13 51 38.7 & -58 35 19 & 310.72 &  3.37 & S& & & ?   & 32x&\phantom{0}12 & 17.8 & 1.00 & 3867& & 646& & 682& & 25.61&:&  3.4 & & &        &      & \phantom{0}3650  \\[0.1cm]
WKK2597++ & AM1348-524    &M&I& 13 51 30.5 & -52 55 22 & 312.03 &  8.88 & S& & &5    & 59x&\phantom{0}47 & 15.0 & 0.39 & 3886&:& 171&:& 268&:& 23.30&:&  3.7 & & & 1.2    & 23.7 &   \\
WKK2640+  &               & & & 13 53 38.8 & -52 55 03 & 312.35 &  8.81 & I& & &     & 51x&\phantom{0}42 & 15.9 & 0.40 & 3705& &  49& & 109& &  4.84& &  4.8 &n&*&        &      &   \\
WKK2644+  &               & & & 13 53 47.1 & -52 50 58 & 312.38 &  8.87 & S& & &M    & 26x&\phantom{00}9 & 17.8 & 0.40 & 9404& & 254&:& 341&:&  6.72&:&  3.8 & & & 4.3    &  8.4 &10100  \\
WKK2670   & ESO174-G002   &M&I& 13 54 33.4 & -53 18 41 & 312.38 &  8.40 & S& & &L    &121x&\phantom{0}19 & 15.7 & 0.43 & 3821& & 346& & 379& &  6.77& &  3.7 & & &        &      &   \\
WKK2693   & ESO066-G003   &M& & 13 56 37.1 & -70 55 38 & 308.25 & -8.73 & S& & &5    &118x&\phantom{0}16 & 15.7 & 0.29 & 7081& & 432& & 452& &  6.36& &  3.9 & & &        &      &   \\[0.1cm]
WKK2722   &               &M& & 13 57 21.0 & -54 33 27 & 312.47 &  7.09 & S& & &5    & 78x&\phantom{0}15 & 16.2 & 0.40 & 3984& & 251& & 285& &  9.43& &  2.3 & & &        &      &   \\
WKK2804   & CSRG0755      &M&I& 14 00 26.4 & -53 43 31 & 313.13 &  7.77 & S& & &1    & 34x&\phantom{0}32 & 15.9 & 0.44 & 4803& & 323& & 353& &  6.30& &  6.9 & & & 0.3    &  6.3 &   \\
WKK2844+  &               &M& & 14 01 41.0 & -53 25 14 & 313.39 &  8.02 & I& & &     & 30x&\phantom{0}15 & 17.4 & 0.46 &$>\!3802$&&...&& ...&&  ... & &  4.6 & & &        &      &   \\
WKK2850   &               &M& & 14 01 56.4 & -55 47 30 & 312.78 &  5.73 & S& & &L    & 38x&\phantom{0}24 & 16.5 & 0.46 & 3968& & 270& & 288& & 18.02& &  4.9 & & &        &      &   \\
WKK2863+  & ESO174-G005   &M&I& 14 02 13.7 & -53 32 28 & 313.44 &  7.88 & S& & &5    & 98x&\phantom{0}83 & 14.2 & 0.47 & 3802&:& 308&:& 333&:& 15.28&:&  4.6 & &*& 8.7    & 38.8 &   \\
\noalign{\smallskip}
\hline
\noalign{\smallskip}
\end{tabular*}
 \normalsize
\end{table}
\addtocounter{table}{-1}
\clearpage
\begin{table}[t]
 \normalsize
 \renewcommand{\baselinestretch}{0.85}
\caption{continued.}
\scriptsize  
\begin{tabular*}{21.5cm}{
  l  @{\extracolsep{3mm}} l @{\extracolsep{5mm}} l@{\extracolsep{-1mm}} l@{\extracolsep{5mm}}      
  l@{\extracolsep{3mm}} l @{\extracolsep{3mm}}  r @{\extracolsep{2mm}} r @{\extracolsep{3mm}} 
  r @{\extracolsep{0mm}}  l @{\extracolsep{0mm}}  c @{\extracolsep{0mm}}   l@{\extracolsep{1mm}}
  r @{\extracolsep{0mm}} l @{\extracolsep{2mm}}
  l @{\extracolsep{2mm}} c @{\extracolsep{0mm}}   
 r @{\extracolsep{0.mm}} c @{\extracolsep{2mm}} r @{\extracolsep{0.mm}} c @{\extracolsep{2mm}}
 r @{\extracolsep{0.mm}} c @{\extracolsep{1mm}} r @{\extracolsep{0.mm}} c @{\extracolsep{1mm}}
 r @{\extracolsep{1mm}} c @{\extracolsep{1mm}} 
 c @{\extracolsep{1mm}} r @{\extracolsep{1mm}} r @{\extracolsep{3mm}} l @{\extracolsep{0mm}}
}
\noalign{\smallskip}
\hline
\noalign{\smallskip}
 \multicolumn{1}{c}{Ident.} & \multicolumn{1}{c}{Other} & IR & & \multicolumn{1}{c}{R.A.} & \multicolumn{1}{c}{Dec.}
& gal $\ell$ \ & gal $b$ & \multicolumn{4}{c}{Type} & \multicolumn{2}{c}{$D$ x $d$} & 
 $B_{J}$ &  $E_{(B-V)}$ & 
 {$V_{hel}$} & & {$\Delta V_{50}$} & & {$\Delta V_{20}$} & & {$I \ \ $} & &
 {rms} & hann & 
 N & dist & {$I_c \ $} &  excised RFI \\
& &  &  & (h\,\, m\,\, s) & \ ($\deg$\,\, $\arcmin$\,\, $\arcsec$) & ($\deg$) \ \ & ($\deg$) \ &
& & & & \multicolumn{2}{c}{($\arcsec$)} & ($^{\rm m}$) & ($^{\rm m}$) & 
km/s & & km/s & & km/s & & {Jy\,km/s} & & m\,Jy & & & {($\arcmin$)} \ & {Jy\,km/s} & km/s \\
\vspace{-1mm} \\
\multicolumn{1}{c}{(1)} & \multicolumn{1}{c}{(2)} & (3) && \multicolumn{1}{c}{(4)} & \multicolumn{1}{c}{(5)} 
& \multicolumn{1}{c}{(6)} & (7) \ & \multicolumn{4}{c}{(8)} \ & \multicolumn{2}{c}{(9)} & (10) &
\multicolumn{1}{c}{(11)} & (12) & & (13) & & (14) & & (15) & & (16)  & (17) & (18) & (19) & (20) & (21) \ \ \\
\noalign{\smallskip}
\hline
\noalign{\smallskip}
WKK2903   &               & & & 14 04 39.6 & -69 04 13 & 309.42 & -7.13 & S& & &6    &114x&\phantom{0}12 & 15.9 & 0.29 & 3015& & 270& & 289& & 10.71& &  3.0 & & &        &      &   \\
WKK2924+  &               & & & 14 05 13.4 & -65 34 30 & 310.47 & -3.79 & S& & &8:   & 58x&\phantom{0}22 & 16.1 & 0.60 & 3410& & 205& & 218& &  2.12& &  2.5 & & &        &      &   \\
WKK2938+  &               &M& & 14 05 45.0 & -65 41 04 & 310.49 & -3.91 & L& & &     & 34x&\phantom{0}22 & 16.3 & 0.56 & 2864& & 128& & 173& &  2.58&:&  2.5 & & & 7.3    &  5.0 &   \\
WKK2993   & ESO175-G003   &M&I& 14 08 41.1 & -55 14 06 & 313.86 &  5.99 & S& & &M    & 62x&\phantom{0}28 & 15.8 & 0.45 & 4347& & 254& & 295& & 18.51& &  2.9 & & &        &      &   \\
WKK3002?+ &               & & & 14 09 17.1 & -55 32 13 & 313.85 &  5.68 & S& & &L?   & 56x&\phantom{0}20 & 16.5 & 0.50 & 3436& &  91& & 120& &  3.59& &  4.8 &n& &        &      &   \\[0.1cm]
WKK3006?+ &               & & & 14 09 46.4 & -55 39 16 & 313.88 &  5.54 &  & & &     & 13x&\phantom{00}8 & 19.1 & 0.50 & 2820& &  78& &  95& &  1.99&:&  4.8 &n& & 8.2    &  4.6 &   \\
WKK3045   & ESO175-G004   & &I& 14 12 32.8 & -56 34 33 & 313.97 &  4.55 & S& & &L    &116x&\phantom{0}56 & 14.5 & 0.66 & 3776& & 293& & 334& &  6.48& &  3.4 & & &        &      &   \\
WKK3068   &               & & & 14 14 12.7 & -53 55 18 & 315.03 &  6.99 & S& & &L    & 60x&\phantom{0}27 & 16.1 & 0.67 & 5501& & 355& & 385& &  8.42& &  5.9 & & &        &      &   \\
WKK3069   &               & & & 14 14 19.7 & -56 41 31 & 314.17 &  4.36 &?I& & &     & 74x&\phantom{0}27 & 15.8 & 0.76 & 3995& & 168& & 203& &  7.56& &  4.0 & & &        &      &   \\
WKK3107   & ESO175-G005   &M&I& 14 17 46.9 & -52 49 48 & 315.90 &  7.85 & S& & &L    &120x&\phantom{0}19 & 15.6 & 0.47 & 3088& & 287& & 309& & 14.72& &  5.3 & & &        &      &   \\[0.1cm]
WKK3139   &               & &I& 14 20 23.0 & -55 04 06 & 315.51 &  5.62 & S& &R&     & 43x&\phantom{0}26 & 15.8 & 0.60 & 2826& & 216& & 241& & 12.21& &  5.5 & & & 0.4    & 12.2 &   \\
WKK3191   &               & &I& 14 26 30.9 & -54 21 59 & 316.59 &  5.96 & S& & &L    & 62x&\phantom{0}19 & 16.3 & 0.77 & 3192& & 220& & 251& & 19.59& &  4.8 & & &        &      &   \\
WKK3221   &               & & & 14 28 59.3 & -54 45 01 & 316.78 &  5.48 & S& & &5    & 54x&\phantom{0}43 & 15.6 & 0.64 & 4902& & 192& & 218& &  3.11& &  2.9 & & &        &      &   \\
WKK3278   & ESO067-G002   & & & 14 37 03.3 & -70 28 51 & 311.59 & -9.41 & S& & &6    & 74x&\phantom{0}17 & 16.1 & 0.29 & 2918& & 160& & 183& &  2.95& &  2.6 & & &        &      &   \\
WKK3285   &               & & & 14 36 48.1 & -53 34 23 & 318.30 &  6.13 & I& & & ?   & 24x&\phantom{0}17 & 17.7 & 0.81 & 3010& & 125&:& 185&:&  3.12&:&  2.9 & & &20.1    &453.1 &   \\[0.1cm]
WKK3323   &               &M&I& 14 39 27.9 & -55 25 04 & 317.91 &  4.28 & S& &D& ?   & 54x&\phantom{0}39 & 15.6 & 0.71 & 4628& & 337& & 362& &  5.90& &  2.8 & & &        &      &   \\
WKK3503   &               & & & 14 56 56.0 & -67 09 29 & 314.69 & -7.20 & S& & &6    & 51x&\phantom{0}24 & 16.3 & 0.31 & 4423& & 139& & 242& &  7.56& &  4.4 & & &        &      &   \\
WKK3672   & ESO067-G007   & & & 15 06 27.6 & -69 21 27 & 314.39 & -9.55 & S& & &7    & 74x&\phantom{0}17 & 16.1 & 0.16 & 5019& & 205& & 222& &  2.14& &  2.3 & & &        &      &   \\
WKK3880   & ESO221-G028   &M& & 14 09 02.2 & -51 10 07 & 315.14 &  9.86 & S& & &5    & 73x&\phantom{0}56 & 15.0 & 0.39 & 4533& & 140& & 166& & 11.05&:&  7.6 & & & 3.7    & 13.1 & 3650/4700  \\
WKK4016+  & ESO222-G008   & & & 14 29 37.3 & -51 54 60 & 317.93 &  8.07 & S& & &L    & 67x&\phantom{0}48 & 15.5 & 0.37 & 4641&:& 109&:& 295&:&  ... & &  7.0 & &*&12.1    &  ... & \phantom{0}4100  \\[0.1cm]
WKK4022+  &               & & & 14 30 20.4 & -51 47 26 & 318.08 &  8.15 & S& & &5    & 91x&\phantom{0}34 & 15.7 & 0.38 & 4621&:& 179&:& 232&:& 12.48&:&  7.0 & & & 2.2    & 13.2 & \phantom{0}4100  \\
WKK4171   &               & & & 14 42 18.5 & -54 05 32 & 318.83 &  5.32 & S& & &7    &106x&\phantom{0}16 & 16.3 & 0.70 & 2779& & 183& & 206& &  5.39& &  4.0 & & &        &      & \phantom{0}2050  \\
WKK4231   & ESO222-G015   & & & 14 44 27.2 & -49 23 59 & 321.13 &  9.44 & S&Y& &5:   &159x&\phantom{0}60 & 14.2 & 0.21 & 2306& & 221& & 237& & 10.64&:&  6.3 & & & 5.4    & 15.2 &   \\
WKK4272   & ESO176-G003   & & & 14 46 11.0 & -53 17 43 & 319.69 &  5.80 & S& & &5    & 66x&\phantom{0}17 & 16.4 & 0.64 & 4914& & 258& & 293& &  5.73& &  2.4 & & &        &      &   \\
WKK4338   &               &M& & 14 48 55.6 & -55 38 49 & 319.03 &  3.51 & S& & &5:   & 51x&\phantom{0}17 & 16.5 & 0.80 & 5655& & 420& & 465& &  5.00& &  1.7 & & &        &      &   \\[0.1cm]
WKK4470   & ESO176-G006   &M&I& 14 57 09.9 & -54 23 31 & 320.65 &  4.10 & S& & &M    & 87x&\phantom{0}23 & 15.6 & 0.85 & 2879& & 476& & 499& & 23.30& &  4.3 & & &        &      &   \\
WKK4482   &               &M&I& 14 57 49.6 & -51 58 16 & 321.88 &  6.19 & S& & &M    & 60x&\phantom{0}24 & 16.1 & 0.60 & 4159& & 546&:& 564&:&  5.48&:&  3.2 & & & 2.1    &  5.8 & \phantom{0}4350  \\
WKK4486   & HIZOAJ1458-55 &M&I& 14 58 14.3 & -55 11 37 & 320.42 &  3.31 & S& & &L    & 79x&\phantom{0}38 & 15.4 & 0.88 & 5223& & 207& & 268& &  8.44& &  4.8 & & &        &      &   \\
WKK4585   & ESO176-G008   &M& & 15 03 48.7 & -53 09 14 & 322.11 &  4.72 & S& & &6    &114x&\phantom{0}20 & 15.7 & 0.70 & 4717& & 490& & 516& &  8.31& &  2.7 & & &        &      &   \\
WKK4748   & HIZSS096      & &I& 15 14 34.4 & -52 59 20 & 323.59 &  4.04 & S& & &5    &212x&\phantom{0}56 & 13.9 & 0.99 & 1449&&445&&470&&152.2\phantom{0}&& 8.0 & & & 0.2 &152.3 &   \\[0.1cm]
WKK4760   &               & & & 15 15 58.5 & -61 03 46 & 319.53 & -2.94 & S& & &     & 28x&\phantom{0}13 & 16.8 & 1.34 & 4739& & 118& & 212& &  3.13& &  3.2 & & &        &      &   \\
WKK5229   & HIZOAJ1543-60 & & & 15 43 28.0 & -60 16 11 & 322.76 & -4.20 & S& & &L    & 58x&\phantom{0}19 & 15.8 & 0.74 & 5187& & 235& & 266& &  8.88& &  3.2 & & &        &      &   \\
WKK5240   & ESO099-G008   & &I& 15 45 11.3 & -66 17 32 & 319.19 & -9.07 & S& & &     &157x&\phantom{0}13 & 15.1 & 0.14 & 4785&:& 297&:& 372&:&  4.26&:&  2.0 & & &11.7    & 23.0 &   \\
WKK5253   &               &M&I& 15 45 26.9 & -60 59 31 & 322.51 & -4.92 & S& & &M    & 47x&\phantom{0}19 & 15.9 & 0.71 & 5261& & 377& & 405& &  9.45& &  5.0 & & & 0.3    &  9.5 &   \\
WKK5260   & ESO099-G009   & & & 15 46 01.4 & -63 18 44 & 321.12 & -6.79 & S& & &L    & 82x&\phantom{0}20 & 15.4 & 0.26 & 4657& & 293& & 309& &  8.93& &  6.0 & & &        &      & \phantom{0}4950  \\[0.1cm]
WKK5266   & ESO136-G002   &M& & 15 46 23.7 & -62 11 24 & 321.85 & -5.93 & S& & &5    & 90x&\phantom{0}22 & 15.0 & 0.37 & 4939& & 297& & 321& & 10.17& &  3.1 & & &        &      &   \\
WKK5285   & HIZOAJ1547-59 & &I& 15 47 10.8 & -59 04 07 & 323.87 & -3.54 &  & & &     & 19x&\phantom{0}12 & 17.1 & 0.63 & 5635& & 339& & 357& & 14.21&:&  3.5 & & & 2.7    & 15.5 &   \\
WKK5299   &               & & & 15 48 04.5 & -63 37 13 & 321.11 & -7.17 & S& & &L    & 70x&\phantom{0}12 & 16.2 & 0.27 & 4402&:& 189&:& 209&:&  4.97&:&  3.1 & &*&        &      &   \\
WKK5366?  & HIZOAJ1550-58 &M& & 15 50 22.6 & -58 22 36 & 324.63 & -3.26 & S& & &5    & 43x&\phantom{0}15 & 16.4 & 0.78 & 2064& & 115& & 139& &  5.40& &  4.2 & & &        &      &   \\
WKK5378   &               & & & 15 51 01.0 & -59 30 41 & 323.97 & -4.19 & S& & &     & 23x&\phantom{0}12 & 17.1 & 0.52 & 5016& & 153& & 220& & 10.05& &  5.1 & & &        &      &   \\[0.1cm]
WKK5404   & ESO100-G001   &M&I& 15 52 19.4 & -62 48 38 & 322.00 & -6.84 & S& & &5:   & 71x&\phantom{0}20 & 15.3 & 0.27 & 6404& & 407& & 462& &  8.47& &  4.1 & & &        &      &   \\
WKK5413   &               &M& & 15 52 29.9 & -60 38 07 & 323.41 & -5.18 & S& & &2    & 47x&\phantom{0}12 & 16.0 & 0.42 & 6106& & 164& & 178& &  1.87& &  3.3 & & &        &      & 5850/6250  \\
WKK5416   &               &D&I& 15 52 33.7 & -58 23 37 & 324.84 & -3.45 &?I& & &?    & 24x&\phantom{00}9 & 17.4 & 0.74 & 5580& & 424& & 474& &  6.57& &  5.0 & & & 0.9    &  6.6 &   \\
WKK5430   &               &M& & 15 53 18.8 & -61 08 13 & 323.16 & -5.63 & S&B& &M    & 36x&\phantom{0}23 & 16.1 & 0.40 & 4350& & 262& & 282& & 17.32& &  4.7 & & &        &      &   \\
WKK5443{\tiny OFF}& HIZSS100      & & & 15 43 10   & -58 44 24:& 323.66 & -2.96 &  & & &     &    &              &      & 1.19 & 2907& & 147& & 161& &  9.14&:&  7.5 & & & 1.7    &  9.5 &   \\[0.1cm]
WKK5456   & ESO136-G007   &M& & 15 54 14.2 & -59 49 55 & 324.09 & -4.70 & S& & &5    & 77x&\phantom{0}30 & 15.0 & 0.49 & 5590& & 292& & 319& &  8.22& &  3.4 & & &        &      &   \\
WKK5459   & ESO136-G008   &M&I& 15 54 22.9 & -61 20 24 & 323.13 & -5.87 & S& & &3    & 78x&\phantom{0}32 & 14.5 & 0.40 & 4388& & 398& & 432& & 13.17&:&  3.3 & & & 6.3    & 21.5 &   \\
WKK5470   &               &M& & 15 54 44.7 & -60 56 06 & 323.43 & -5.58 & S& & &5    & 38x&\phantom{0}26 & 15.8 & 0.48 & 5120& & 369& & 400& &  5.56& &  3.6 & & &        &      & 4300/4500/4750  \\
WKK5562+  &               & & & 15 57 52.7 & -61 02 53 & 323.64 & -5.91 & S& & &L    & 67x&\phantom{0}26 & 15.5 & 0.42 & 4443& & 228& & 249& &  4.81& &  2.9 & & &        &      &   \\
WKK5584   & ESO136-G010   &M& & 15 58 29.4 & -60 17 43 & 324.19 & -5.39 & S&Y& &5    & 73x&\phantom{0}43 & 14.6 & 0.36 & 4936& & 269& & 288& &  7.32&:&  4.4 & & &        &      & \phantom{0}4850  \\[0.1cm]
WKK5616+  &               & & & 15 59 27.3 & -61 07 03 & 323.74 & -6.09 & S& & &     & 19x&\phantom{00}5 & 18.4 & 0.41 & 4167& &  41& &  55& &  1.95&:&  5.1 &n&*& 5.3    &  2.8 &   \\
WKK5636   &               &M& & 16 00 04.0 & -61 21 51 & 323.64 & -6.33 & S& & &L    & 32x&\phantom{0}30 & 15.9 & 0.36 & 5866& &  93& & 117& &  5.97& &  3.2 & &*&        &      &   \\
WKK5642?++&               &M&I& 16 00 10.8 & -61 06 54 & 323.81 & -6.15 & S& & &M    & 48x&\phantom{0}17 & 15.9 & 0.40 & 6446& & 122& & 222& &  2.72& &  3.6 & & &        &      &   \\
WKK5659?++&               & & & 16 00 38.0 & -61 06 03 & 323.86 & -6.17 & S& & &6    & 44x&\phantom{0}15 & 16.3 & 0.38 & 4418& & 146& & 157& &  2.22&:&  3.4 & & & 3.4    &  2.6 &   \\
IC4584+   & ESO100-G004   & &I& 16 00 13.7 & -66 22 42 & 320.31 &-10.11 & S&A&S&7    &102x&\phantom{0}90 & 15.8N& 0.11 & 3671&:& 330&:& 377&:&  9.76&:&  3.4 & & & 9.4    & 29.0 &   \\[0.1cm]
IC4585+   & ESO100-G005   &M&I& 16 00 17.6 & -66 19 20 & 320.35 &-10.07 & S&Y&S&5    &126x&\phantom{0}36 & 13.0R& 0.11 & 3671&:& 330&:& 377&:&  9.76&:&  3.4 & & & 8.6    & 24.3 &   \\
WKK5729+  &               & & & 16 02 47.6 & -61 03 14 & 324.09 & -6.31 & S& & &L    & 48x&\phantom{0}16 & 16.3 & 0.32 &$>\!5729$&&...&& ...&&  ... & &  4.4 & & & 6.9    &  ... &   \\
WKK5760   & ESO136-208    &M& & 16 03 29.2 & -59 39 47 & 325.08 & -5.33 & S&Y& &5    &116x&\phantom{0}27 & 14.7 & 0.35 & 5149& & 374& & 392& &  8.17& &  3.4 & & &        &      &   \\
WKK5768   & ESO136-G016   &M& & 16 03 49.2 & -60 58 41 & 324.23 & -6.34 & S& & &5    &211x&\phantom{0}26 & 14.0 & 0.32 & 5422& & 524&:& 541&:&  6.18&:&  2.9 & & & 9.5    & 18.8 &   \\
WKK5796   & ESO136-G017   &M& & 16 04 25.0 & -60 44 14 & 324.45 & -6.20 & S& & &0:   &109x&\phantom{0}22 & 14.6 & 0.32 & 5372& & 100& & 119& &  1.94&:&  5.2 & & & 2.8    &  2.1 & \phantom{0}4300  \\
\noalign{\smallskip}
\hline
\noalign{\smallskip}
\end{tabular*}
 \normalsize
\end{table}
\addtocounter{table}{-1}
\clearpage
\begin{table}[t]
 \normalsize
 \renewcommand{\baselinestretch}{0.85}
\caption{continued.}
\scriptsize  
\begin{tabular*}{21.5cm}{
  l  @{\extracolsep{3mm}} l @{\extracolsep{5mm}} l@{\extracolsep{-1mm}} l@{\extracolsep{5mm}}      
  l@{\extracolsep{3mm}} l @{\extracolsep{3mm}}  r @{\extracolsep{2mm}} r @{\extracolsep{3mm}} 
  r @{\extracolsep{0mm}}  l @{\extracolsep{0mm}}  c @{\extracolsep{0mm}}   l@{\extracolsep{1mm}}
  r @{\extracolsep{0mm}} l @{\extracolsep{2mm}}
  l @{\extracolsep{2mm}} c @{\extracolsep{0mm}}   
 r @{\extracolsep{0.mm}} c @{\extracolsep{2mm}} r @{\extracolsep{0.mm}} c @{\extracolsep{2mm}}
 r @{\extracolsep{0.mm}} c @{\extracolsep{1mm}} r @{\extracolsep{0.mm}} c @{\extracolsep{1mm}}
 r @{\extracolsep{1mm}} c @{\extracolsep{1mm}} 
 c @{\extracolsep{1mm}} r @{\extracolsep{1mm}} r @{\extracolsep{3mm}} l @{\extracolsep{0mm}}
}
\noalign{\smallskip}
\hline
\noalign{\smallskip}
 \multicolumn{1}{c}{Ident.} & \multicolumn{1}{c}{Other} & IR & & \multicolumn{1}{c}{R.A.} & \multicolumn{1}{c}{Dec.}
& gal $\ell$ \ & gal $b$ & \multicolumn{4}{c}{Type} & \multicolumn{2}{c}{$D$ x $d$} & 
 $B_{J}$ &  $E_{(B-V)}$ & 
 {$V_{hel}$} & & {$\Delta V_{50}$} & & {$\Delta V_{20}$} & & {$I \ \ $} & &
 {rms} & hann & 
 N & dist & {$I_c \ $} &  excised RFI \\
& &  &  & (h\,\, m\,\, s) & \ ($\deg$\,\, $\arcmin$\,\, $\arcsec$) & ($\deg$) \ \ & ($\deg$) \ &
& & & & \multicolumn{2}{c}{($\arcsec$)} & ($^{\rm m}$) & ($^{\rm m}$) & 
km/s & & km/s & & km/s & & {Jy\,km/s} & & m\,Jy & & & {($\arcmin$)}& {Jy\,km/s} & km/s \\
\vspace{-1mm} \\
\multicolumn{1}{c}{(1)} & \multicolumn{1}{c}{(2)} & (3) && \multicolumn{1}{c}{(4)} & \multicolumn{1}{c}{(5)} 
& \multicolumn{1}{c}{(6)} & (7) \ & \multicolumn{4}{c}{(8)} \ & \multicolumn{2}{c}{(9)} & (10) &
\multicolumn{1}{c}{(11)} & (12) & & (13) & & (14) & & (15) & & (16)  & (17) & (18) & (19) & (20) & (21) \ \ \\
\noalign{\smallskip}
\hline
\noalign{\smallskip}
WKK5829   &               & & & 16 05 34.8 & -61 15 28 & 324.21 & -6.69 & S& & &L    & 91x&\phantom{0}12 & 15.6 & 0.25 & 4409& & 167&:& 186&:&  9.58& &  7.2 &n&*&        &      &   \\
WKK5882   &               &M& & 16 06 27.4 & -57 33 52 & 326.78 & -4.02 & S& & &5    & 60x&\phantom{0}16 & 16.3 & 0.58 & 5056& & 401& & 426& &  9.83& &  3.6 & & &        &      &   \\
WKK5891   &               & &I& 16 07 04.2 & -61 27 56 & 324.20 & -6.96 & S& & &L    & 65x&\phantom{0}34 & 15.0 & 0.25 & 5255& & 141& & 210& &  2.48& &  2.5 & & &        &      &   \\
WKK5904   & ESO100-G010   & & & 16 07 39.5 & -63 57 36 & 322.54 & -8.84 & S&Y& &5:   & 65x&\phantom{0}35 & 15.1 & 0.18 & 7403& & 264& & 301& &  7.15& &  4.0 & & &        &      & \phantom{0}6850  \\
WKK5993+  & ESO136-G019   &M& & 16 09 50.4 & -60 11 27 & 325.32 & -6.25 & S& & &L    & 90x&\phantom{0}32 & 14.7 & 0.25 & 3443&:& 235&:& 303&:&  4.30&:&  2.3 & &*&        &      &   \\[0.1cm]
WKK5999   & ESO136-G020   &M& & 16 09 56.0 & -60 19 18 & 325.24 & -6.36 & S&Y&R&5    & 63x&\phantom{0}54 & 14.2 & 0.25 & 3244& & 176& & 195& & 21.33&:&  7.0 & & &        &      & \phantom{0}3250  \\
WKK6100   &               &M& & 16 11 55.8 & -60 43 33 & 325.14 & -6.82 & S& &P&     & 34x&\phantom{0}28 & 15.7 & 0.21 & 5011& &  91& & 100& &  0.97& &  3.3 &n& & 5.7    &  1.4 &   \\
WKK6181   & ESO100-G013   &M&I& 16 13 44.5 & -63 24 26 & 323.42 & -8.90 & S& & &E    & 60x&\phantom{0}20 & 15.0 & 0.14 & 3331& & 258& & 289& &  6.05& &  4.4 & & & 0.1    &  6.1 &   \\
WKK6219   &               & & & 16 13 46.0 & -56 16 25 & 328.39 & -3.75 & S& & &     & 32x&\phantom{0}24 & 16.5 & 0.60 & 2627& & 108& & 136& &  3.94& &  4.8 & & &        &      & \phantom{0}3200  \\
WKK6331   &               & &I& 16 16 14.4 & -61 50 49 & 324.73 & -7.98 & I& & &9    & 43x&\phantom{0}15 & 16.4 & 0.18 & 4376& & 161& & 184& &  2.69& &  1.8 & & &        &      &   \\[0.1cm]
WKK6353   & ESO100-G015   & & & 16 16 35.0 & -62 41 26 & 324.16 & -8.61 & S& & &5    & 70x&\phantom{0}16 & 15.8 & 0.14 & 9753& & 438& & 460& &  5.27&:&  3.0 & & &        &      & 9700/10150  \\
WKK6483   & ESO100-G018   &M&I& 16 19 01.2 & -63 03 11 & 324.10 & -9.07 & S& & &7    & 85x&\phantom{0}66 & 13.9 & 0.14 & 3273& & 149& & 184& &  5.34& &  4.1 & & &        &      &   \\
WKK6535?+ &               &M& & 16 20 08.6 & -63 28 13 & 323.89 & -9.45 & S& & &5    & 39x&\phantom{00}9 & 17.2 & 0.14 & 4159& & 183& & 230& &  3.68&:&  3.1 & & & 6.5    &  6.2 &   \\
WKK6570?+ &               &M& & 16 20 40.1 & -63 22 45 & 324.00 & -9.43 & S&Y& &3    & 60x&\phantom{0}27 & 15.5 & 0.15 & 2938& & 148& & 192& &  2.42& &  1.8 & & &        &      & \phantom{0}3300  \\
WKK6594   & ESO137-G018   &M&I& 16 20 59.2 & -60 29 13 & 326.11 & -7.43 & S& & &7    &239x&\phantom{0}82 & 12.6 & 0.25 &  605& & 141& & 164& & 45.55& &  3.6 & & &        &      &   \\[0.1cm]
WKK6680   &               & & & 16 22 46.0 & -62 50 05 & 324.56 & -9.22 & S& & &7    & 46x&\phantom{0}34 & 15.5 & 0.20 & 6462& &  74& &  90& &  1.22& &  5.4 &n&*&        &      &   \\
WKK6689+  & ESO137-G020   & & & 16 22 47.6 & -60 18 56 & 326.39 & -7.46 & S&B& &5    & 82x&\phantom{0}60 & 14.2 & 0.24 & 3184&:& 136&:& 176&:&  4.86&:&  2.6 & & &        &      &   \\
WKK6732   &               & & & 16 23 36.1 & -60 11 29 & 326.55 & -7.45 & S&B& &5    & 44x&\phantom{0}36 & 15.5 & 0.23 & 3226& &  93& & 112& &  5.23&:&  5.5 &n& &        &      &   \\
WKK6872   &               & & & 16 27 21.3 & -60 27 22 & 326.68 & -7.97 & I& & &9:   &109x&\phantom{0}62 & 14.1 & 0.20 & 1155& &  91& & 108& & 24.64& &  5.9 &n& &        &      &   \\
WKK6913   &               & & & 16 28 36.0 & -59 56 50 & 327.16 & -7.73 & I& & &     & 30x&\phantom{0}27 & 16.2 & 0.23 & 5540& & 130& & 144& &  2.90&:&  4.1 & & & 3.2    &  3.3 &   \\[0.1cm]
WKK6993   &               & & & 16 30 06.3 & -57 41 16 & 328.96 & -6.33 & S& & &L    & 99x&\phantom{0}38 & 14.7 & 0.45 & 2683& & 219& & 238& & 10.50& &  4.5 & & &        &      & \phantom{0}3050  \\
WKK7149   & NGC6159       &M&I& 16 34 52.2 & -60 37 06 & 327.19 & -8.76 & S& & &4    & 98x&\phantom{0}83 & 13.2 & 0.17 & 3275& & 145& & 235& & 23.39& &  7.1 & & &        &      &   \\
WKK7198   & CSRG0806      &M&I& 16 36 51.9 & -60 16 36 & 327.61 & -8.71 & S& &R&5    & 91x&\phantom{0}70 & 14.0 & 0.23 & 3407& & 250& & 267& & 12.12& &  8.0 & & &        &      &   \\
WKK7287+  &               & & & 16 40 51.4 & -60 20 23 & 327.89 & -9.13 & I& & &     & 30x&\phantom{0}20 & 16.5 & 0.25 & 5740& &  66& &  74& &  1.14&:&  4.7 & & & 3.3    &  1.3 & 4750/5900  \\
WKK7289+  & ESO137-G038   &M&I& 16 40 52.5 & -60 23 40 & 327.85 & -9.17 & S&Y&R&5    &191x&\phantom{0}73 & 13.1 & 0.25 & 5278& & 435& & 454& & 12.19& &  4.7 & & &        &      & 4750/5900  \\[0.1cm]
WKK7293   & ESO137-G039   & & & 16 41 06.7 & -60 58 55 & 327.42 & -9.57 & S& & &7    & 69x&\phantom{0}11 & 16.2 & 0.26 & 3502& & 174& & 210& &  5.24& &  3.6 & & &        &      & \phantom{0}3200  \\
WKK7294   &               & & & 16 40 52.7 & -56 24 21 & 330.90 & -6.56 & S& & &5    & 79x&\phantom{0}19 & 15.4 & 0.29 & 5013& & 301& & 321& &  4.12& &  2.0 & & &        &      &   \\
WKK7328   & FGCE 1256     &M&I& 16 42 49.6 & -61 05 25 & 327.47 & -9.80 & S& & &6    &126x&\phantom{0}17 & 15.2 & 0.30 & 4485& & 328& & 402& &  5.69& &  2.6 & & &        &      &   \\
WKK7377   & ESO179-G012   &M& & 16 44 21.6 & -55 29 39 & 331.92 & -6.33 & S& & &L    &145x&\phantom{0}20 & 14.7 & 0.35 & 5117& & 450& & 481& & 25.45& &  3.5 & & &        &      &   \\
WKK7460+  & ESO179-G013   &M&I& 16 47 20.0 & -57 26 28 & 330.68 & -7.90 & S&Y& &L    &198x&          105 & 12.7 & 0.27 &  842&:& 179&:& 222&:&113.8:& &  3.7 & & & 1.4    &116.6 &   \\[0.1cm]
WKK7463+  &               & & & 16 47 27.8 & -57 25 36 & 330.70 & -7.90 & S& & &     & 82x&\phantom{0}67 & 14.3 & 0.28 &  842&:& ...& & ...& &  ... & &  3.7 & & &        &      &   \\
WKK7465+  & ESO137-G042   & &I& 16 47 40.5 & -60 08 58 & 328.59 & -9.66 & S& & &4    &112x&\phantom{0}66 & 13.6 & 0.19 &$>\!3256$&&...&& ...&&  ... & &  5.9 & & &        &      &   \\
WKK7652   & NGC6221       & &I& 16 52 46.0 & -59 13 07 & 329.74 & -9.57 & S&Y& &3    &302x&          179 & 11.2 & 0.16 & 1519& & 300& & 340& & 26.43&:&  6.3 & & &11.6    &138.7 &   \\
WKK7776   &               &M& & 16 57 29.7 & -58 34 27 & 330.63 & -9.66 & S&Y& &5    &130x&\phantom{0}95 & 13.6 & 0.18 & 2791& &  49& &  67& & 15.71& &  7.8 &n& &        &      &   \\
WKK7863   &               & & & 17 02 31.6 & -52 50 24 & 335.67 & -6.74 & S& & &     & 74x&\phantom{0}54 & 14.6 & 0.38 & 4436& & 134& & 169& & 10.05& &  3.4 & & &        &      &   \\[0.1cm]
WKK7949   & ESO180-G003   & & & 17 08 36.1 & -56 35 11 & 333.15 & -9.67 & S& & &4    & 65x&\phantom{0}56 & 14.6 & 0.18 & 4677& & 113& & 137& &  4.70& &  3.1 & & &        &      &   \\
Y395-4    & ESO395-G004   &M&I& 18 24 23.3 & -34 10 55 & 359.40 & -9.76 & S& & &     & 72x&\phantom{0}42 & 15.8N& 0.13 & 4904& & 324& & 352& &  7.02& &  3.1 & & & 0.1    &  7.0 & \phantom{0}4350  \\
\noalign{\smallskip}
\hline
\noalign{\smallskip}
\end{tabular*}
\newline
{\bf Notes:} {\it WKK0969:} see plot of WKK1117; {\it WKK2576:} the signal
is blended with WKK2595 and WKK2597 (cf. text), the width and velocity of
the narrow peak are given here; {\it WKK2597:} see plot of WKK2595; {\it
WKK2863:} see plot of WKK2844; {\it WKK4016:} the signal is blended with
WKK4022, the velocity measurements refer to the narrow peak at
$v\!\simeq\!4640$\,km\,s$^{-1}$ only; {\it WKK5299:} the parameters are uncertain due
to an RFI at the edge of the profile at $v\!\simeq\!4500$\,km\,s$^{-1}$; {\it
WKK5616:} see plot of WKK5659; {\it WKK5636:} Hanning-smoothed once despite
the narrow line, due to ringing in the spectrum; {\it WKK5829:} not
Hanning-smoothed due to an RFI close to the profile at $v\!=\!4550$\,km\,s$^{-1}$;
{\it WKK5993:} the profile of WKK5999 (see separate pointing) has been
excluded from the profile measurement; {\it WKK6680:} once Hanning-smoothed
for the plot. 
\normalsize
\end{table}
\end{landscape}

{\it Column 10:} Apparent magnitude $B_J$. These magnitudes are
eye-estimates from the ESO/SERC IIIaJ film copies. They compare well with
the Lauberts \& Valentijn (1989) ${\rm B_{25}}$ magnitudes and have a
1$\sigma$-dispersion of less than $0\fm5$.

{\it Column 11:} The Galactic reddening at the position of the galaxy, as
given by the DIRBE/IRAS extinction maps (Schlegel et al. 1998). See the
catalogue paper (Woudt \& Kraan-Korteweg 2001) for a more detailed
discussion.

{\it Column 12:} Heliocentric \ion{H}{i} radial velocity in km\,s$^{-1}$
taken at the midpoint of the \ion{H}{i} profile at the 20\% level. The
velocity is given in the optical convention $V = c \cdot
(\lambda-\lambda_o)/\lambda_o$. The uncertainty on the velocities,
$\sigma_{V}$, can be determined following Schneider et al. (1986) using
$1.5 (\Delta V_{20} - \Delta V_{50}) ({\rm S/N})^{-1}$, where S/N is the
ratio of the peak signal to rms noise level. The median error over all
detections is 3.6\,km\,s$^{-1}$.

{\it Column 13:} Velocity width in km\,s$^{-1}$ of the \ion{H}{i} profile
measured at the $50\%$ level of the peak intensity. The expected error is
$2.0 \sigma_{V}$ (Schneider et al. 1986), and the median error derived from
the table is 7.3\,km\,s$^{-1}$.

{\it Column 14:} Velocity width in km\,s$^{-1}$ of the \ion{H}{i} profile
measured at the $20\%$ level of the peak intensity. The expected error is
$3.1 \sigma_{V}$ (Schneider et al. 1986), and the median error derived from
the table is 11.3\,km\,s$^{-1}$.

{\it Column 15:} Integrated \ion{H}{i} flux density, in Jy\,km\,s$^{-1}$,
uncorrected for finite beam size. The uncertainty on the flux densities,
$\sigma_{I}$, can be determined following Schneider et al. (1990) using $2
(1.2\Delta V_{20}/R)^{0.5} R\sigma$, where R is the velocity resolution of
the data (see Sect.~\ref{obs}) and $\sigma$ is the rms noise level. The
median error over all defined flux densities is 0.65\,Jy.

{\it Column 16:} RMS noise level in mJy measured over the region used to
fit a baseline, typically of a width of 1600\,km\,s$^{-1}$ centred on, but
not including, the detection.

{\it Column 17:} Most spectra have been Hanning-smoothed, except when the
line width was smaller than 100\,km\,s$^{-1}$ or for other reasons (see
footnotes to the table), which is indicated with an `n' for `no
Hanning-smoothing'.

{\it Column 18:} A star indicates a footnote for this entry.

{\it Column 19:} Angular distance of the detected galaxy from the centre of
the beam in arcminutes. Sometimes the telescope was not pointed towards the
galaxy listed as, e.g., in a close pair or when there was a second
detection in the beam. As the sensitivity of the beam decreases with
distance to the beam centre, the fluxes will have higher uncertainties (see
Col.~20).

{\it Column 20:} Corrected flux densities for off-centre detections. The
sensitivity decreases as a function of the distance from the beam
centre. The observed flux density of galaxies detected away from the
central pointing will be underestimated. We have provided a rough
correction for such galaxies by assuming that the beam sensitivity can be
approximated by a circular Gaussian. This correction becomes uncertain for
distances above the beam radius, i.e., $7\farcm5$.

{\it Column 21:} Excised RFI (radio frequency interference) on or near the
detected \ion{H}{i} profile.

The galaxy density in the Crux and particular in the GA region is quite
high. In 16\% of the pointings, more than one galaxy was found within the
$1\sigma$ beam radius. In most cases the proper identification of the
detected galaxy was straightforward (by comparing size, morphological type,
optical velocity if available, and distance from the beam centre). In
questionable cases we have made use of the HIPASS public data
release\footnote{Data provided by the ATNF under the auspices of the
Multibeam Survey Working Group, see {\tt
http://www.atnf.csiro.au/research/multibeam/release/}} (as a blind
survey it is independent of our pointed observations) to specify the origin
of a signal. In a few cases the identification of the detection remains
ambiguous or the counterpart could not be found at all. Due to the high
extinction in these regions, low surface brightness (late-type) spiral
galaxies are often too obscured to be visible. In other cases the detection
is a combination of the signals from two or even more galaxies and the
individual \ion{H}{i} parameters are uncertain or could not be derived.
The problematic cases are discussed in more detail in Appendix~\ref{notes}.

We have compared our detected velocities with independent velocity
determinations in the literature. Table~\ref{veltab} gives the galaxy IDs
for which we have found velocity determinations, their observed velocity
(from Col.~12 in Table~\ref{cxgadet}), the velocity from the literature and
its error, the origin of the measurement (optical or \ion{H}{i}), and the
reference (as explained in Table~\ref{refveltab}).

Since we did not repeat observations at Parkes for galaxies with already
existing \ion{H}{i} data, most of the given independent velocities
originate from optical spectroscopy. A number of strong \ion{H}{i} sources
were subsequently detected with HIZSS, JS00 and/or with HIPASS.


\begin{table*}[ht]
 \normalsize
 \renewcommand{\baselinestretch}{0.77}
\caption{Comparison of velocities.}
\label{veltab}
\scriptsize  
\begin{tabular}{l@{\extracolsep{2mm}}r@{\extracolsep{0mm}}cll@{\extracolsep{5mm}}l@{\extracolsep{10mm}}l@{\extracolsep{10mm}}|l@{\extracolsep{2mm}}r@{\extracolsep{0mm}}cll@{\extracolsep{5mm}}l}
\hline
 Ident. & {$V_{hel}$} & & \multicolumn{1}{c}{$V_{other}$} & origin & Reference &&  Ident. & {$V_{hel}$} & & \multicolumn{1}{c}{$V_{other}$} & origin & Reference \\
& km/s & & \multicolumn{1}{c}{km/s} & & && & km/s & & \multicolumn{1}{c}{km/s} & & \\
\vspace{-1mm} \\
\ \ \ (1) & (2) & & \multicolumn{1}{c}{(3)} & (4) & (5) \ \ && \ \ \ (1) & (2) & & \multicolumn{1}{c}{(3)} & (4) & (5) \ \ \\
\hline
WKK0204   &  4349& &$\phantom{0}4332\pm\phantom{00}8$ & \ion{H}{i}  & HIPASS  && WKK3880   &  4533& &$\phantom{0}4572\pm\phantom{0}85$ & opt         & RS06    \\
WKK0207   &  9290& &$\phantom{0}9178\pm\phantom{0}70$ & opt         & VY96    && WKK4231   &  2306& &$\phantom{0}2254\pm\phantom{0}10$ &opt \& \ion{H}{i}&CF97 \\
WKK0304   &  3813& &$\phantom{0}3807\pm\phantom{00}7$ & \ion{H}{i}  & HIPASS  &&           &      & &$\phantom{0}2303\pm\phantom{00}6$ & \ion{H}{i}  & HIPASS  \\
WKK0491   &  7353& &$\phantom{0}7523\pm250$           & opt         & FW98    && WKK4272   &  4914& &$\phantom{0}4940\pm\phantom{00}9$ & \ion{H}{i}  & HIPASS  \\
WKK0539   &  7562& &$\phantom{0}7651\pm\phantom{0}85$ & opt         & RS06    && WKK4470   &  2879& &$\phantom{0}2875\pm\phantom{0}10$ & \ion{H}{i}  & HIZSS   \\
          &      & &$\phantom{0}7534\pm\phantom{00}8$ & \ion{H}{i}  & HIPASS  &&           &      & &$\phantom{0}2877\pm\phantom{0}10$ & \ion{H}{i}  & HIZOA   \\
WKK0662   &  5603& &$\phantom{0}6526\pm231$           & opt         & FW98    &&           &      & &$\phantom{0}2890\pm\phantom{00}6$ & \ion{H}{i}  & HIPASS  \\
          &      & &$\phantom{0}5760\pm\phantom{0}76$ & opt         & WK04    && WKK4486   &  5223& &$\phantom{0}5200\pm\phantom{0}10$ & \ion{H}{i}  & HIZOA   \\
          &      & &$\phantom{0}5630\pm\phantom{0}85$ & opt         & RS06    && WKK4585   &  4717& &$\phantom{0}4728\pm137$           & opt         & WK99    \\
WKK1045   &  5411& &$\phantom{0}5268\pm250$           & opt         & FW98    && WKK4748   &  1449& &$\phantom{0}1449\pm\phantom{0}10$ & \ion{H}{i}  & HIZSS   \\
WKK1089   &  1712& &$\phantom{0}3939\pm150$           & opt         & FW98    &&           &      & &$\phantom{0}1445\pm\phantom{0}10$ & \ion{H}{i}  & HIZOA   \\
          &      & &$\phantom{0}1714\pm\phantom{00}5$ & \ion{H}{i}  & HIPASS  &&           &      & &$\phantom{0}1456\pm\phantom{00}3$ & \ion{H}{i}&HIPASS-BGC \\
WKK1294   &  1919& &$\phantom{0}1910\pm\phantom{0}10$ &opt \& \ion{H}{i}&CF97 &&           &      & &$\phantom{0}1456\pm\phantom{00}5$ & \ion{H}{i}  & HIPASS  \\
WKK1352   &  5437& &$\phantom{0}5528\pm100$           & opt         & FW98    && WKK5229   &  5187& &$\phantom{0}5193\pm\phantom{0}10$ & \ion{H}{i}  & HIZOA   \\
WKK1446   &  3677& &$\phantom{0}3679\pm\phantom{00}9$ & \ion{H}{i}  & HIPASS  && WKK5240   &  4785&:&$\phantom{0}4787\pm\phantom{00}8$ & \ion{H}{i}  & HIPASS  \\
WKK1455   &  3692& &$\phantom{0}3684\pm100$           & opt         & WK04    && WKK5253   &  5261& &$\phantom{0}5093\pm\phantom{0}70$ & opt         & VY96    \\
WKK1972   &  5651& &$\phantom{0}5630\pm\phantom{0}85$ & opt         & RS06    && WKK5260   &  4657& &$\phantom{0}4963$\phantom{000000} & opt         & D91     \\
WKK2029   &  2349& &$\phantom{0}2355\pm\phantom{0}50$ & opt         & FW98    &&           &      & &$\phantom{0}4740\pm\phantom{0}10$ & \ion{H}{i}  & HIPASS  \\
          &      & &$\phantom{0}2337\pm\phantom{0}10$ & \ion{H}{i}  & HIZSS   && WKK5266   &  4939& &$\phantom{0}4944$\phantom{000000} & \ion{H}{i}  & MH91    \\
          &      & &$\phantom{0}2340\pm\phantom{00}5$ & \ion{H}{i}  & HIPASS  &&           &      & &$\phantom{0}4940\pm\phantom{00}7$ & \ion{H}{i}  & HIPASS  \\
WKK2147   &  4180& &$\phantom{0}4153\pm\phantom{0}44$ & opt         & FH95    && WKK5285   &  5635& &$\phantom{0}5631\pm\phantom{0}30$ & \ion{H}{i}  & HIZOA   \\
WKK2171   &  3709& &$\phantom{0}3633\pm\phantom{0}10$ & \ion{H}{i}  & HIPASS  && WKK5366?  &  2064& &$\phantom{0}2059\pm\phantom{0}10$ & \ion{H}{i}  & HIZOA   \\
WKK2172   &  4029& &$\phantom{0}4069\pm\phantom{0}11$ & \ion{H}{i}  & HIPASS  &&           &      & &$\phantom{0}4822\pm\phantom{0}82$ & opt         & WK04    \\
WKK2222   &  2582& &$\phantom{0}2570\pm\phantom{00}7$ & \ion{H}{i}  & HIPASS  && WKK5404   &  6404& &$\phantom{0}6450\pm\phantom{0}70$ & opt         & DN97    \\
WKK2245   &  2912&:&$\phantom{0}2915\pm\phantom{0}10$ & \ion{H}{i}  & HIZSS   && WKK5413   &  6106& &$\phantom{0}6057\pm174$           & opt         & WK99    \\
          &      & &$\phantom{0}2903\pm\phantom{0}10$ & \ion{H}{i}  & HIZOA   && WKK5416   &  5580& &$\phantom{0}5524\pm114$           & opt         & WK04    \\
          &      & &$\phantom{0}2992\pm\phantom{0}85$ & opt         & RS06    &&           &      & &$12403\pm\phantom{0}70$           & opt         & VY96    \\
WKK2254   &  5518& &$\phantom{0}5653\pm\phantom{0}58$ & opt         & FW06    && WKK5443OFF&  2907& &$\phantom{0}2905\pm\phantom{0}10$ & \ion{H}{i}  & HIZSS   \\
WKK2372   &  4058& &$\phantom{0}4058\pm\phantom{00}3$ & \ion{H}{i}  & HK01    &&           &      & &$\phantom{0}2897\pm\phantom{0}10$ & \ion{H}{i}  & HIZOA   \\
WKK2388   &  3938& &$\phantom{0}3976\pm\phantom{0}40$ & opt         & FH95    && WKK5459   &  4388& &$\phantom{0}4390\pm\phantom{0}60$ & opt         & RC3     \\
WKK2390   &  3659&:&$\phantom{0}3790\pm\phantom{0}70$ & opt         & VY96    &&           &      & &$\phantom{0}4396\pm\phantom{00}7$ & \ion{H}{i}  & HIPASS  \\
          &      & &$\phantom{0}3586\pm\phantom{0}10$ & \ion{H}{i}  & HIPASS  && WKK5470   &  5120& &$\phantom{0}5209\pm214$           & opt         & WK99    \\
WKK2392   &  3659&:&$\phantom{0}3790\pm\phantom{0}70$ & opt         & VY96    && WKK5584   &  4936& &$\phantom{0}5027$\phantom{000000} & opt         & D91     \\
WKK2402   &  3947& &$\phantom{0}3956\pm\phantom{00}6$ & \ion{H}{i}  & HIPASS  && WKK5642?  &  6446& &$\phantom{0}6045\pm\phantom{0}42$ & opt         & SH92    \\
WKK2433   &  5335& &$\phantom{0}5367\pm\phantom{0}53$ & opt         & FH95    &&           &      & &$\phantom{0}6118\pm100$           & opt         & WK04    \\
WKK2503?  &  2794&:&$\phantom{0}2789\pm\phantom{0}10$ & \ion{H}{i}  & HIZSS   && IC4584    &  3671&:&$\phantom{0}3700\pm\phantom{0}44$ & opt         & SH92    \\
          &      & &$\phantom{0}2769\pm\phantom{0}10$ & \ion{H}{i}  & HIZOA   && IC4585    &  3671&:&$\phantom{0}3638\pm\phantom{0}40$ & opt         & RC3     \\
          &      & &$\phantom{0}2774\pm\phantom{00}6$ & \ion{H}{i}  & HIPASS  && WKK5768   &  5422& &$\phantom{0}5426\pm\phantom{0}10$ & \ion{H}{i}  & RC3     \\
WKK2542   &   684& &$\phantom{00}694\pm\phantom{00}6$ & \ion{H}{i}  & HK01    &&           &      & &$\phantom{0}5428\pm\phantom{00}6$ & \ion{H}{i}  & HIPASS  \\
          &      & &$\phantom{00}680$\phantom{000000} & \ion{H}{i}  & BD99    && WKK5796   &  5372& &$\phantom{0}5260\pm\phantom{0}60$ & opt         & RC3     \\
          &      & &$\phantom{00}688\pm\phantom{00}3$ & \ion{H}{i}&HIPASS-BGC && WKK5993   &  3443&:&$\phantom{0}3487\pm108$           & opt         & WK04    \\
          &      & &$\phantom{00}687\pm\phantom{00}5$ & \ion{H}{i}  & HIPASS  && WKK5999   &  3244& &$\phantom{0}3250\pm\phantom{0}38$ & opt         & WK04    \\
WKK2576   &  3883&:&$\phantom{0}3948\pm\phantom{0}70$ & opt         & DN97    &&           &      & &$\phantom{0}3246\pm\phantom{00}6$ & \ion{H}{i}  & HIPASS  \\
          &      & &$\phantom{0}3876\pm\phantom{0}85$ & opt         & RS06    && WKK6181   &  3331& &$\phantom{0}3278\pm\phantom{0}70$ & opt         & VY96    \\
          &      & &$\phantom{0}3872\pm\phantom{00}5$ & \ion{H}{i}  & HIPASS  &&           &      & &$\phantom{0}3308\pm\phantom{0}70$ & opt         & WK99    \\
WKK2595   &  3886&:&$\phantom{0}3873\pm\phantom{0}85$ & opt         & RS06    && WKK6353   &  9753& &$\phantom{0}9900\pm\phantom{0}70$ & opt         & DN97    \\
WKK2596   &  3867& &$\phantom{0}3869\pm\phantom{0}10$ & \ion{H}{i}  & HIZSS   && WKK6483   &  3273& &$\phantom{0}3367$\phantom{000000} & opt         & D91     \\
          &      & &$\phantom{0}3848\pm\phantom{0}10$ & \ion{H}{i}  & HIZOA   &&           &      & &$\phantom{0}3268\pm\phantom{0}17$ & \ion{H}{i}  & HIPASS  \\
          &      & &$\phantom{0}3881\pm\phantom{00}7$ & \ion{H}{i}  & HIPASS  && WKK6594   &   605& &$\phantom{00}606\pm\phantom{0}20$ & opt         & HG95    \\
WKK2597   &  3886&:&$\phantom{0}3973\pm\phantom{0}43$ & opt         & SH92    &&           &      & &$\phantom{00}605\pm\phantom{00}3$ & \ion{H}{i}&HIPASS-BGC \\
          &      & &$\phantom{0}3954\pm\phantom{0}85$ & opt         & RS06    &&           &      & &$\phantom{00}605\pm\phantom{00}5$ & \ion{H}{i}  & HIPASS  \\
WKK2640   &  3705& &$\phantom{0}3574\pm\phantom{0}85$ & opt         & RS06    && WKK6689   &  3184&:&$\phantom{0}3239\pm\phantom{0}88$ & opt         & WK04    \\
WKK2644   &  9404& &$\phantom{0}9406\pm100$           & opt         & WK04    && WKK6872   &  1155& &$\phantom{0}1157\pm\phantom{00}3$ & \ion{H}{i}&HIPASS-BGC \\
          &      & &$\phantom{0}9276\pm\phantom{0}85$ & opt         & RS06    &&           &      & &$\phantom{0}1157\pm\phantom{00}5$ & \ion{H}{i}  & HIPASS  \\
WKK2670   &  3821& &$\phantom{0}3758\pm\phantom{0}40$ & opt         & FH95    && WKK7149   &  3275& &$\phantom{0}3300\pm\phantom{0}30$ & opt         & SE95    \\
          &      & &$\phantom{0}3798\pm\phantom{0}85$ & opt         & RS06    &&           &      & &$\phantom{0}3263\pm\phantom{00}6$ & \ion{H}{i}  & HIPASS  \\
WKK2693   &  7081& &$\phantom{0}7389\pm250$           & opt         & FW98    && WKK7198   &  3407& &$\phantom{0}3405\pm\phantom{00}6$ & \ion{H}{i}  & HIPASS  \\
          &      & &$\phantom{0}7013\pm\phantom{0}80$ & opt         & WK04    && WKK7289   &  5278& &$\phantom{0}2100\pm100$           & opt         & F83     \\
WKK2804   &  4803& &$\phantom{0}4779\pm\phantom{0}39$ & opt         & FH95    && WKK7377   &  5127& &$\phantom{0}5122\pm\phantom{00}7$ & \ion{H}{i}  & HIPASS  \\
WKK2850   &  3968& &$\phantom{0}3959\pm\phantom{00}6$ & \ion{H}{i}  & HIPASS  && WKK7460   &   842&:&$\phantom{00}775\pm\phantom{0}36$ & opt         & SH92    \\
WKK2863   &  3802&:&$\phantom{0}3775\pm\phantom{0}37$ & opt         & SH92    &&           &      & &$\phantom{00}842\pm\phantom{00}4$ & \ion{H}{i}&HIPASS-BGC \\
          &      & &$\phantom{0}3778\pm\phantom{0}30$ & opt         & SE95    &&           &      & &$\phantom{00}843\pm\phantom{00}5$ & \ion{H}{i}  & HIPASS  \\
          &      & &$\phantom{0}3768\pm\phantom{00}6$ & \ion{H}{i}  & HIPASS  && WKK7465 &$>\!3256$&&$\phantom{0}3255\pm\phantom{0}39$ & opt         & SH92    \\
WKK2938   &  2864& &$\phantom{0}3024\pm157$           & opt         & FW98    &&           &      & &$\phantom{0}3265\pm\phantom{00}4$ & \ion{H}{i}  & DN96    \\
WKK2993   &  4347& &$\phantom{0}4313\pm\phantom{0}37$ & opt         & FH95    &&           &      & &$\phantom{0}3283\pm\phantom{0}85$ & opt         & RS06    \\
          &      & &$\phantom{0}4364\pm\phantom{0}45$ & opt         & FW98    && WKK7652   &  1519& &$\phantom{0}1350\pm\phantom{0}31$ & opt         & RC3     \\
          &      & &$\phantom{0}4345\pm\phantom{00}7$ & \ion{H}{i}  & HIPASS  &&           &      & &$\phantom{0}1478\pm\phantom{0}38$ & opt         & WK04    \\
WKK3107   &  3088& &$\phantom{0}3090\pm\phantom{00}7$ & \ion{H}{i}  & HIPASS  &&           &      & &$\phantom{0}1340\pm\phantom{0}85$ & opt         & RS06    \\
WKK3139   &  2826& &$\phantom{0}2844\pm\phantom{0}70$ & opt         & VY96    &&           &      & &$\phantom{0}1499\pm\phantom{00}5$ & \ion{H}{i}  & HIPASS  \\
          &      & &$\phantom{0}2823\pm\phantom{00}7$ & \ion{H}{i}  & HIPASS  && WKK7776   &  2791& &$\phantom{0}2790\pm\phantom{00}3$ & \ion{H}{i}&HIPASS-BGC \\
WKK3191   &  3192& &$\phantom{0}3187\pm\phantom{00}6$ & \ion{H}{i}  & HIPASS  &&           &      & &$\phantom{0}2790\pm\phantom{00}5$ & \ion{H}{i}  & HIPASS  \\
WKK3278   &  2918& &$\phantom{0}3118\pm\phantom{0}40$ & opt         & FW98    && Y395-4    &  4904& &$\phantom{0}4896\pm\phantom{0}43$ & opt         & FH95    \\
WKK3285   &  3010& &$\phantom{0}3016$\phantom{000000} & \ion{H}{i}  & RK02    &&           &      & &$\phantom{0}4895\pm\phantom{00}9$ & \ion{H}{i}  & TH07    \\
          &      & &$\phantom{0}3016\pm\phantom{00}3$ & \ion{H}{i}&HIPASS-BGC &\phantom{00}&&&&&& \\
          &      & &$\phantom{0}3017\pm\phantom{00}5$ & \ion{H}{i}  & HIPASS  &\phantom{00}&&&&&& \\
\hline
\end{tabular}
\normalsize
\end{table*}

\begin{table}[tb]
\normalsize{\caption{References for independent velocity determinations. }\label{refveltab}}
\begin{tabular}{ll}
\noalign{\smallskip}
\hline
\noalign{\smallskip}
BD99:	& Banks et al. (1999) \\
CF97:	& Cot\'e et al. (1997) \\
DN96:	& Di Nella et al. (1996) \\
DN97:	& Di Nella et al. (1997) \\
DT90:	& Djorgovski et al. (1990) \\
D91:    & Dressler (1991) \\
F83:	& Fairall (1983) \\
FW98:	& Fairall et al. (1998) \\
FW06:   & Fairall \& Woudt (2006) \\
FH95:   & Fisher et al. (1995) \\
HW00:	& Hasegawa et al. (2000) \\
HIZSS:  & Henning et al. (2000) \\
HIPASS: & HIPASS (2006) \\
HG95:   & Huchra et al. (1995) \\
HK01:   & Huchtmeier et al. (2001) \\
JS00:   & Juraszek et al. (2000) \\
KD04:   & Koribalski \& Dickey (2004) \\
HIPASS-BGC & Koribalski et al. (2004) \\
MF96:	& Mathewson \& Ford (1996) \\
MH91:   & Mould et al. (1991) \\
PT03:   & Paturel et al. (2003) \\
RS06:   & Radburn-Smith et al. (2006) \\
RK02:   & Ryan-Weber et al. (2002) \\
SE95:   & Sanders et al. (1995) \\
SH92:	& Strauss et al. (1992) \\
TH07:   & Theureau et al. (2007) \\
RC3:    & de Vaucouleurs et al. (1991) \\
Vv92:   & Visvanathan \& van den Bergh (1992) \\
VY96:	& Visvanathan \& Yamada (1996) \\
WK99:	& Woudt et al. (1999) \\
WK04:	& Woudt et al. (2004) \\
\noalign{\smallskip}
\hline
\noalign{\smallskip}
\end{tabular}
\end{table}

The \ion{H}{i} velocities measured by HIPASS with the Multibeam (MB)
receiver compared to our single beam observations are in very good
agreement. The 1$\sigma$-dispersion of 31 galaxies in common is
11\,km\,s$^{-1}$ (including only velocities with HIPASS error measurements
$<10$\,km\,s$^{-1}$). A comparison with HIZSS gives a dispersion of
6.5\,km\,s$^{-1}$ for 7 galaxies. The agreement with optical velocities is
also satisfactory, giving a dispersion of 100\,km\,s$^{-1}$ for 50
measurements with errors $<100$\,km\,s$^{-1}$ (excluding one wrong
measurement for WKK\,5416).

The line width measurements also agree well with HIPASS albeit with a
larger scatter of 28\,km\,s$^{-1}$ for both 50\% and 20\% line widths
(using 29 measurements, excluding all uncertain measurements).

\begin{figure}[tb]
\addtocounter{figure}{-1}
\vspace{-2.cm}
\resizebox{\hsize}{!}{\includegraphics{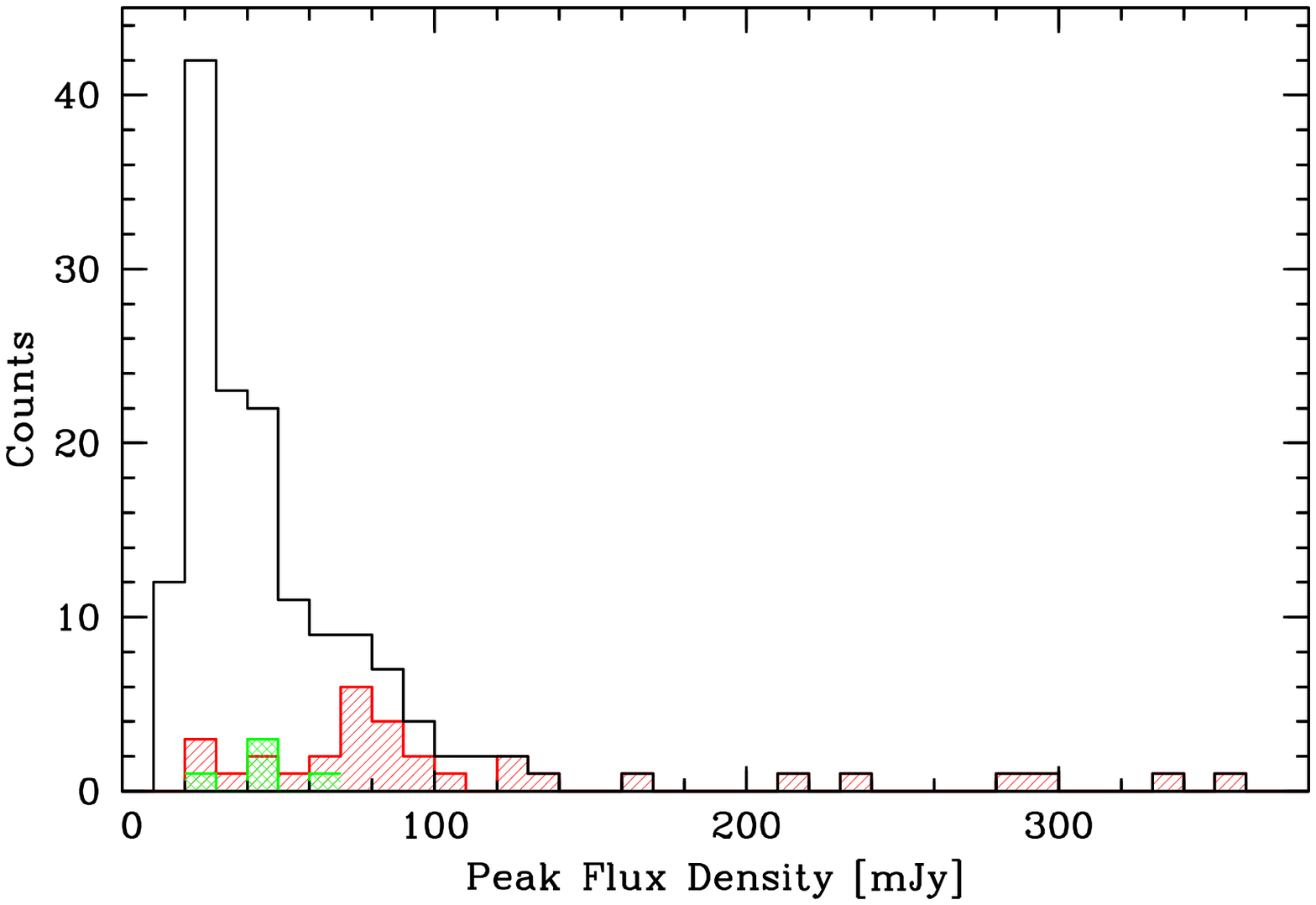}} 
\caption[]{Histogram of peak flux density in mJy of our detections (open) and of
  the detections in common with HIPASS (hashed). Cross-hashed detections
  are HIPASS detections with errors $\ge 10$\,km\,s$^{-1}$. Note that there are
  three more detections by both surveys beyond the frame of the plot
  between 400\,mJy and 800\,mJy. }
\label{histpeakplot}
\end{figure}

While the shallower HIPASS survey detected all galaxies with peak flux
density $>\!120$\,mJy, the majority of our detections have peak flux
densities $<\!70$\,mJy, see Fig.~\ref{histpeakplot}. Most of the HIPASS
detections below $70$\,mJy have larger uncertainties, while our survey
becomes incomplete only at $<\!20$\,mJy. This compares well with the
distribution of rms noise in our survey: the median rms level lies at
3.8\,mJy. The majority of detections have an rms noise level between~2 and
6\,mJy. For comparison, the rms noise for HIPASS typically is 13\,mJy,
though expected to be slightly higher in the Galactic plane. This
indicates that we are sensitive enough to detect $M^*$ galaxies in the GA
region -- contrary to the blind HIPASS.

Comparing integrated flux densities with HIPASS gives a
1$\sigma$-dispersion of 25\% of the flux for 23 measurements (uncertain
measurements excluded). Reducing the acceptable HIPASS error on velocities
from 9\,km\,s$^{-1}$ to 8\,km\,s$^{-1}$ (cf. the comparison of \ion{H}{i}
velocities above) the 1$\sigma$-dispersion is 15\% of the flux (for 19
measurements), which is more comparable with what we find for the
calibrators (cf. Sect.~\ref{obs}). Table~\ref{cxgadet} lists 14 galaxies
offset from the beam centre. A comparison of the corrected fluxes with
HIPASS confirms that corrections for distances up to half the beam width
(i.e., $7\farcm5$) are acceptable.

\section{The non-detections}	\label{ndet}

A further 152 galaxies that were observed in the Crux and GA regions were
not detected. They are listed in Table~\ref{cxgandet} with the searched
velocity range as well as the rms within that interval. Some spectra did
reveal a signal, but careful investigation showed them to be due to close
neighbours rather than the targeted galaxies. These cases are marked with a
plus in Col.~1 of Table~\ref{cxgandet} (see also Table~\ref{cxgadet}), and
the identification of the signal is given in the footnote. Some of these
cases are discussed in detail in Appendix~\ref{notes}.

{\it Column 1 -- 11:} Same as in Table~\ref{cxgadet}. CGMW4\# in Col.~2
stands for the 4th Catalog of Galaxies Behind the Milky Way (Roman et al.
2000).

{\it Column 12:} The searched velocity range in km\,s$^{-1}$.

{\it Column 13:} The rms noise of the searched velocity range in mJy,
typically of the order of 4\,mJy. These values were determined after
baseline fitting over a width of 1600\,km\,s$^{-1}$ -- hence similar to the
determination for detections -- centred at increasingly higher redshifts in
order to obtain values for the whole velocity range. The quoted values
represent the highest rms for the velocity intervals, the rms for the
nearer velocities are on average slightly lower.

{\it Column 14:} Perturbed velocity intervals, mainly due to recurring RFI
around 800, 1250 and $\sim\!7400$\,km\,s$^{-1}$, very strong GPS (Global
Positioning System) signals around $8300-8500$\,km\,s$^{-1}$, and
detections of other galaxies. In these intervals a signal would have gone
undetected.

{\it Column 15:} A star indicates a footnote for this entry.

{\it Column 16:} The distance in arcminutes of the targeted galaxy to the
centre of the beam. Given the falloff of sensitivity with distance from the
centre of the beam, any upper limits for flux densities calculated from the
rms of such cases are underestimated.

{\it Column 17 -- 19:} Independent velocity determinations and error for
the non-detected galaxies. The reference coding (Col.~18) is as in
Table~\ref{refveltab}, and Col.~19 gives the origin of the measurement
(optical or \ion{H}{i}).

Based on this table, only 42\% of the 149 pointings have an unperturbed
velocity range, and 52\% of the pointings are RFI free (i.e., 18 pointings
are affected by the detection of a galaxy, either in the ON or in the OFF
scan). Table~\ref{rfitab} lists the frequencies of the noted velocity
ranges affected by RFIs which affect the possible detection of a
galaxy. Note that the single 10-minute scans show many more RFIs, most of
which, however, could be excised successfully. The worst affected velocity
ranges are either very low ($800-1350$\,km\,s$^{-1}$ for 23\% of pointings)
or around $8300-8500$\,km\,s$^{-1}$ for 14\% of the cases. In 8\% of the
cases the velocity range $7000-7700$\,km\,s$^{-1}$ shows some problems.

WKK\,3836 is the only galaxy that we have not detected but was subsequently
detected by HIPASS. On the one hand, our rms of 7\,mJy is fairly high (only
13\% of our detection have an rms of 7\,mJy and higher) and the pointing is
$d=8\farcm5$ away from the galaxy position (which means the signal would be
reduced by a factor of 2.4). On the other hand, the HIPASS detection is
weak (the peak flux density is 37\,mJy\ and the error on the velocity at
12\,km\,s$^{-1}$ is the second highest value in our sample of HIPASS
galaxies).

\begin{landscape}  
\begin{table}[h]
 \normalsize
 \renewcommand{\baselinestretch}{0.65}
\caption{\ion{H}{i} non-detections in the Crux and Great Attractor region}
\label{cxgandet}
\scriptsize  
\begin{tabular*}{23.5cm}{
 l  @{\extracolsep{2mm}} l @{\extracolsep{3mm}} l@{\extracolsep{-1mm}} l@{\extracolsep{3mm}} 
  l@{\extracolsep{3mm}} l @{\extracolsep{3mm}}  r @{\extracolsep{2mm}} r @{\extracolsep{3mm}} 
  r @{\extracolsep{0mm}}  l @{\extracolsep{0mm}}  c @{\extracolsep{0mm}}   l@{\extracolsep{1mm}}
  r @{\extracolsep{0mm}} l @{\extracolsep{2mm}}
  l @{\extracolsep{0mm}} c @{\extracolsep{1mm}}   
 r @{\extracolsep{2mm}} r @{\extracolsep{3mm}} r @{\extracolsep{2mm}} 
 c @{\extracolsep{1mm}} r @{\extracolsep{3mm}} l @{\extracolsep{2mm}} l @{\extracolsep{3mm}} l                     
}
\noalign{\smallskip}
\hline
\noalign{\smallskip}
 \multicolumn{1}{c}{Ident.} & \multicolumn{1}{c}{Other} & IR & & \multicolumn{1}{c}{R.A.} & \multicolumn{1}{c}{Dec.}
& gal $\ell$ \ & gal $b$ & \multicolumn{4}{l}{Type} & \multicolumn{2}{c}{$D$ x $d$} & 
 $B_{J}$ &  $E_{(B-V)}$ & \multicolumn{1}{c}{$V_{{\rm range}}^{{\rm obs}}$} & rms &
 \multicolumn{1}{c}{$V_{{\rm range}}^{{\rm pert}}$} & N & dist & \multicolumn{1}{c}{$V_{other}$} & Ref & origin\\
& &  &  & (h\,\, m\,\, s) & \ ($\deg$\,\, $\arcmin$\,\, $\arcsec$) & ($\deg$) \ \ & ($\deg$) \ &
& & & & \multicolumn{2}{c}{($\arcsec$)} & ($^{\rm m}$) & ($^{\rm m}$) & 
\multicolumn{1}{c}{km/s} & m\,Jy & \multicolumn{1}{c}{km/s} & & {($\arcmin$)} \ &\multicolumn{1}{c}{km/s} & & \\
\vspace{-1mm} \\
\multicolumn{1}{c}{(1)} & \multicolumn{1}{c}{(2)} & (3) && \multicolumn{1}{c}{(4)} & \multicolumn{1}{c}{(5)} 
& \multicolumn{1}{c}{(6)} & (7) \ & \multicolumn{4}{c}{(8)} \ & \multicolumn{2}{c}{(9)} & (10) &
\multicolumn{1}{c}{(11)} & \multicolumn{1}{c}{(12)} & (13) & \multicolumn{1}{c}{(14)} 
& (15) & (16) & \multicolumn{1}{c}{(17)} & (18) & (19) \\
\noalign{\smallskip}
\hline
\noalign{\smallskip}
WKK0199     &               & & & 11 07 24.8 & -70 22 40 & 294.35 & -9.26 & S&Y& &5  & 58x&\phantom{0}30 & 15.9 & 0.23 &$  600 -\phantom{0}5800 $& 4.8    &                          & &     & & & \\
WKK0331     &ESO063-G019    &M& & 11 44 00.8 & -68 21 35 & 296.76 & -6.30 & S& & &5  & 73x&\phantom{0}23 & 16.0 & 0.35 &$  600 -          13600 $& 4.4    &                          & &     &$12565\pm250$ & FW98 & opt \\
WKK0338     &ESO064-G001    &M& & 11 45 09.1 & -71 50 27 & 297.76 & -9.64 & S& & &4  & 70x&\phantom{0}23 & 16.0 & 0.30 &$  500 -          13350 $& 4.1    &                          & &     &$\phantom{0}7313\pm100$ & FW98 & opt \\
WKK0411     &ESO171-G002    & &I& 11 54 15.3 & -54 09 39 & 294.54 &  7.78 & S& &R&   & 60x&\phantom{0}40 & 15.6 & 0.27 &$  600 -          13600 $& 3.6    &                          & &     &$\phantom{0}5359\pm\phantom{0}77$ & VV92 & opt \\
WKK0425     &               &M& & 11 55 15.5 & -55 20 28 & 294.94 &  6.66 & S&B& &5  & 42x&\phantom{0}38 & 16.3 & 0.27 &$  700 -          10450 $& 3.6    &  $7500 -\phantom{0}7750$ & &     & & & \\
WKK0428     &               & &I& 11 55 24.9 & -56 54 58 & 295.31 &  5.12 &? & & &   & 54x&\phantom{0}16 & 16.8 & 0.44 &$  500 -          10450 $& 2.5    &  $1150 -\phantom{0}1300$ & &     & & & \\
WKK0585     &ESO171-G006    &M& & 12 04 28.0 & -53 56 09 & 295.98 &  8.30 & S& & &5  & 87x&\phantom{0}11 & 16.3 & 0.20 &$  700 -          10350 $& 3.1    &                          & &     &$\phantom{0}8382\pm134$ & WK04 & opt \\
            &               & & &            &           &        &       &  & & &   &    &\phantom{0}   &      &      &$                       $&        &                          & &     &$\phantom{0}8112\pm\phantom{0}85$ & RS06 & opt \\
Y217- 21    &ESO217-G021    &M&I& 12 04 43.7 & -52 28 49 & 295.75 &  9.74 & S& & &4  & 63x&\phantom{00}9 & 17.5L& 0.14 &$  450 -          10450 $& 3.0    &  $7400 -\phantom{0}8750$ & &     &$\phantom{0}8074\pm\phantom{0}39$ & FH95 & opt \\
            &               & & &            &           &        &       &  & & &   &    &\phantom{0}   &      &      &$                       $&        &                          & &     &$\phantom{0}7963\pm\phantom{0}85$ & RS06 & opt \\
WKK0633     &               &M& & 12 06 37.6 & -68 18 44 & 298.79 & -5.80 & S& & &5  & 40x&\phantom{00}5 & 18.1 & 0.35 &$ 3950 -          11100 $& 4.6    &                          & &     & & & \\
WKK0975     &               &M& & 12 23 48.8 & -59 56 41 & 299.48 &  2.74 & S& & &5  & 43x&\phantom{00}7 & 18.3 & 1.22 &$  600 -\phantom{0}5700 $& 4.4    &                          & &     & & & \\
WKK1584     &               &M& & 12 57 51.2 & -56 01 18 & 303.83 &  6.84 & S& & &3  & 40x&\phantom{00}9 & 17.3 & 0.35 &$  600 -\phantom{0}5700 $& 4.2    &                          & &     & & & \\
WKK1597     &               &M& & 12 58 38.9 & -71 35 09 & 303.51 & -8.72 & S& & &   & 56x&\phantom{00}9 & 17.6 & 0.30 &$ 5000 -          10450 $& 3.0    &  $8100 -\phantom{0}8950$ & &     & & & \\
WKK1694+    &ESO173-G002    &M&I& 13 02 22.9 & -56 17 35 & 304.46 &  6.55 & S& & &4: & 78x&\phantom{0}17 & 16.1 & 0.33 &$  600 -          10350 $& 3.1    &  $6350 -\phantom{0}6700$ &*&     &$\phantom{0}5957\pm\phantom{0}85$ & RS06 & opt \\
WKK1707     &               &M& & 13 02 57.6 & -66 03 50 & 304.10 & -3.22 & S& & &   & 26x&\phantom{0}11 & 17.5 & 0.79 &$  450 -          10350 $& 3.9    &                          & &     & & & \\
WKK1806     &               &M& & 13 07 00.4 & -67 19 23 & 304.44 & -4.50 & S& & &   & 54x&\phantom{0}24 & 16.3 & 0.53 &$  600 -          10000 $& 3.7    & $10000 -          10350$ & &     & & & \\
WKK1822     &ESO173-G006    &M&I& 13 07 17.9 & -54 07 35 & 305.28 &  8.67 & S& &R&5  & 60x&\phantom{0}30 & 15.9 & 0.38 &$  400 -          13600 $& 5.3    &                          & &0.1  &$\phantom{0}7730\pm\phantom{0}30$ & WK04 & opt \\
            &               & & &            &           &        &       &  & & &   &    &\phantom{0}   &      &      &$                       $&        &                          & &     &$\phantom{0}8016\pm\phantom{0}70$ & VY96 & opt \\
            &               & & &            &           &        &       &  & & &   &    &\phantom{0}   &      &      &$                       $&        &                          & &     &$\phantom{0}7777\pm\phantom{0}85$ & RS06 & opt \\
WKK1827     &ESO065-G005    &M& & 13 07 52.5 & -71 55 17 & 304.22 & -9.09 & S& & &6  & 51x&\phantom{0}38 & 16.2 & 0.29 &$  400 -          10250 $& 3.1    &  $6950 -\phantom{0}7200$ & &     & & & \\
            &               & & &            &           &        &       &  & & &   &    &\phantom{0}   &      &      &$                       $&        &  $8200 -\phantom{0}8750$ & &     & & & \\
WKK1883     &               &M&I& 13 09 30.4 & -57 28 16 & 305.37 &  5.32 & S& & &3  & 32x&\phantom{0}12 & 17.3 & 0.46 &$  400 -\phantom{0}5700 $& 3.4    &                          & &2.2  &$\phantom{0}6801\pm\phantom{0}70$ & VY96 & opt \\
            &               & & &            &           &        &       &  & & &   &    &\phantom{0}   &      &      &$                       $&        &                          & &     &$\phantom{0}5657\pm\phantom{0}38$ & WK04 & opt \\
            &               & & &            &           &        &       &  & & &   &    &\phantom{0}   &      &      &$                       $&        &                          & &     &$\phantom{0}5666\pm\phantom{0}85$ & RS06 & opt \\
WKK1884     &               & & & 13 09 59.3 & -71 44 07 & 304.40 & -8.92 & S&B& &5  & 51x&\phantom{0}40 & 16.2 & 0.30 &$  600 -          10450 $& 3.8    &  $6950 -\phantom{0}7350$ & &     & & & \\
WKK1888     &               & & & 13 09 44.8 & -57 52 43 & 305.37 &  4.91 & S& & &   & 40x&\phantom{0}22 & 16.6 & 0.44 &$  600 -          10250 $& 3.9    &  $9250 -\phantom{0}9450$ & &     &$\phantom{0}6413\pm 146$ & WK04 & opt \\
WKK1891     &               &M& & 13 10 24.4 & -72 01 37 & 304.41 & -9.21 & S& & &L  & 54x&\phantom{00}8 & 17.9 & 0.25 &$  350 -          10450 $& 4.3    &                          & &     & & & \\
WKK1909     &               & & & 13 10 28.3 & -56 18 13 & 305.59 &  6.47 & S& & &L  & 51x&\phantom{0}28 & 16.5 & 0.59 &$  700 -          10450 $& 3.8    &                          & &     & & & \\
WKK2042     &ESO173-G009    & &I& 13 15 47.3 & -56 29 40 & 306.31 &  6.22 & S& & &L  & 81x&\phantom{0}30 & 15.6 & 0.62 &$  600 -          10450 $& 3.5    & $10050 -          10250$ & &     &$\phantom{0}6012\pm\phantom{0}70$ & FW06 & opt \\
WKK2049     &               & & & 13 15 58.5 & -57 47 29 & 306.21 &  4.92 & S& & &6: & 58x&\phantom{0}11 & 17.2 & 0.59 &$  700 -          10450 $& 4.1    &  $7150 -\phantom{0}7700$ & &     & & & \\
WKK2060     &               &M&I& 13 16 28.6 & -58 34 55 & 306.20 &  4.13 & S& & &7? & 50x&\phantom{0}16 & 16.8 & 0.85 &$  600 -\phantom{0}9850 $& 4.1    &  $7800 -\phantom{0}8750$ & &     & & & \\
WKK2101     &               & &I& 13 18 50.2 & -57 36 09 & 306.61 &  5.07 & I& & &9: & 47x&\phantom{0}24 & 16.7 & 0.75 &$  400 -          10250 $& 3.4    & $10250 -          12550$ &*&     & & & \\
WKK2134     &               & & & 13 21 07.8 & -69 00 07 & 305.60 & -6.28 & S& & &L  & 43x&\phantom{00}9 & 17.8 & 0.52 &$  600 -          10350 $& 3.7    &                          & &     & & & \\
WKK2143     &               &M& & 13 21 46.3 & -56 49 23 & 307.09 &  5.80 & S& & &4  & 59x&\phantom{0}11 & 16.9 & 0.77 &$  200 -          10250 $& 5.6    &   $800 -\phantom{0}1100$ & &     &$\phantom{0}5369\pm199$& FW98 & opt \\
WKK2296     &               &M& & 13 31 02.2 & -55 09 39 & 308.61 &  7.27 & S& & &L  & 58x&\phantom{0}31 & 15.8 & 0.47 &$  600 -          10550 $& 4.6    &                          & &     & & & \\
WKK2300     &               &M& & 13 31 33.1 & -57 50 05 & 308.27 &  4.62 & S& & &5  & 38x&\phantom{0}12 & 17.0 & 0.65 &$  100 -\phantom{0}7050 $&10.7    &                          & &     &$\phantom{0}5897\pm\phantom{0}85$ & RS06 & opt \\
WKK2333     &               & & & 13 33 31.8 & -55 28 34 & 308.92 &  6.91 & S& & &6  & 55x&\phantom{0}13 & 17.0 & 0.50 &$  600 -          10350 $& 5.3    &                          & &     & & & \\
WKK2337     &               &M& & 13 33 44.0 & -55 14 28 & 308.98 &  7.13 & I& & &   & 38x&\phantom{0}20 & 16.9 & 0.49 &$  300 -          10350 $& 4.1    &   $800 -\phantom{0}1100$ & &     & & & \\
            &               & & &            &           &        &       &  & & &   &    &\phantom{0}   &      &      &$                       $&        &  $3700 -\phantom{0}3900$ & &     & & & \\
WKK2376     &               &M& & 13 36 35.1 & -57 19 59 & 309.02 &  5.00 & S& & &L  & 60x&\phantom{0}23 & 16.3 & 0.68 &$  600 -          10450 $& 3.5    &                          & &     & & & \\
WKK2453     &               & & & 13 40 53.2 & -53 15 57 & 310.38 &  8.89 & S& & &4  & 40x&\phantom{0}15 & 16.7 & 0.37 &$  400 -          10250 $& 3.2    &  $8200 -\phantom{0}9050$ & &     & & & \\
WKK2493     &               & &I& 13 44 10.8 & -71 16 13 & 307.18 & -8.84 & S& & &2  & 36x&\phantom{0}30 & 16.1 & 0.37 &$  500 -\phantom{0}5600 $& 4.4    &                          & &0.6  &$\phantom{0}7259\pm\phantom{0}35$ & FH95 & opt \\
            &               & & &            &           &        &       &  & & &   &    &\phantom{0}   &      &      &$                       $&        &                          & &     &$\phantom{0}7220\pm\phantom{0}45$ & FW98 & opt \\
WKK2519     &ESO066-G002    &M& & 13 46 26.7 & -69 42 26 & 307.70 & -7.35 & S&B& &4  & 51x&\phantom{0}28 & 15.8 & 0.30 &$  800 -          10550 $& 2.6    &  $7500 -\phantom{0}7900$ & &     &$\phantom{0}2071\pm250$ & FW98 & opt \\
WKK2564     &               & & & 13 49 49.6 & -57 37 26 & 310.71 &  4.36 & S& & &L  & 43x&\phantom{0}19 & 16.7 & 0.60 &$ 1050 -          10250 $& 4.9    &                          & &     & & & \\
WKK2873     &               & & & 14 03 03.5 & -65 31 58 & 310.26 & -3.69 &  & & &   & 31x&\phantom{0}11 & 17.6 & 0.70 &$  700 -          10350 $& 4.3    &  $1200 -\phantom{0}1350$ & &     & & & \\
            &               & & &            &           &        &       &  & & &   &    &\phantom{0}   &      &      &$                       $&        &  $7700 -\phantom{0}8550$ & &     & & & \\
WKK2874     &               &M&I& 14 02 36.3 & -53 05 52 & 313.62 &  8.29 & S& & &7: & 35x&\phantom{0}19 & 16.8 & 0.49 &$  600 -\phantom{0}5800 $& 4.7    &                          & &1.1  &$\phantom{0}6000\pm\phantom{0}70$ & VY96 & opt \\
            &               & & &            &           &        &       &  & & &   &    &\phantom{0}   &      &      &$                       $&        &                          & &     &$\phantom{0}5819\pm\phantom{0}85$ & RS06 & opt \\
WKK2892     &               &M& & 14 04 07.6 & -52 54 30:& 313.89 &  8.41 & S& & &7  & 67x&\phantom{0}38 & 16.0 & 0.43 &$  300 -          10350 $& 3.9    &  $3550 -\phantom{0}3800$ &*&7.5  &$13844\pm\phantom{0}85$ & RS06 & opt \\
WKK2893     &               &M& & 14 04 14.4 & -71 07 46 & 308.80 & -9.09 & S& & &4  & 55x&\phantom{0}19 & 16.0 & 0.29 &$  700 -          10450 $& 3.9    &  $7150 -\phantom{0}7450$ & &     & & & \\
WKK2895     &               & & & 14 03 56.6 & -66 03 03 & 310.21 & -4.21 & S& & &   & 46x&\phantom{00}9 & 17.4 & 0.51 &$  300 -          10250 $& 4.0    &   $900 -\phantom{0}1150$ & &     & & & \\
WKK2958     &               & & & 14 06 46.0 & -65 06 52 & 310.75 & -3.39 & S&Y& &   & 31x&\phantom{0}17 & 17.1 & 0.71 &$  300 -          10250 $& 6.0    &                          & &     & & & \\
WKK3023     &               & & & 14 10 48.3 & -60 19 07 & 312.61 &  1.05 & S& & &E? & 43x&\phantom{0}15 & 16.6 & 4.32 &$  500 -          10450 $& 4.4    & $10050 -          10250$ & &     & & & \\
WKK3041     &               &M&I& 14 13 08.6 & -71 00 21 & 309.54 & -9.19 & S& &P&   & 47x&\phantom{0}28 & 16.1 & 0.34 &$  600 -\phantom{0}9800 $& 4.1    &  $8400 -\phantom{0}8900$ & &1.3  &$\phantom{0}7533\pm\phantom{0}70$ & VY96 & opt \\
            &               & & &            &           &        &       &  & & &   &    &\phantom{0}   &      &      &$                       $&        &                          & &     &$\phantom{0}7604\pm\phantom{0}70$ & FW98 & opt \\
WKK3072     &               &M&I& 14 15 21.8 & -67 31 31 & 310.84 & -5.95 & S& & &   & 66x&\phantom{00}8 & 17.4 & 0.46 &$  400 -          10450 $& 4.7    &  $8200 -\phantom{0}8550$ & &     & & & \\
WKK3128+    &               & & & 14 19 24.5 & -55 13 42 & 315.32 &  5.51 & S& & &L  & 56x&\phantom{0}15 & 16.7 & 0.63 &$  600 -          10550 $& 3.6    &  $4150 -\phantom{0}4500$ &*&     & & & \\
WKK3136     &               & & & 14 21 09.2 & -71 20 39 & 310.04 & -9.72 &  & & &   & 54x&\phantom{0}22 & 16.3 & 0.25 &$  600 -          10450 $& 3.8    &                          & &     & & & \\
WKK3163     &               & & & 14 24 29.2 & -70 37 27 & 310.56 & -9.14 & S& & &L  & 56x&\phantom{00}7 & 17.6 & 0.32 &$  600 -          10350 $& 3.4    &  $8000 -\phantom{0}8500$ & &     & & & \\
WKK3266     &               &M& & 14 34 37.0 & -52 38 42 & 318.36 &  7.11 & S& & &7: & 55x&\phantom{00}9 & 17.3 & 0.53 &$  400 -          10450 $& 4.6    &                          & &     & & & \\
WKK3279     &               &M&I& 14 36 06.5 & -54 24 51 & 317.87 &  5.39 & S& & & ? & 38x&\phantom{0}12 & 17.2 & 0.75 &$  600 -\phantom{0}9800 $& 5.2    &  $3100 -\phantom{0}3350$ & &1.1  &$\phantom{0}3118\pm\phantom{0}40$ & FH95 & opt \\
WKK3296+    &               &M& & 14 37 31.0 & -53 53 25 & 318.27 &  5.79 & S& & &L  & 60x&\phantom{0}24 & 16.2 & 0.77 &$  400 -          10400 $& 3.7    &  $2950 -\phantom{0}3200$ &*&     & & & \\
WKK3346     &ESO067-G003    &M& & 14 43 45.3 & -70 07 49 & 312.26 & -9.32 & S& & &3: & 63x&\phantom{0}16 & 16.4 & 0.26 &$  400 -          10350 $& 3.1    &  $8000 -\phantom{0}8350$ & &     &$\phantom{0}5147\pm100$ & FW98 & opt \\
\noalign{\smallskip}
\hline
\noalign{\smallskip}
\end{tabular*}
 \normalsize
\end{table}
\addtocounter{table}{-1}
\clearpage
\begin{table}[h]
 \normalsize
 \renewcommand{\baselinestretch}{0.65}
\caption{continued.}
\scriptsize  
\begin{tabular*}{23.5cm}{
 l  @{\extracolsep{2mm}} l @{\extracolsep{3mm}} l@{\extracolsep{-1mm}} l@{\extracolsep{3mm}} 
  l@{\extracolsep{3mm}} l @{\extracolsep{3mm}}  r @{\extracolsep{2mm}} r @{\extracolsep{3mm}} 
  r @{\extracolsep{0mm}}  l @{\extracolsep{0mm}}  c @{\extracolsep{0mm}}   l@{\extracolsep{1mm}}
  r @{\extracolsep{0mm}} l @{\extracolsep{2mm}}
  l @{\extracolsep{0mm}} c @{\extracolsep{1mm}}   
 r @{\extracolsep{2mm}} r @{\extracolsep{3mm}} r @{\extracolsep{2mm}} 
 c @{\extracolsep{1mm}} r @{\extracolsep{3mm}} l @{\extracolsep{2mm}} l @{\extracolsep{3mm}} l                     
}
\noalign{\smallskip}
\hline
\noalign{\smallskip}
 \multicolumn{1}{c}{Ident.} & \multicolumn{1}{c}{Other} & IR & & \multicolumn{1}{c}{R.A.} & \multicolumn{1}{c}{Dec.}
& gal $\ell$ \ & gal $b$ & \multicolumn{4}{l}{Type} & \multicolumn{2}{c}{$D$ x $d$} & 
 $B_{J}$ &  $E_{(B-V)}$ & \multicolumn{1}{c}{$V_{{\rm range}}^{{\rm obs}}$} & rms &
 \multicolumn{1}{c}{$V_{{\rm range}}^{{\rm pert}}$} & N & dist & \multicolumn{1}{c}{$V_{other}$} & Ref & origin\\
& &  &  & (h\,\, m\,\, s) & \ ($\deg$\,\, $\arcmin$\,\, $\arcsec$) & ($\deg$) \ \ & ($\deg$) \ &
& & & & \multicolumn{2}{c}{($\arcsec$)} & ($^{\rm m}$) & ($^{\rm m}$) & 
\multicolumn{1}{c}{km/s} & m\,Jy & \multicolumn{1}{c}{km/s} & & {($\arcmin$)} \ &\multicolumn{1}{c}{km/s} & & \\
\vspace{-1mm} \\
\multicolumn{1}{c}{(1)} & \multicolumn{1}{c}{(2)} & (3) && \multicolumn{1}{c}{(4)} & \multicolumn{1}{c}{(5)} 
& \multicolumn{1}{c}{(6)} & (7) \ & \multicolumn{4}{c}{(8)} \ & \multicolumn{2}{c}{(9)} & (10) &
\multicolumn{1}{c}{(11)} & \multicolumn{1}{c}{(12)} & (13) & \multicolumn{1}{c}{(14)} 
& (15) & (16) & \multicolumn{1}{c}{(17)} & (18) & (19) \\
\noalign{\smallskip}
\hline
\noalign{\smallskip}
WKK3357     &               &M& & 14 44 44.8 & -67 34 22 & 313.44 & -7.04 & S&B& &5  & 56x&\phantom{0}46 & 15.4 & 0.34 &$  550 -          10450 $& 4.0    &                          & &     & & & \\
WKK3372     &               &M& & 14 46 38.7 & -66 26 58 & 314.09 & -6.11 & S& & &5  & 67x&\phantom{0}32 & 15.2 & 0.30 &$  550 -          10450 $& 3.5    &  $7050 -\phantom{0}7700$ & &     & & & \\
WKK3823     &               & & & 13 57 33.1 & -52 06 43 & 313.13 &  9.45 & S& & &M: & 20x&\phantom{00}6 & 18.6 & 0.44 &$  300 -          10550 $& 7.3    &   $650 -\phantom{0}1000$ &*&8.7  &$27483\pm\phantom{0}85$ & RS06 & opt \\
WKK3836     &               & & & 14 00 06.4 & -52 03 02 & 313.53 &  9.40 & S& & &L  & 83x&\phantom{0}15 & 16.4 & 0.42 &$  300 -          10550 $& 7.3    &   $650 -\phantom{0}1000$ &*&8.5  &$\phantom{0}3612\pm\phantom{0}12$ & HIPASS & \ion{H}{i} \\
WKK3991     &               &M&I& 14 27 30.8 & -52 07 15 & 317.55 &  8.00 & S& & &M  & 34x&\phantom{00}8 & 18.0 & 0.40 &$  600 -\phantom{0}5700 $& 6.1    &                          & &0.8  &$17164\pm\phantom{0}52$&  FH95 & opt \\
            &               & & &            &           &        &       &  & & &   &    &\phantom{0}   &      &      &$                       $&        &                          & &     &$16968\pm\phantom{0}85$ & RS06 & opt \\
WKK4751     &               &M& & 15 14 36.0 & -48 12 41 & 326.11 &  8.11 & S&Y& &5  & 56x&\phantom{0}19 & 16.5 & 0.40 &$  400 -          10250 $& 5.1    &   $650 -\phantom{0}1100$ &*&2.9  & & & \\
WKK4755     &               &M& & 15 14 50.4 & -48 12 44 & 326.15 &  8.09 & S& & &M: & 58x&\phantom{0}35 & 15.8 & 0.40 &$  400 -          10250 $& 5.1    &   $650 -\phantom{0}1100$ &*&2.9  & & & \\
Y68-3L      &ESO068-G003    & &I& 15 20 15   & -69 02 52:& 315.61 & -9.92 & S& & &   & 72x&\phantom{0}24 & 15.6N& 0.11 &$  300 -\phantom{0}5500 $& 5.4    &                          & &1.1  &$\phantom{0}4981\pm\phantom{0}37$ & FH95 & opt \\
WKK4907     &               &M&I& 15 25 06.2 & -48 49 03 & 327.26 &  6.66 & S& & &5  & 40x&\phantom{0}16 & 16.9 & 0.39 &$  600 -\phantom{0}9700 $& 7.9    &                          & &0.8  &$10525\pm\phantom{0}70$ & VY96 & opt \\
WKK4989     &               & & & 15 29 48.8 & -53 59 34 & 324.95 &  1.97 & S& & &   & 43x&\phantom{0}11 & 17.1 & 2.69 &$  300 -          10250 $& 4.0    &  $1100 -\phantom{0}1350$ & &     & & & \\
WKK5039     &               & & & 15 32 36.7 & -62 52 33 & 320.16 & -5.53 & S& & &L? & 42x&\phantom{0}20 & 16.3 & 0.48 &$  400 -          10250 $& 3.6    &  $1100 -\phantom{0}1300$ & &     & & & \\
WKK5061     &               & & & 15 33 29.7 & -60 21 38 & 321.71 & -3.54 & S& & &   & 23x&\phantom{0}13 & 17.0 & 0.83 &$  600 -          10350 $& 4.0    &                          & &     & & & \\                   
WKK5124     &               & & & 15 37 12.1 & -60 25 09 & 322.05 & -3.85 &?I& & &   & 24x&\phantom{0}20 & 16.8 & 0.71 &$  400 -          10250 $& 5.6    &                          & &     & & & \\
WKK5131     &               &M& & 15 37 41.6 & -62 33 06 & 320.83 & -5.61 &  & & &   & 20x&\phantom{0}12 & 17.4 & 0.50 &$  300 -          10250 $& 5.0    &                          & &     &$14498\pm\phantom{0}54$ & WK04 & opt \\
WKK5138     &               &M& & 15 38 05.3 & -59 47 13 & 322.51 & -3.41 & S& & &   & 46x&\phantom{0}35 & 15.5 & 1.02 &$  200 -          10250 $& 3.6    &   $650 -\phantom{0}1000$ & &     & & & \\
WKK5186     &               &M&I& 15 40 44.8 & -60 05 22 & 322.60 & -3.85 & S& & &6  & 46x&\phantom{0}26 & 15.5 & 0.74 &$  600 -          10250 $& 5.7    &                          & &0.7  &$\phantom{0}5151\pm\phantom{0}70$ & VY96 & opt \\
WKK5195     &               &M&I& 15 41 11.6 & -59 58 50 & 322.71 & -3.80 & S& & &   & 28x&\phantom{0}17 & 16.7 & 0.73 &$  300 -          10250 $& 5.9    &                          & &     &$\phantom{0}5918\pm\phantom{0}70$ & VY96 & opt \\
WKK5196     &               & & & 15 41 14.9 & -60 38 34 & 322.31 & -4.33 & S& & &   & 24x&\phantom{00}7 & 17.6 & 0.59 &$  300 -          10250 $& 5.7    &                          & &     & & & \\
WKK5251     &               & & & 15 45 16.8 & -59 21 45 & 323.49 & -3.62 & S& & &   & 20x&\phantom{00}8 & 17.9 & 0.80 &$  300 -          10250 $& 6.1    &                          & &     & & & \\
WKK5267+    &               & & & 15 46 47.8 & -66 24 04 & 319.25 & -9.26 & S& & &   & 62x&\phantom{00}8 & 16.9 & 0.13 &$  600 -          10450 $& 3.4    &  $4550 -\phantom{0}5050$ &*&     & & & \\
WKK5268     &               &M& & 15 46 19.5 & -58 53 20 & 323.89 & -3.33 & S& & &   & 24x&\phantom{0}11 & 16.9 & 0.61 &$  300 -          10350 $& 6.3    &                          & &     &$\phantom{0}5541\pm\phantom{0}46$ & WK04 & opt \\
WKK5288     &               &M& & 15 47 21.5 & -60 14 18 & 323.16 & -4.47 & S& & &L  & 52x&\phantom{0}19 & 15.9 & 0.58 &$  600 -          10350 $& 5.4    &                          & &     & & & \\
WKK5297+    &               & & & 15 47 41.7 & -58 59 27 & 323.97 & -3.52 & S& & &7  & 51x&\phantom{0}19 & 15.8 & 0.60 &$  600 -          10450 $& 5.3    &  $5300 -\phantom{0}5650$ &*&     &$\phantom{0}4343\pm\phantom{0}52$ & WK04 & opt \\
            &               & & &            &           &        &       &  & & &   &    &\phantom{0}   &      &      &$                       $&        &  $5950 -\phantom{0}6150$ & &     & & & \\
WKK5330     &               & & & 15 49 06.3 & -60 42 39 & 323.03 & -4.97 & S& & &9  & 63x&\phantom{0}34 & 15.1 & 0.61 &$  600 -          10550 $& 3.8    &                          & &     & & & \\
WKK5334     &               &M& & 15 49 19.8 & -61 55 59 & 322.29 & -5.94 & S& & & ? & 44x&\phantom{0}20 & 15.8 & 0.35 &$  300 -          10350 $& 4.2    &  $8350 -\phantom{0}8650$ & &     &$\phantom{0}6038\pm\phantom{0}33$ & WK04 & opt \\
WKK5349     &               &M&I& 15 49 42.0 & -61 00 34 & 322.90 & -5.25 & S& & &5: & 38x&\phantom{00}9 & 16.9 & 0.48 &$  300 -          10250 $& 4.2    &                          & &     & & & \\
WKK5357     &               &M& & 15 49 52.4 & -58 42 56 & 324.36 & -3.48 & S& & &M  & 44x&\phantom{0}23 & 15.8 & 0.63 &$  600 -          10350 $& 4.3    &                          & &     &$\phantom{0}5447\pm\phantom{0}44$ & WK04 & opt \\
WKK5381+    &               &D& & 15 51 05.4 & -58 35 28 & 324.56 & -3.48 & S& & &   & 32x&\phantom{0}15 & 16.7 & 0.64 &$  300 -          10150 $& 3.8    &  $2000 -\phantom{0}2150$ &*&     &$\phantom{0}5518\pm\phantom{0}98$ & WK04 & opt \\
WKK5424     &               & &I& 15 52 54.2 & -58 34 53 & 324.75 & -3.62 &?S& & &   & 19x&\phantom{0}16 & 17.1 & 0.62 &$  300 -          10350 $& 4.7    &  $4350 -\phantom{0}4600$ & &     & & & \\
WKK5443+    &               &M& & 15 53 37.7 & -58 41 30 & 324.76 & -3.77 & S& & &5: & 38x&\phantom{0}12 & 16.6 & 0.75 &$  150 -          10350 $& 4.0    &  $2850 -\phantom{0}3050$ &*&     &$\phantom{0}5069\pm\phantom{0}40$ & WK04 & opt \\
WKK5482     &               & &I& 15 54 51.3 & -56 52 08 & 326.05 & -2.47 &?I& & & ? & 71x&\phantom{0}60 & 14.9 & 1.14 &$  200 -          10350 $& 4.2    &  $1150 -\phantom{0}1450$ & &     & & & \\
WKK5490     &               &M& & 15 55 23.9 & -58 14 30 & 325.22 & -3.57 & S& & &   & 27x&\phantom{0}12 & 17.1 & 0.74 &$  600 -          10550 $& 4.5    &                          & &     & & & \\
WKK5528     &               &M& & 15 56 49.3 & -61 48 59 & 323.04 & -6.42 & S& & &5: & 60x&\phantom{00}9 & 16.7 & 0.47 &$  600 -          10250 $& 4.2    &                          & &     & & & \\
WKK5534+    &               & & & 15 56 58.6 & -59 06 26 & 324.82 & -4.36 & S& & &9: & 43x&\phantom{0}26 & 15.8 & 0.56 &$  300 -          10350 $& 4.1    &  $5300 -\phantom{0}5800$ &*&     & & & \\
WKK5544+    &               &M& & 15 57 33.1 & -63 21 41 & 322.10 & -7.65 & S& & &M  & 51x&\phantom{0}13 & 16.1 & 0.26 &$  100 -          10250 $& 3.5    &  $4450 -\phantom{0}4850$ &*&     & & & \\
            &               & & &            &           &        &       &  & & &   &    &\phantom{0}   &      &      &$                       $&        &  $6850 -\phantom{0}7150$ & &     & & & \\
            &               & & &            &           &        &       &  & & &   &    &\phantom{0}   &      &      &$                       $&        &  $8000 -\phantom{0}8700$ & &     & & & \\
WKK5556+    &               &M& & 15 57 33.3 & -59 01 06 & 324.93 & -4.34 & S& & &   & 22x&\phantom{0}15 & 16.6 & 0.52 &$ 1100 -\phantom{0}8100 $& 4.6    &  $5400 -\phantom{0}5900$ &*&     &$\phantom{0}4658\pm162$ & WK99 & opt \\
WKK5563     &               &M& & 15 57 54.4 & -60 53 18 & 323.75 & -5.80 & S& & &   & 35x&\phantom{0}15 & 16.5 & 0.45 &$  600 -          10550 $& 3.6    &                          & &     &$13060\pm102$ & WK04 & opt \\
WKK5568     &               &M& & 15 58 13.2 & -62 53 13 & 322.47 & -7.34 & S& & &3: & 65x&\phantom{0}16 & 15.6 & 0.34 &$  300 -          10250 $& 3.5    &   $850 -\phantom{0}1000$ & &     &$\phantom{0}7460\pm\phantom{0}70$ & DN97 & opt \\
WKK5581+    &               & & & 15 58 53.3 & -66 17 55 & 320.26 & -9.96 & S&B& &4: & 58x&\phantom{0}39 & 14.8 & 0.11 &$  400 -          10250 $& 4.1    &  $3450 -\phantom{0}3900$ &*&     &$\phantom{0}9750\pm\phantom{0}70$ & DN97 & opt \\
            &               & & &            &           &        &       &  & & &   &    &\phantom{0}   &      &      &$                       $&        &  $6800 -\phantom{0}7150$ & &     & & & \\
            &               & & &            &           &        &       &  & & &   &    &\phantom{0}   &      &      &$                       $&        &  $8350 -\phantom{0}8700$ & &     & & & \\
WKK5595     &               & & & 15 58 40.7 & -57 27 10 & 326.07 & -3.25 & I& & & ? & 30x&\phantom{0}19 & 16.8 & 0.66 &$  300 -          10250 $& 5.5    &   $800 -\phantom{0}1100$ & &0.7  & & & \\
WKK5596     &               & & & 15 59 16.0 & -65 02 13 & 321.13 & -9.03 & S& & &L  & 50x&\phantom{0}26 & 15.4 & 0.13 &$  400 -          10250 $& 3.2    &  $1100 -\phantom{0}1300$ & &     & & & \\
WKK5597     &               & & & 15 58 45.7 & -57 27 26 & 326.07 & -3.26 & S& & &L? & 28x&\phantom{0}11 & 17.3 & 0.67 &$  300 -          10250 $& 5.5    &   $800 -\phantom{0}1100$ & &     & & & \\
WKK5615     &               &M&I& 15 59 22.2 & -60 01 24 & 324.45 & -5.26 & S& & &M  & 42x&\phantom{0}12 & 16.5 & 0.38 &$  600 -\phantom{0}7700 $& 5.7    &                          & &     &$\phantom{0}3936\pm\phantom{0}37$ & FH95 & opt \\
            &               & & &            &           &        &       &  & & &   &    &\phantom{0}   &      &      &$                       $&        &                          & &     &$\phantom{0}3867\pm\phantom{0}70$ & WK99 & opt \\
WKK5647     &               &M& & 16 00 14.4 & -60 22 22 & 324.31 & -5.59 & S& & &L  & 44x&\phantom{0}31 & 15.4 & 0.35 &$  400 -          10250 $& 4.6    &                          & &     & & & \\
WKK5650     &               & &I& 16 00 18.1 & -60 25 20 & 324.28 & -5.63 & S& & &M  & 52x&\phantom{0}19 & 15.6 & 0.34 &$  300 -          10350 $& 4.5    &  $8300 -\phantom{0}8600$ & &     & & & \\
WKK5693     &               &M& & 16 01 36.8 & -59 04 44 & 325.29 & -4.73 & S& & & ? & 42x&\phantom{0}24 & 15.6 & 0.37 &$  300 -          10250 $& 5.9    &   $650 -\phantom{0}1000$ & &     & & & \\
WKK5694+    &               &D&I& 16 01 48.3 & -61 08 54 & 323.94 & -6.30 & S& & &M  & 66x&\phantom{0}38 & 14.7 & 0.33 &$  400 -\phantom{0}9700 $& 4.0    &  $8100 -\phantom{0}8950$ &*&     &$\phantom{0}3412\pm\phantom{0}36$ & FH95 & opt \\
            &               & & &            &           &        &       &  & & &   &    &\phantom{0}   &      &      &$                       $&        &                          & &     &$\phantom{0}3505\pm\phantom{0}50$ & WK04 & opt \\
WKK5709+    &               &M& & 16 02 12.6 & -60 59 10 & 324.08 & -6.21 & S&D& & ? & 46x&\phantom{0}26 & 15.5 & 0.32 &$  200 -          10250 $& 5.0    &  $5600 -\phantom{0}5800$ &*&     & & & \\
WKK5733++   &               &M& & 16 02 50.2 & -61 10 07 & 324.02 & -6.40 & S& & &3  & 54x&\phantom{00}9 & 16.4 & 0.31 &$ 3650 -\phantom{0}8950 $& 3.8    &  $4150 -\phantom{0}4500$ &*&     &$\phantom{0}6215\pm\phantom{0}92$ & WK99 & opt \\
            &               & & &            &           &        &       &  & & &   &    &\phantom{0}   &      &      &$                       $&        &  $5500 -\phantom{0}5850$ & &     & & & \\
WKK5780+    &               & & & 16 04 11.7 & -61 07 46 & 324.17 & -6.48 & S& & &6  & 56x&\phantom{0}35 & 15.2 & 0.29 &$  300 -          12950 $& 4.4    &  $5100 -\phantom{0}5700$ &*&     & & & \\
WKK5791     &               & & & 16 04 11.9 & -59 09 28 & 325.49 & -5.01 & S& & &5: & 95x&\phantom{0}16 & 15.2 & 0.32 &$  200 -          10250 $& 4.7    &   $800 -\phantom{0}1000$ & &     & & & \\
WKK5801     &               &M& & 16 04 24.8 & -59 40 19 & 325.16 & -5.41 & S& & &M  & 44x&\phantom{0}26 & 15.3 & 0.34 &$  600 -          10450 $& 4.5    &                          & &     & & & \\
WKK5805+    &ESO136-G018    &M& & 16 04 37.5 & -60 46 32 & 324.44 & -6.25 & S& & &5  & 71x&\phantom{0}16 & 15.5 & 0.30 &$ 3450 -\phantom{0}8950 $& 5.6    &                          & &     &$\phantom{0}6207\pm230$ & WK99 & opt \\
WKK5923     &               &M&I& 16 07 53.1 & -59 13 37 & 325.79 & -5.38 & S& &R&3  & 50x&\phantom{0}28 & 15.2 & 0.38 &$ 2400 -\phantom{0}7900 $& 5.3    &                          & &     &$\phantom{0}4514\pm\phantom{0}42$ & SH92 & opt \\
WKK5928     &ESO100-G011    & & & 16 08 20.8 & -62 57 39 & 323.28 & -8.16 & S& & &5  & 90x&\phantom{0}28 & 14.6 & 0.18 &$  400 -          10250 $& 4.2    &                          & &     &$\phantom{0}8354\pm130$ & WK04 & opt \\
WKK5971     &               &M& & 16 09 24.3 & -62 41 18 & 323.56 & -8.04 &  & & &   & 43x&\phantom{0}23 & 15.9 & 0.20 &$  300 -          10250 $& 4.1    &  $8350 -\phantom{0}8650$ & &     & & & \\
\noalign{\smallskip}
\hline
\noalign{\smallskip}
\end{tabular*}
 \normalsize
\end{table}
\addtocounter{table}{-1}
\clearpage
\begin{table}[h]
 \normalsize
 \renewcommand{\baselinestretch}{0.65}
\caption{continued.}
\scriptsize  
\begin{tabular*}{23.5cm}{
 l  @{\extracolsep{2mm}} l @{\extracolsep{3mm}} l@{\extracolsep{-1mm}} l@{\extracolsep{3mm}} 
  l@{\extracolsep{3mm}} l @{\extracolsep{3mm}}  r @{\extracolsep{2mm}} r @{\extracolsep{3mm}} 
  r @{\extracolsep{0mm}}  l @{\extracolsep{0mm}}  c @{\extracolsep{0mm}}   l@{\extracolsep{1mm}}
  r @{\extracolsep{0mm}} l @{\extracolsep{2mm}}
  l @{\extracolsep{0mm}} c @{\extracolsep{1mm}}   
 r @{\extracolsep{2mm}} r @{\extracolsep{3mm}} r @{\extracolsep{2mm}} 
 c @{\extracolsep{1mm}} r @{\extracolsep{3mm}} l @{\extracolsep{2mm}} l @{\extracolsep{3mm}} l                     
}
\noalign{\smallskip}
\hline
\noalign{\smallskip}
 \multicolumn{1}{c}{Ident.} & \multicolumn{1}{c}{Other} & IR & & \multicolumn{1}{c}{R.A.} & \multicolumn{1}{c}{Dec.}
& gal $\ell$ \ & gal $b$ & \multicolumn{4}{l}{Type} & \multicolumn{2}{c}{$D$ x $d$} & 
 $B_{J}$ &  $E_{(B-V)}$ & \multicolumn{1}{c}{$V_{{\rm range}}^{{\rm obs}}$} & rms &
 \multicolumn{1}{c}{$V_{{\rm range}}^{{\rm pert}}$} & N & dist & \multicolumn{1}{c}{$V_{other}$} & Ref & origin\\
& &  &  & (h\,\, m\,\, s) & \ ($\deg$\,\, $\arcmin$\,\, $\arcsec$) & ($\deg$) \ \ & ($\deg$) \ &
& & & & \multicolumn{2}{c}{($\arcsec$)} & ($^{\rm m}$) & ($^{\rm m}$) & 
\multicolumn{1}{c}{km/s} & m\,Jy & \multicolumn{1}{c}{km/s} & & {($\arcmin$)} \ &\multicolumn{1}{c}{km/s} & & \\
\vspace{-1mm} \\
\multicolumn{1}{c}{(1)} & \multicolumn{1}{c}{(2)} & (3) && \multicolumn{1}{c}{(4)} & \multicolumn{1}{c}{(5)} 
& \multicolumn{1}{c}{(6)} & (7) \ & \multicolumn{4}{c}{(8)} \ & \multicolumn{2}{c}{(9)} & (10) &
\multicolumn{1}{c}{(11)} & \multicolumn{1}{c}{(12)} & (13) & \multicolumn{1}{c}{(14)} 
& (15) & (16) & \multicolumn{1}{c}{(17)} & (18) & (19) \\
\noalign{\smallskip}
\hline
\noalign{\smallskip}
WKK5983     &               &M& & 16 09 52.5 & -65 00 48 & 321.99 & -9.77 & S& & &   & 47x&\phantom{0}30 & 15.5 & 0.14 &$  400 -          10250 $& 4.3    &  $1100 -\phantom{0}1300$ & &     &$15050\pm\phantom{0}70$ & DN97 & opt \\
            &               & & &            &           &        &       &  & & &   &    &\phantom{0}   &      &      &$                       $&        &                          & &     &$14838\pm178$ & WK99 & opt \\
WKK6007     &ESO136-G021    &M& & 16 10 09.5 & -61 25 60 & 324.49 & -7.19 & S&B& &2  & 75x&\phantom{0}58 & 14.1 & 0.23 &$  100 -          10250 $& 4.2    &  $5750 -\phantom{0}5950$ & &     & & & \\
WKK6055     &               & & & 16 10 50.2 & -56 25 59 & 327.98 & -3.59 & S& & &L  & 43x&\phantom{0}17 & 16.7 & 0.58 &$  200 -          10250 $& 3.8    &   $900 -\phantom{0}1050$ & &     & & & \\
WKK6090     &               &D& & 16 11 51.8 & -61 11 42 & 324.81 & -7.15 & S&Y& &5  & 31x&\phantom{0}20 & 16.2 & 0.23 &$  600 -          10350 $& 5.1    &                          & &     & & & \\
WKK6092+    &               &M& & 16 11 51.4 & -60 37 55 & 325.20 & -6.74 & S&B& &2  & 56x&\phantom{0}47 & 14.7 & 0.19 &$  600 -          10350 $& 2.1    &  $4900 -\phantom{0}5100$ &*&     &$\phantom{0}4688\pm\phantom{0}38$ & WK99 & opt \\
            &               & & &            &           &        &       &  & & &   &    &\phantom{0}   &      &      &$                       $&        &  $8200 -\phantom{0}8750$ & &     &$\phantom{0}4711\pm\phantom{0}58$ & WK04 & opt \\
WKK6154     &               & & & 16 12 41.8 & -56 54 09 & 327.85 & -4.11 & S& & &L  & 32x&\phantom{0}19 & 16.7 & 0.50 &$  200 -          10350 $& 3.7    &   $900 -\phantom{0}1050$ & &     & & & \\
            &               & & &            &           &        &       &  & & &   &    &\phantom{0}   &      &      &$                       $&        &  $8050 -\phantom{0}8850$ & &     & & & \\
WKK6162     &               &M& & 16 13 13.8 & -62 34 54 & 323.96 & -8.27 & S&Y& &   & 44x&\phantom{0}17 & 15.9 & 0.17 &$  100 -          10350 $& 2.1    &                          & &     & & & \\
WKK6176     &ESO136-G001    &M&I& 16 13 27.2 & -60 45 51 & 325.25 & -6.98 & S& & &5  & 86x&\phantom{0}31 & 14.6 & 0.21 &$  300 -          10350 $& 6.8    &                          & &     &$\phantom{0}4630\pm\phantom{0}58$ & WK99 & opt \\
WKK6187     &               &D& & 16 13 42.4 & -62 17 14 & 324.20 & -8.09 &  & & &   & 22x&\phantom{0}22 & 16.8 & 0.15 &$  600 -          10550 $& 5.2    &  $7400 -\phantom{0}7600$ & &0.9  && & \\
WKK6189     &               & & & 16 13 44.9 & -62 17 04 & 324.21 & -8.10 & E& & &   & 13x&\phantom{00}8 & 17.3 & 0.14 &$  600 -          10550 $& 5.2    &  $7400 -\phantom{0}7600$ & &     & & & \\
WKK6249     &               &M& & 16 14 53.5 & -62 53 09 & 323.88 & -8.62 & S& & &5  & 51x&\phantom{0}20 & 15.7 & 0.14 &$  300 -          10350 $& 3.1    &  $8350 -\phantom{0}8650$ & &     & & & \\      
WKK6251     &               &M& & 16 14 45.0 & -60 55 36 & 325.25 & -7.20 & S& & &4  & 36x&\phantom{0}18 & 15.9 & 0.20 &$ 1350 -          10250 $&27.6    &  $8350 -\phantom{0}8750$ &*&     &$\phantom{0}5689\pm100$ & WK99 & opt \\
WKK6286     &               & & & 16 15 14.1 & -60 16 59 & 325.74 & -6.78 & S& & &   & 20x&\phantom{00}5 & 18.5 & 0.27 &$ 2400 -\phantom{0}7900 $& 4.0    &                          & &7.4  & & & \\
WKK6368     &ESO137-G013    &M& & 16 16 51.5 & -62 21 06 & 324.42 & -8.40 & S&Y& &5  & 74x&\phantom{0}19 & 15.5 & 0.13 &$  600 -          10450 $& 4.4    &                          & &     &$13242\pm\phantom{0}34$ & WK04 & opt \\
WKK6459     &               &M& & 16 18 26.7 & -60 46 47 & 325.68 & -7.41 & S& & &5: & 62x&\phantom{0}43 & 14.8 & 0.26 &$  200 -          10250 $& 6.5    &  $1100 -\phantom{0}1400$ & &     &$\phantom{0}5268\pm\phantom{0}35$ & WK08 & opt \\
WKK6541     &   	    &M& & 16 20 00.9 & -60 00 47 & 326.36 & -7.01 & S&Y& &M  & 42x&\phantom{0}27 & 15.6 & 0.21 &$  300 -          10250 $& 2.6    &  $8350 -\phantom{0}8550$ & &     & & & \\
WKK6640     &CSRG 0801      &M& & 16 21 46.8 & -61 25 31 & 325.50 & -8.15 & S& &R&   & 56x&\phantom{0}43 & 14.9 & 0.20 &$  400 -          10250 $& 3.3    &  $1150 -\phantom{0}1450$ & &     &$\phantom{0}4147\pm\phantom{0}71$ & WK04 & opt \\
WKK6663     &               &M& & 16 22 20.3 & -61 21 28 & 325.60 & -8.15 & S& & &2  & 85x&\phantom{0}20 & 15.0 & 0.21 &$  300 -          10350 $& 5.0    &  $1150 -\phantom{0}1300$ & &     &$\phantom{0}3062\pm\phantom{0}70$ & WK99 & opt \\
WKK6768     &               & & & 16 24 25.1 & -56 18 41 & 329.42 & -4.82 & S& & &5: & 39x&\phantom{0}19 & 16.2 & 0.36 &$  300 -          10350 $&11.5    &   $700 -\phantom{0}1000$ & &     & & & \\
WKK6830     &               & & & 16 25 47.8 & -56 10 55 & 329.65 & -4.87 & S& & &M: & 35x&\phantom{0}19 & 16.2 & 0.37 &$  300 -          10350 $& 4.7    &   $800 -\phantom{0}1050$ & &     & & & \\
            &               & & &            &           &        &       &  & & &   &    &\phantom{0}   &      &      &$                       $&        &  $8350 -\phantom{0}8700$ & &     & & & \\
WKK7037     &               &M& & 16 31 05.3 & -59 49 06 & 327.47 & -7.87 & S& & &5  & 42x&\phantom{0}15 & 16.5 & 0.24 &$ 2500 -\phantom{0}7800 $& 5.3    &                          & &     &$18722\pm\phantom{0}71$ & WK04 & opt \\
WKK7042     &               &M& & 16 31 00.4 & -56 45 35 & 329.72 & -5.79 & S& & &5: & 78x&\phantom{0}42 & 14.6 & 0.35 &$  300 -          10250 $& 5.2    &   $850 -\phantom{0}1050$ & &     & & & \\
            &               & & &            &           &        &       &  & & &   &    &\phantom{0}   &      &      &$                       $&        &  $4800 -\phantom{0}5100$ & &     & & & \\
WKK7175+    &               &M& & 16 35 41.2 & -60 09 36 & 327.60 & -8.53 & S& & &5: & 39x&\phantom{00}9 & 16.8 & 0.24 &$ 1600 -\phantom{0}6750 $& 6.7    &  $3250 -\phantom{0}3500$ &*&     & & & \\
WKK7177     &ESO137-G035    &M& & 16 35 50.8 & -61 27 60 & 326.63 & -9.41 & S& & &3  &112x&\phantom{0}36 & 14.3 & 0.25 &$ 2400 -\phantom{0}7900 $& 4.9    &                          & &     &$\phantom{0}4780\pm\phantom{0}60$ & RC3 & opt \\
WKK7248+    &ESO137-G037    &M& & 16 39 18.5 & -59 53 11 & 328.11 & -8.69 & S& & &3  &106x&\phantom{0}22 & 14.8 & 0.28 &$ 2400 -\phantom{0}7900 $& 4.4    &  $5450 -\phantom{0}5600$ &*&     &$\phantom{0}5430\pm\phantom{0}50$ & RC3 & opt \\
WKK7320     &               & & & 16 42 05.3 & -55 49 11 & 331.46 & -6.30 & S& & &5  & 67x&\phantom{0}24 & 15.3 & 0.31 &$  200 -          10250 $& 3.5    &                          & &     & & & \\
WKK7375     &               &M& & 16 44 37.9 & -60 51 05 & 327.80 & -9.81 & S& & &   & 70x&\phantom{0}22 & 15.1 & 0.26 &$  300 -          10250 $& 4.6    &   $800 -\phantom{0}1000$ & &     & & & \\
Y453-3      &ESO453-G003    &M&I& 16 46 35.2 & -31 35 21 & 350.44 &  8.86 & S& & &3  & 60x&\phantom{0}30 & 13.6I& 0.43 &$  600 -\phantom{0}5800 $& 4.6    &                          & &0.1  &$\phantom{0}6644\pm\phantom{0}10$ & MF96 & opt \\
WKK7625     &               &M& & 16 52 14.6 & -59 40 10 & 329.34 & -9.80 & S&B& &5  & 65x&\phantom{0}54 & 14.8 & 0.21 &$  100 -          10250 $& 2.8    &  $1200 -\phantom{0}1350$ & &     &$14278\pm\phantom{0}40$ & WK04 & opt \\
            &               & & &            &           &        &       &  & & &   &    &\phantom{0}   &      &      &$                       $&        &                          & &     &$14322\pm\phantom{0}85$ & RS06 & opt \\
WKK7635     &               & & & 16 52 14.4 & -55 16 39 & 332.80 & -7.06 &?I& & & ? & 78x&\phantom{0}54 & 14.7 & 0.26 &$  300 -          10250 $& 4.7    &                          & &     & & & \\
WKK7689+    &               & & & 16 53 48.6 & -59 05 24 & 329.92 & -9.60 & I& & &   & 74x&\phantom{0}43 & 15.2 & 0.13 &$  300 -          10250 $& 6.5    &  $1300 -\phantom{0}1700$ &*&     &$\phantom{0}1559\pm\phantom{00}3$ & HK01 & \ion{H}{i} \\
            &               & & &            &           &        &       &  & & &   &    &\phantom{0}   &      &      &$                       $&        &                          & &     &$\phantom{0}1655\pm\phantom{000}$ & KD04 & \ion{H}{i} \\
WKK7726     &               & & & 16 54 50.5 & -55 17 03 & 333.03 & -7.35 & S&B& &   &105x&\phantom{0}74 & 14.1 & 0.30 &$  500 -          10250 $& 3.8    &  $1200 -\phantom{0}1350$ & &     & & & \\
Y453-11     &ESO453-G011    &M&I& 16 56 42.5 & -31 37 16 & 351.78 &  7.16 & S& & &3  & 78x&\phantom{0}18 & 15.8N& 0.33 &$  500 -\phantom{0}9700 $& 6.3    &                          & &0.1  &$\phantom{0}7028\pm\phantom{00}1$ & HW00 & opt \\
WKK7794+    &               &M&I& 16 58 40.8 & -58 29 06 & 330.80 & -9.73 & S&B& &4  & 51x&\phantom{0}44 & 15.1 & 0.19 &$  500 -\phantom{0}5800 $& 5.4    &                          &*&0.8  &$\phantom{0}5854\pm\phantom{0}70$ & VY96 & opt \\
WKK7853     &               & & & 17 01 44.2 & -52 57 14 & 335.51 & -6.72 & S& & &5  & 52x&\phantom{0}15 & 16.2 & 0.39 &$  300 -          10350 $& 8.8    &   $700 -\phantom{0}1050$ & &     & & & \\
Y17030-48   &               &M&I& 17 06 43.5 & -48 23 59 & 339.64 & -4.61 & S& & &   & 48x&\phantom{000} &      & 0.86 &$  400 -\phantom{0}9800 $& 5.5    &  $8750 -\phantom{0}8900$ & &0.5  & & & \\
Y519-9      &ESO519-G009    & &I& 17 11 44.2 & -25 44 48 & 358.51 &  8.03 & S& & &5  & 73x&\phantom{00}8 & 15.6N& 0.49 &$  500 -\phantom{0}9800 $& 4.9    &  $8300 -\phantom{0}8800$ & &     &$\phantom{0}6259\pm\phantom{0}50$ & DT90 & opt \\
            &               & & &            &           &        &       &  & & &   &    &\phantom{0}   &      &      &$                       $&        &                          & &     &$\phantom{0}6335\pm\phantom{0}15$ & PT03 & \ion{H}{i} \\
WKK8074     &               & & & 17 15 00.5 & -54 08 26 & 335.71 & -9.01 & S& &R&5  & 55x&\phantom{0}24 & 15.9 & 0.20 &$  300 -          10250 $& 5.8    &  $1100 -\phantom{0}1300$ & &     & & & \\
WKK8090     &               &M& & 17 16 04.6 & -54 46 55 & 335.26 & -9.50 & S& & &5  & 59x&\phantom{0}12 & 16.4 & 0.21 &$  200 -          10350 $& 4.9    &                          & &     & & & \\                   
WKK8163     &               &M& & 17 22 28.6 & -54 19 45 & 336.17 &-10.02 & S& & &7: & 50x&\phantom{0}15 & 16.3 & 0.20 &$  600 -          10550 $& 5.8    &                          & &     & & & \\
WKK8165     &               &M&I& 17 22 34.7 & -52 42 04 & 337.56 & -9.14 & S& & &3  & 58x&\phantom{0}13 & 16.0 & 0.23 &$  500 -\phantom{0}9800 $& 4.6    &                          & &0.7  & & & \\
Y17477-40   &               &M&I& 17 51 18   & -40 26 25:& 350.70 & -6.90 & S& & &   & 24x&\phantom{000} & 16.8N& 0.32 &$  500 -\phantom{0}9800 $& 5.4    &  $8200 -\phantom{0}9050$ & &2.4  &$12808\pm\phantom{0}38$& FH95 & opt \\
Y18139-37   &CGMW4-0433     &M&I& 18 17 21   & -37 21 08:& 355.87 & -9.89 & S& & &   & 42x&\phantom{0}18 & 16.5N& 0.10 &$  500 -\phantom{0}9800 $& 4.8    &  $1200 -\phantom{0}1300$ & &0.5  &$\phantom{0}7214\pm\phantom{0}70$ & VY96 & opt \\
\noalign{\smallskip}
\hline
\noalign{\smallskip}
\end{tabular*}
\newline
{\bf Notes}: {\it WKK1694:} detected WKK1696; {\it WKK2101:} the velocity
range $10250-12550$ has an rms of 6.1; {\it WKK2892:} the pointing lies
between the galaxies WKK2892 at $d=7\farcm5$ and WKK2934 at $d=11\farcm3$
(not listed); {\it WKK3128:} detected WKK2993 in the OFF pointing; {\it
WKK3296:} detected WKK3285; {\it WKK3823} and {\it WKK3836} are in the same
beam at a similar distance from the centre; {\it WKK4751} and {\it WKK4755}
are in the same beam at a similar distance from the centre; {\it WKK5267:}
detected WKK5240; {\it WKK5297:} detected WKK5285; {\it WKK5381:} detetced
WKK5366; {\it WKK5443:} detected HIZSS100 (WKK5443OFF) in the OFF pointing;
{\it WKK5534:} detetced WKK5285 in the OFF pointing; {\it WKK5544:}
detected WKK5260 in the OFF pointing; {\it WKK5556:} detected WKK5285 in
the OFF pointing; {\it WKK5581:} detected IC4584 and IC4585; {\it WKK5694:}
detected WKK5768 and WKK5729; {\it WKK5709:} detected WKK5768; {\it
WKK5733:} detected WKK5768 and WKK5729; {\it WKK5733:} detetced WKK5768 and
WKK5729; {\it WKK5780:} detetced WKK5768; {\it WKK5805:} detetced WKK5796;
{\it WKK6092:} detected WKK6100; {\it WKK6251} lies close to a hot source
(probably WKK6269, a cD galaxy) which causes strong baseline variations;
{\it WKK7175:} detected WKK7198; {\it WKK7248:} detected WKK6913 in the OFF
pointing; {\it WKK7689:} detected WKK7652; {\it WKK7794:} detected WKK7776.
\normalsize
\end{table}
\end{landscape}

\begin{table}[tb]
\normalsize{\caption{References for independent velocity determinations. }\label{rfitab}}
\small
\begin{tabular}{rr}
\noalign{\smallskip}
\hline
\noalign{\smallskip}
velocity range & counts \\
\noalign{\smallskip}
\hline
\noalign{\smallskip}
$800-1000$   &  19  \\
$1200-1350$  &  15  \\
3200         &   1  \\
$3700-3800$  &   2  \\
4450         &   2  \\
4900         &   1  \\
5800         &   1  \\
6000         &   1  \\
7000         &   3  \\
$7200-7300$  &   3  \\
$7500-7700$  &   6  \\
8100         &   1  \\
$8300-8500$  &  21  \\
$8800-9000$  &   2  \\
9300         &   1  \\
10000        &   1  \\
$10100-10300$&   4  \\
\noalign{\smallskip}
\hline
\noalign{\smallskip}
\end{tabular}
\end{table}

\section{Velocity distribution and detection rate} \label{results}

\begin{figure*}[tb]
\resizebox{\hsize}{!}{\includegraphics[angle=270]{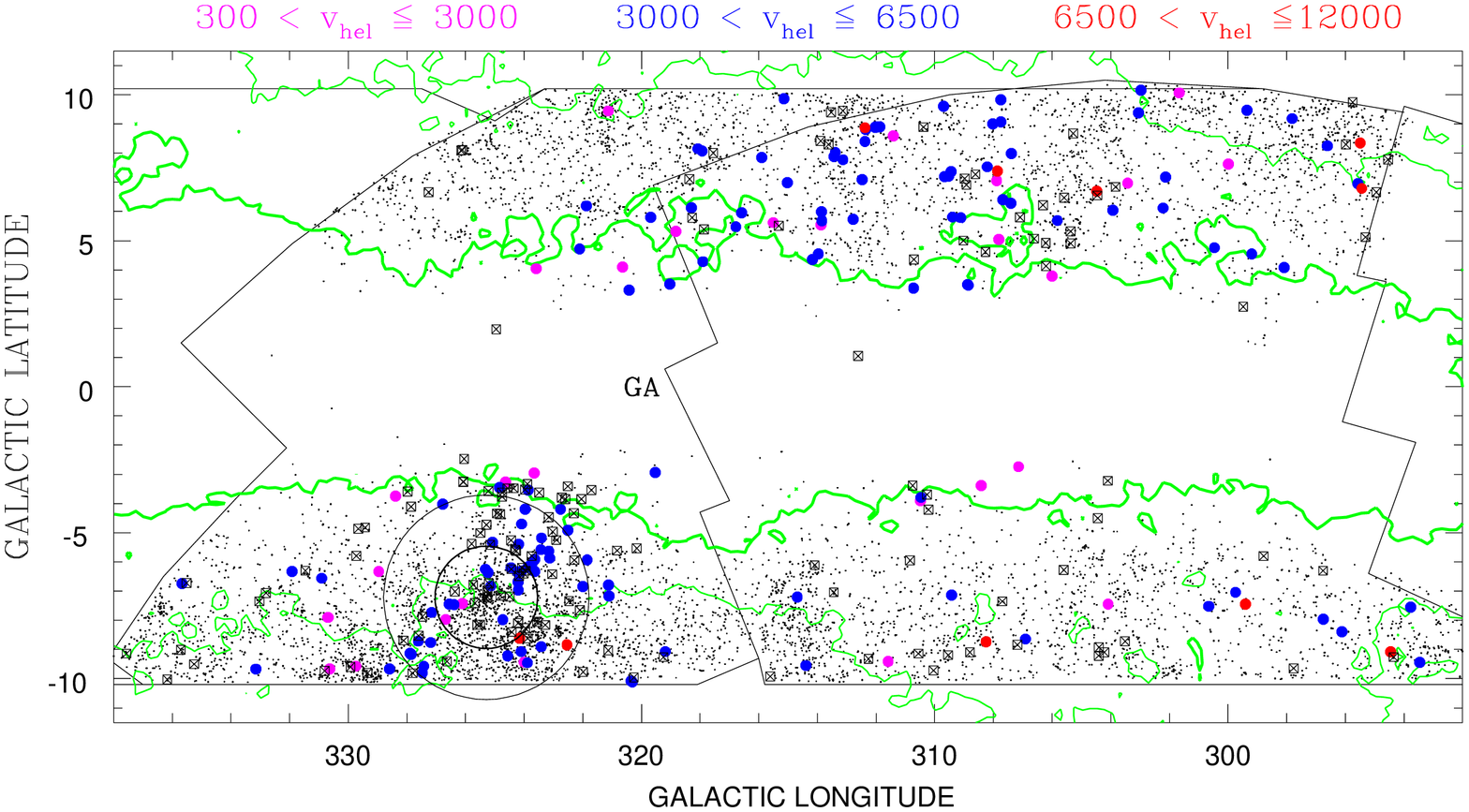}} 
\caption[]{Distribution of galaxies in the Crux and GA ZOA regions. Filled
  circles indicate the 162 21\,cm detections, crosses in squares the 152
  non-detections. The small dots represent the galaxies uncovered within
  the optical search regions (outlined areas). The contours show the dust
  extinction as determined from the 100$\mu$m DIRBE maps (Schlegel et
  al. 1998) at the levels $A_B = 1\fm0$ and 3$\fm$0 (thick line). The
  circles are centred on the Norma cluster with $1 R_{\rm A}$ and $2 R_{\rm
  A}$.
}
\label{allplot}
\end{figure*}

Figure~\ref{allplot} shows the Crux/GA search area (outlined regions) in
Galactic coordinates with the optically discovered galaxies ($D \la 0\fm2$)
plotted as small dots (Woudt \& Kraan-Korteweg 2001). The 314 galaxies
observed with the Parkes radio telescope (indicated with larger symbols)
are distributed fairly homogeneously over the galaxy density distribution
-- leading naturally to a larger number of observations in the high density
area of the Norma cluster. The contours mark extinction levels of
$A_B=1\fm0$ and $3\fm0$ following Schlegel et al. (1998). The former
indicates where previous optical surveys become incomplete, the latter
where the deep optical ZOA survey becomes incomplete (Kraan-Korteweg 2000;
Woudt \& Kraan-Korteweg 2001). Detected galaxies are marked by filled
circles, non-detected galaxies by crossed squares. The detections are
colour-coded for radial velocity in such a way that they emphasise the
velocity space of the GA region (blue dots; $3000 < v_{hel} <
6500$\,km\,s$^{-1}$), showing the Norma cluster and Norma Wall that (is
believed to) encompass the GA overdensity. Magenta dots are at lower
velocities, while the red dots indicate higher velocities.

The majority of the detections are found to lie in the velocity range of
the GA overdensity ($3000 < v_{hel} < 6500$\,km\,s$^{-1}$). Few nearer
galaxies are detected ($V_{hel} < 3000$\,km\,s$^{-1}$), even though we are
sensitive to them. And very few more distant galaxies were detected. This
is partly due to the lower sensitivity at these higher redshifts, but also
indicative of the lack of distant large-scale structures in this area.

\begin{figure}[tb]
\resizebox{\hsize}{!}{\includegraphics{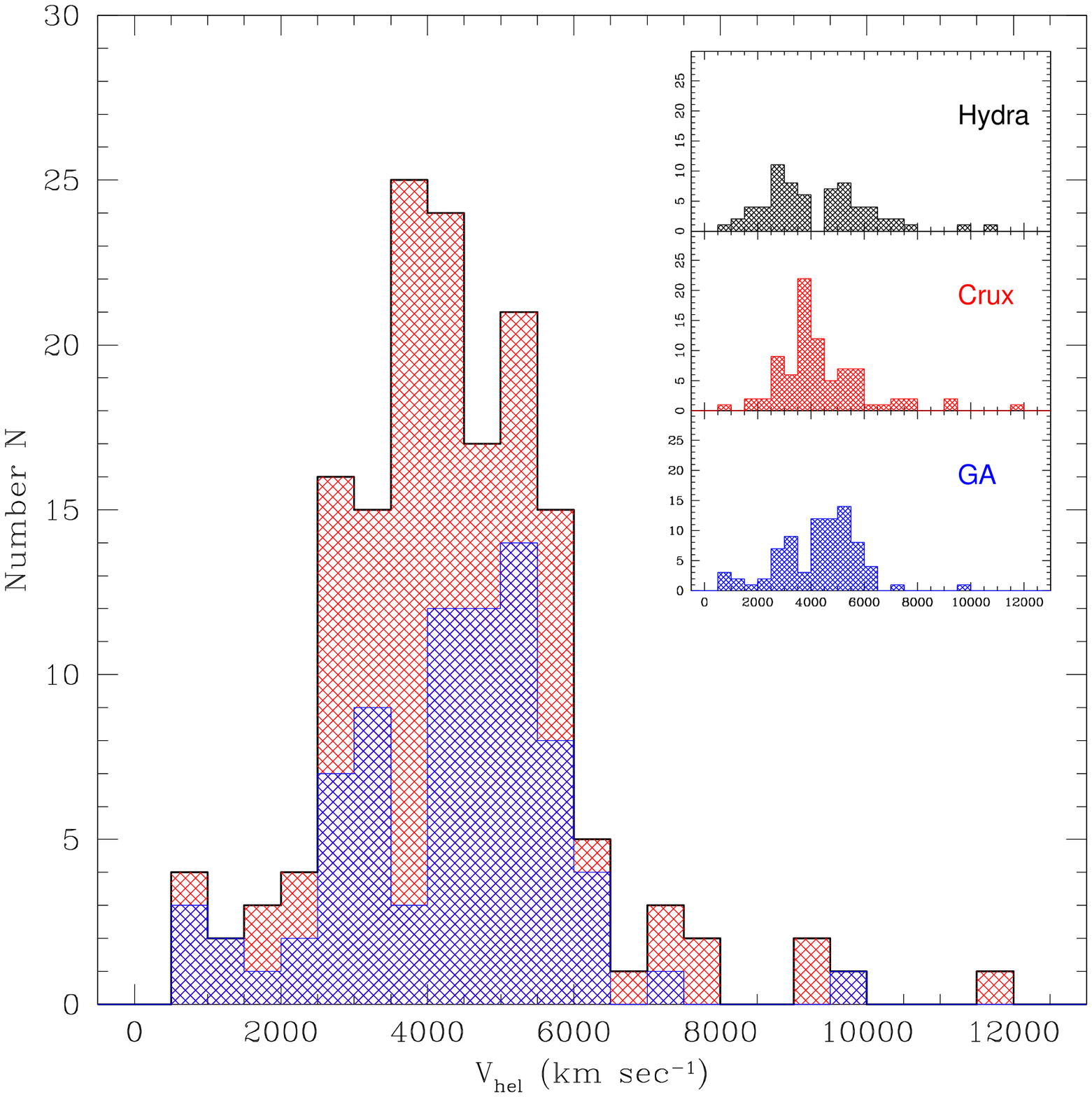}} 
\caption[]{Velocity distribution of the 162 \ion{H}{i} detections in the
  Crux (light grey/red) and GA (dark grey/blue) regions. The inset shows
  the detections separated by regions; the detections in the Hydra/Antlia
  region from Paper I are also shown.
}
\label{vhistplot}
\end{figure}

This is further illustrated by a histogram of the heliocentric radial
velocities for the detections presented in this paper
(Fig.~\ref{vhistplot}). The overall histogram shows all
detections. Detections in the GA survey region are marked in blue, and the
ones in the Crux survey region in red. The inset shows the velocity
histograms for the Crux and GA region separately, including the previous
work in the adjacent Hydra/Antlia survey region.

The histogram shows a markedly different behaviour from what is expected
for a uniform distribution of galaxies in space given our rms, with a sharp
drop-off at about 6000\,km\,s$^{-1}$. The peak -- albeit quite broad -- is
centred roughly at 4500\,km\,s$^{-1}$ ranging from 3000 --
6000\,km\,s$^{-1}$, similar to the velocity range of the Norma cluster
(Woudt et al. 2008) which completely coincides with the predicted velocity
range of the GA (Kolatt et al. 1995) and the mean Norma cluster velocity
(Woudt et al. 2008). It is distinct from the much flatter distribution of
the \ion{H}{i}-survey undertaken under the same observing conditions for
the Hydra/Antlia survey area (top panel of inset; as in Paper I).

This is in agreement with the results by Woudt et al. (2004) who have shown
that the velocity distribution of all survey galaxies for which we obtained
redshifts (about 15\% on average for the deep ZOA surveys, mostly optical
spectra from our dedicated follow-up surveys at the SAAO and ESO, plus some
previously published redshifts), is fairly flat out to
20\,000\,km\,s$^{-1}$ in the Hydra/Antlia region, while the Crux region and
-- much more pronounced -- the GA region show a distinct broad peak of
galaxies at $\sim\,4000-5000$\,km\,s$^{-1}$ (see their Fig.~5), while the
histogram is otherwise similar for the three survey regions. The prominence
of the GA overdensity around $\ell \sim 320\degr$ leads to an overall
higher fraction of nearby galaxies ($V \la 6000$\,km\,s$^{-1}$) over the
sampled volume (out to about 20\,000\,km\,s$^{-1}$). This may well explain
the slightly higher detection rate of 52\% found for this survey compared
to Paper~I (45\%, with a total 148 observed galaxies compared to 314 here).
In general, the non-detections are for fainter and smaller galaxies
(extinction-corrected), implying that they are most likely background
galaxies beyond the velocity search range of our survey, plus some
closer-by galaxies with an \ion{H}{i} flux density below our sensitivity
limit.

Inspection of Fig.~\ref{allplot} indicates that the detections and
non-detections are spread quite evenly with respect to the galaxy density
of the overall galaxy distribution. We subsequently have a larger number of
observations in and around the high density area of the Norma cluster. A
relatively higher fraction of observations were also made at higher
extinction levels: (i) in the suspected extension of the Norma Wall across
the Galactic plane, i.e., extending from the Norma cluster at $(\ell, b,
v$) = ($325.29\degr, -7.21\degr, 4821$\,km\,s$^{-1}$) (Woudt et al. 2008)
towards the low-latitude CIZA\,J1324.7--5736 and Cen-Crux clusters at
$(\ell, b, v$) = ($307.4\degr, +5.0\degr, 5700$\,km\,s$^{-1}$) and
($305\degr, +5\degr, 6214$\,km\,s$^{-1}$), respectively (Radburn-Smith et
al. 2006, see also their Fig.~4), which form part of the GA Wall; (ii) for
some highly obscured galaxy candidates deep into the ZOA (in total 37 were
observed with $A_B > 3\fm0$ of which 20 were detected) as \ion{H}{i}
observations are the only way to obtain a redshift for such heavily
obscured galaxies.

\begin{figure}[tb]
\resizebox{\hsize}{!}{\includegraphics{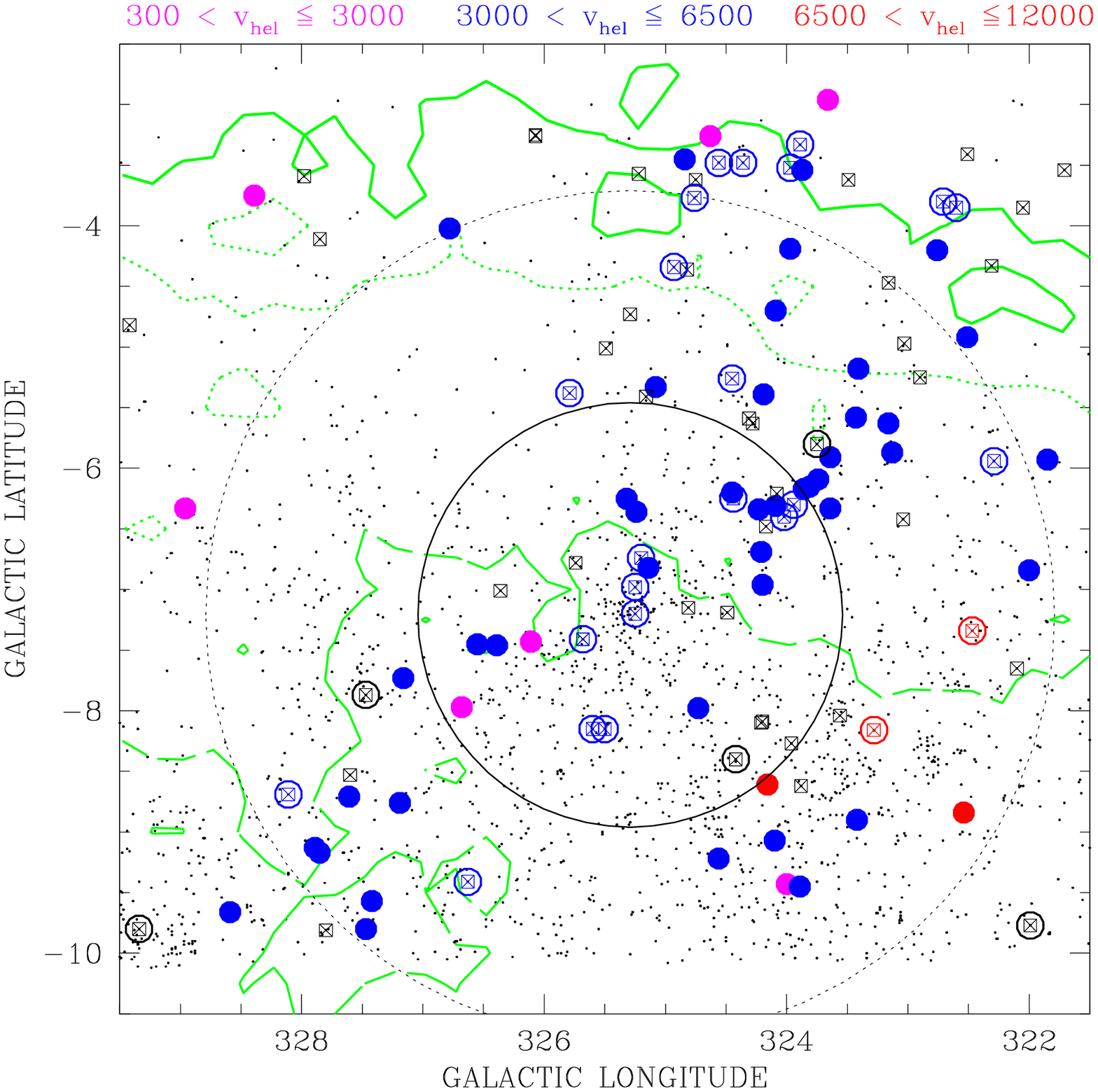}} 
\caption[]{Distribution of the galaxies observed at 21\,cm zoomed in at the
  Norma cluster (cf.\ description of Fig.~\ref{allplot}). A third contour
  level at $A_B = 2\fm0$ is shown as a dotted line. Non-detections with a
  large circle have a known optical velocity (a black circle stands for
  $v>12000$\,km\,s$^{-1}$).
}
\label{normaplot}
\end{figure}

Overall the detections and non-detections seem to be similarly distributed
over the survey area. However, a closer look at the Norma cluster (see also
Fig.~\ref{normaplot}) reveals that the detection rate within one Abell
radius ($R_{\rm A} =1\fdg75$) of the central cD galaxy, WKK\,6292, is only
41\% ($n_{tot}=32$), which is lower than for the whole survey on the
1$\sigma$ level. A bit further out, in the annulus $1 - 2 R_{\rm A}$, the
detection rate is similar to the rest of the GA/Crux survey, namely 53\%
($n_{tot}=53$). Such a trend, if real, would be expected if there are not
many spiral galaxies in the cluster or if the spirals are \ion{H}{i}
deficient. The former is unlikely as morphological distinction between
ellipticals and spirals is largely unaffected at the extinction levels of
the Norma cluster. The latter is probable since rich, massive, and X-ray
strong galaxy clusters like the Norma cluster generally show \ion{H}{i}
deficiencies (Giovanelli \& Haynes 1985). We explore this effect in more
detail in Section~\ref{hidef}.

\section{HI deficiency in the Norma cluster } \label{hidef}

As a first measure of \ion{H}{i}-deficiency we regard the \ion{H}{i}-masses
of the galaxies in and around the cluster as a function of cluster-centric
distance. The \ion{H}{i} masses are calculated using $M_{\rm HI} = 2.356
\times 10^5 D^2 S $, where $S$ is the \ion{H}{i} flux integral in
Jy\,km\,s$^{-1}$, and $D$ the distance in Mpc calculated from the measured
velocity and corrected for the motion with respect to the Local Group
(Yahil et al. 1977). For galaxies assumed to lie in the Norma cluster
(i.e., within $1.5 R_{\rm A}$ and 2096\,km\,s$^{-1}$ $< v <$
7646\,km\,s$^{-1}$, cf.\ Woudt et al.\ 2008) we adopt a distance of
$D=67\,h^{-1}_{70}$\,Mpc.

\begin{figure}[tb]
\resizebox{\hsize}{!}{\includegraphics{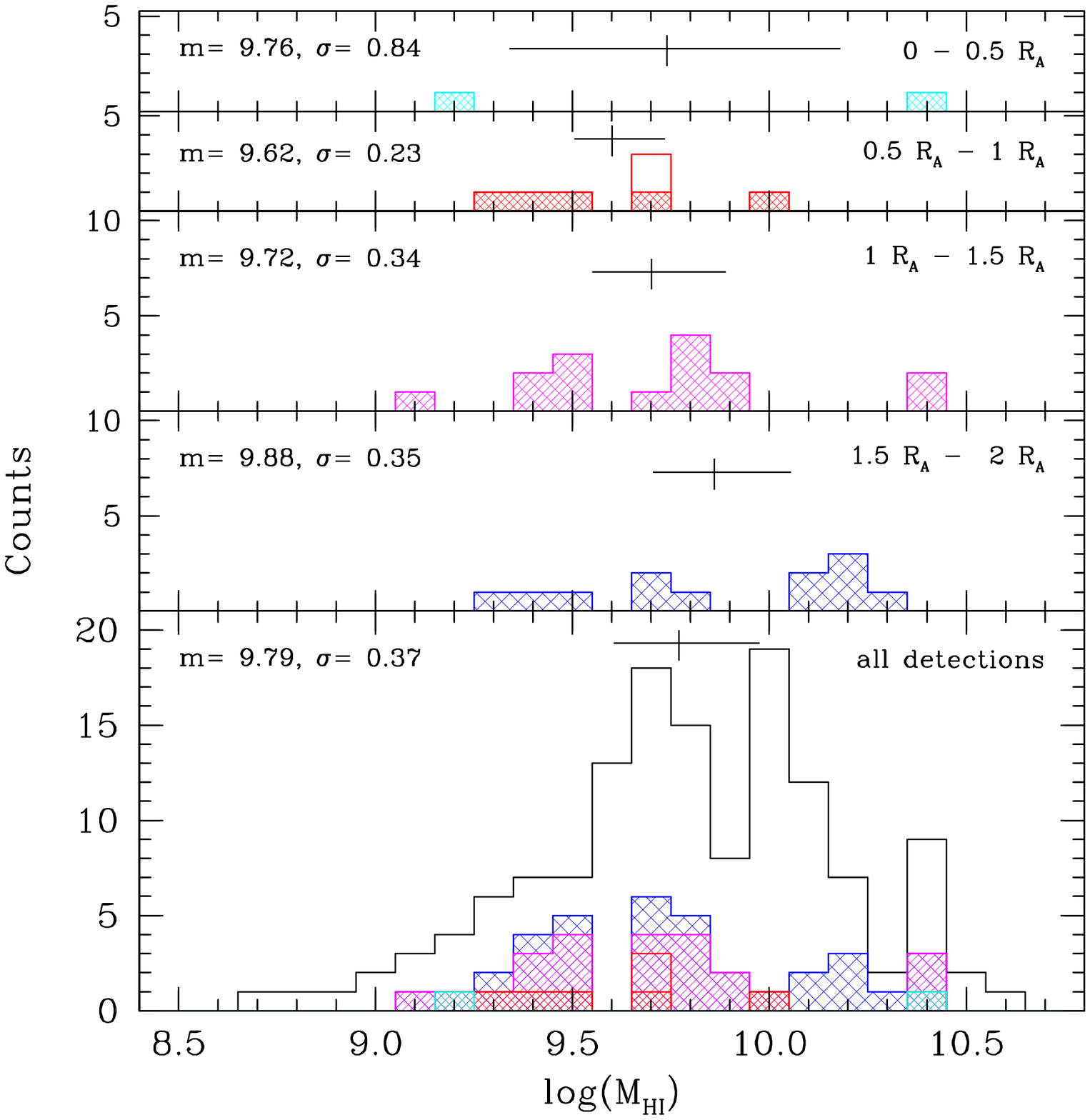}} 
\caption[]{Histogram of the logarithm of \ion{H}{i} mass for various
  regions in the Norma cluster (hashed) and the total region observed
  (non-hashed). The non-hashed region in the second panel from the top
  indicates two galaxies (WKK\,6689 and WKK\,6732) with an uncertain though
  small contribution from each other. The \ion{H}{i} mass in both cases is
  therefore slightly overestimated.
}
\label{mhiplot}
\end{figure}

Figure~\ref{mhiplot} shows histograms of the logarithm of \ion{H}{i} mass
for galaxies in the inner region of the Norma cluster ($<\!0.5R_{\rm A}$,
top panel), three different annuli ($0.5R_{\rm A} - 1R_{\rm A}$; $1R_{\rm
A} - 1.5R_{\rm A}$, and $1.5R_{\rm A} - 2R_{\rm A}$, middle panels),
including for comparison the \ion{H}{i}-mass distribution of all the 162
detections in the survey area (bottom panel). For reference, the hashed
histograms are shown in accumulation in the bottom panel as well. The mean
values and the standard deviations are indicated for each subset and for
the total region in the bottom panel. Our median value of all detections is
$\log M_{\rm HI} = 9.86$ (in units of solar mass) and compares well with
the \ion{H}{i} mass distribution in the Northern Extension of the HIZOA
survey ($\log M_{\rm HI} = 9.7$, Donley et al.\ 2005) which has s similar
sensitivity to our pointed survey. These histograms show that galaxies
closer to the cluster core have lower \ion{H}{i} mass than farther
galaxies, although at only about the $1\sigma$ level.

In the innermost region only two galaxies (out of nine; 22\%) were
detected, one of which actually has a substantial \ion{H}{i}-mass
(WKK\,5999 lies at $d=0.49 R_{\rm A}$ and $v=3244$\,km\,s$^{-1}$). Apart
from this galaxy, the distribution of \ion{H}{i} mass for the galaxies
within $1.5R_{\rm A}$ is shifted towards the lower end as compared to the
field. This is expected if we assume that cluster members have passed at
least once through the centre of the cluster and have undergone
ram-pressure stripping. In contrast, galaxies in the range $1.5R_{\rm A} -
2R_{\rm A}$ have an \ion{H}{i} mass distribution comparable to the field.

A better estimate of the effect of the cluster environment on the
\ion{H}{i} content of the Norma spiral galaxies is the \ion{H}{i}
deficiency parameter (Giovanelli \& Haynes 1985) which compares the
\ion{H}{i} content of a cluster galaxy with the average \ion{H}{i} content
of an isolated field spiral of the same morphological type. Solanes et
al. (1996) have shown that the \ion{H}{i} content of a field spiral also
depends on the diameter. However, due to the uncertainties in diameter of
our highly obscured galaxies we have not included the diameter dependence
in our calculations. Moreover, there is uncertainty in the morphology of
many of the highly obscured galaxies. Such obscured galaxies were
generally labelled as `S' by Woudt \& Kraan-Korteweg (2001) without a
further subtype. We assumed these to be late type spirals (i.e., Scd or
Sd,) as a bulge dominated spiral could have been classified, whereas
strongly obscured low-surface brightness irregular galaxies would most
likely have remained invisible on the optical survey plate.

We note furthermore, that our sample of latest type spirals (Sdm -- I)
shows a systematic offset in \ion{H}{i} mass compared to our other
morphological sub-samples. The mean \ion{H}{i} mass of this latest type
sample (not including the possibly \ion{H}{i}-deficient Norma galaxies
within $2R_{\rm A}$) is $\log M_{\rm HI} = 9.79$ with a standard deviation
of 0.33. This value is typical of Scd and Sd galaxies ($\log M_{\rm
HI}=9.62$, with a standard deviation of 0.31) and consistent with the
fact that Giovanelli \& Haynes (1985) find the value 9.09 for their field
Sm -- Im sample (all \ion{H}{i} masses from Giovanelli \& Haynes were
corrected for the difference of their adopted Hubble constant of
50\,km\,s$^{-1}$,Mpc$^{-1}$ to our value of 70\,km\,s$^{-1}$,Mpc$^{-1}$).
Considering the high foreground extinction of our sample, we assume that
most of the galaxies classified as late type S/Im were probably
misclassified: it is unlikely to find more than a couple of very late-type
galaxies at these foreground extinctions. Hence, we have calculated the
\ion{H}{i} deficiency parameter for the Sdm -- I galaxy sample using the
average \ion{H}{i} content of field galaxies of types Scd and Sd.

\begin{figure}[tb]
\vspace{-2.5cm}
\resizebox{\hsize}{!}{\includegraphics{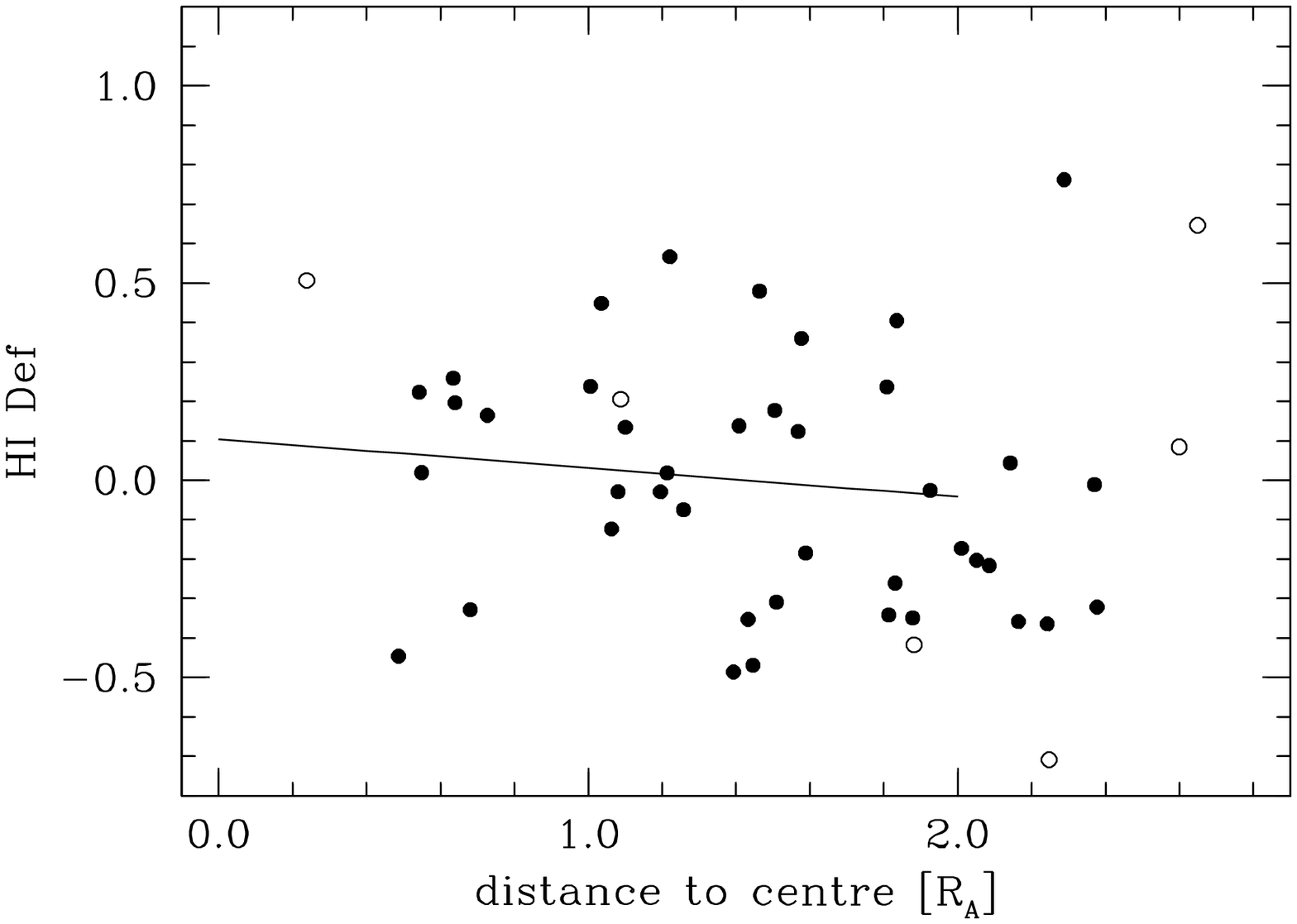}} 
\caption[]{\ion{H}{i} deficiency parameters of Norma cluster are plotted
 versus distance from the centre of the cluster in Abell radii. Open
  circles indicate unknown spiral types or early morphological types which
  are not included in the fitted line. (One $R_{\rm A}$ corresponds to
  $1\fdg75$).
}
\label{hidefplot}
\end{figure}

The derived \ion{H}{i} deficiency parameters are plotted as a function of
cluster-centric distance out to $2R_{\rm A}$, as illustrated in
Fig.~\ref{hidefplot}. The dispersion is quite large and only a very weak
dependence is seen. The least squares fit out to a radius of $2 R_{\rm A}$
has a slope of $-0.07\pm0.12$. A comparison with the respective plot for
the Coma cluster (Bravo-Alfaro et al. 2000, their Fig.~4) which is similar
to the Norma cluster in its cluster specific properties (Kraan-Korteweg et
al. 1996; Woudt 1998), shows that we have only one detection within
$0.4R_{\rm A}$, whereas Coma shows obvious deficiencies for seven central
galaxies. This one detection (WKK\,6100) is of unknown morphological
type. Revisiting the $R$- and $J$-band images and taking account of the
effect of extinction, it is likely that this galaxy is actually an Sc
galaxy. that would make this galaxy more deficient.

\begin{figure}[tb]
\vspace{-2.5cm}
\resizebox{\hsize}{!}{\includegraphics{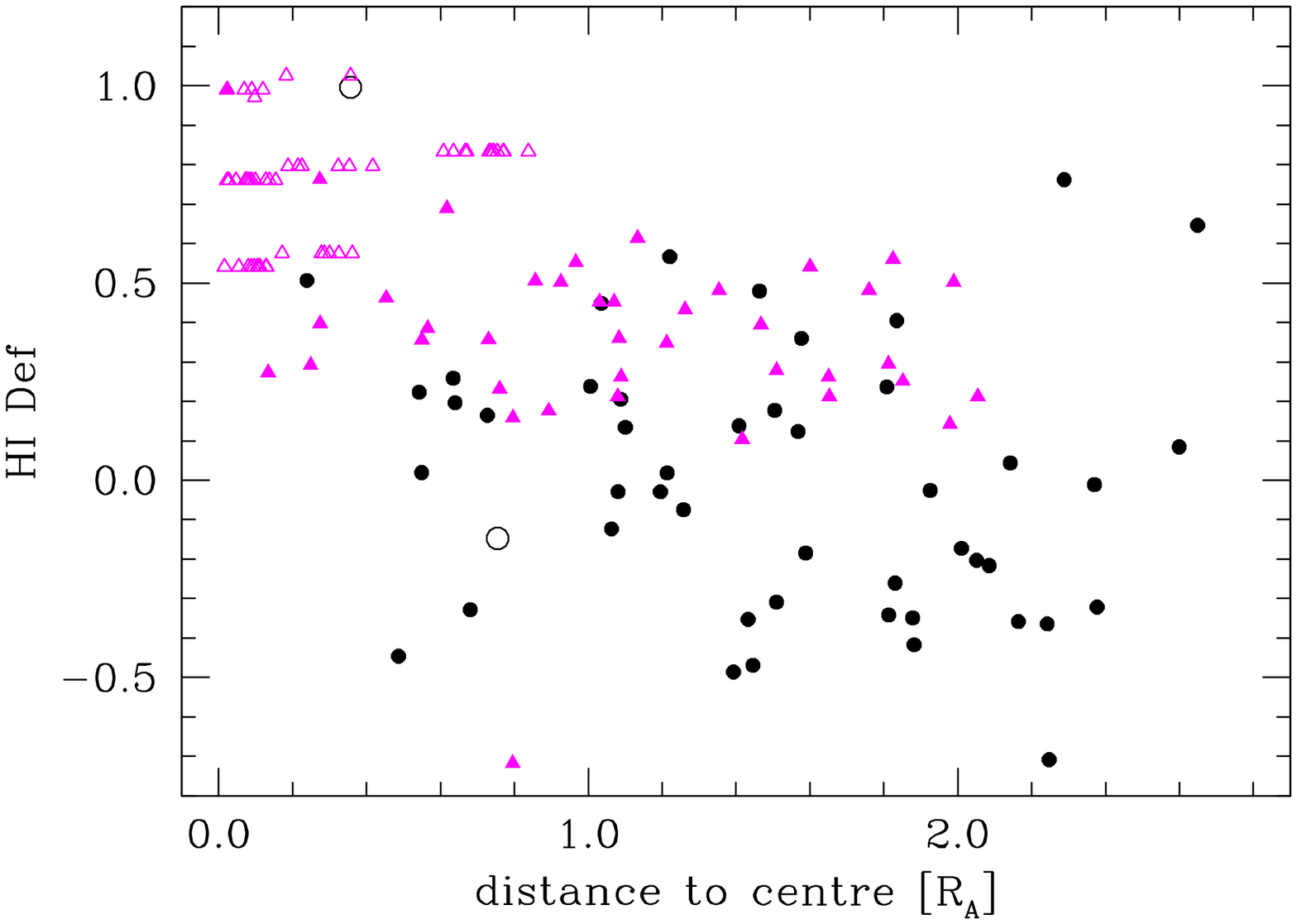}} 
\caption[]{Same as Fig.~\ref{hidefplot} but with lower limits for
  non-detections (triangles). Open symbols stand for ATCA data, filled
  symbols are from the present paper.
}
\label{hidefatcaplot}
\end{figure}

However, the fact that we have six non-detections within $d=0.49 R_{\rm A}$
of the Norma cluster indirectly suggests that the cluster is even more
\ion{H}{i} deficient than suggested by Fig.~\ref{hidefplot}. These galaxies
must lie below our sensitivity limit. To test this, we looked at the
non-detections that have optically determined velocities. Based on that, we
can test whether they are part of the cluster or cluster environment. These
are identified with open circles in Fig.~\ref{normaplot}. Nearly all
non-detections with optically known velocities have velocities consistent
with being part of the cluster (blue open circles with crosses). For these
we have calculated upper limits of the \ion{H}{i}-mass based on our
measured rms, assuming that we would have detected the galaxy if it had an
S/N = 3, and a 50\% line width typical of a spiral galaxy of
200\,km\,s$^{-1}$. These values are added to the sample's \ion{H}{i}
deficiency parameters, and shown in Fig.~\ref{hidefatcaplot}.

This is not all the data available to test the \ion{H}{i} deficiency
hypothesis. Vollmer et al. (2001) obtained ATCA radio synthesis imaging
observations of the centre of the Norma cluster as well as for two fields
just slightly offset from the centre. In all three fields (with a HPBW of
$30\arcmin$) they detected only two galaxies (both were not observed by
us). That alone presents an indication of \ion{H}{i} deficiency. Only one
of their two detected galaxies is clearly \ion{H}{i} deficient: WKK\,6489
lies at $0.36R_{\rm A}$ and has an \ion{H}{i} deficiency of 1.0, while the
other galaxy detected by them, WKK\,6801, at $0.75R_{\rm A}$ has a normal
value of $-0.15$.

\begin{figure}[tb]
\resizebox{\hsize}{!}{\includegraphics{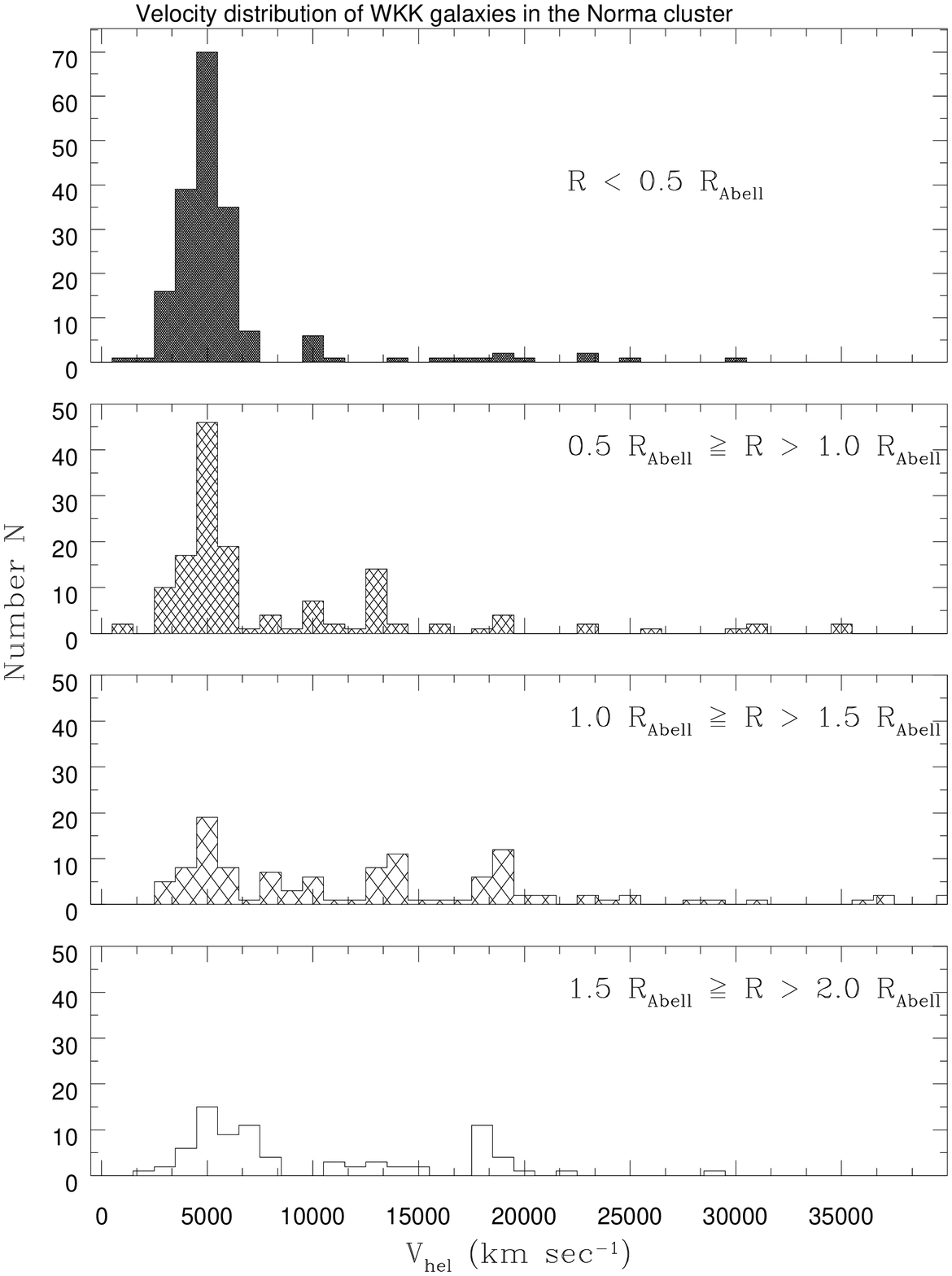}} 
\caption[]{Velocity distribution of WKK (optical) detections in the Norma
  region, separated by Abell radii, as indicated.
}
\label{vhist2plot}
\end{figure}

The ATCA fields contain, however, many more spiral galaxy candidates (the
reason why these fields were chosen for observations). We assume that all
the spiral galaxies in the ATCA fields actually are members of the Norma
cluster. An inspection of the velocity histograms of all known redshifts of
galaxies in the Norma cluster (available from Woudt et al. 2008) in rings
out to $2R_{\rm A}$ confirms that nearly all galaxies within $0.5R_{\rm A}$
are members of the cluster, with just a very low number of outliers (see
Fig.~\ref{vhist2plot}). Even for the outermost ATCA field (at about
$0.65R_{\rm A}$ from the cluster centre) contamination by background or
foreground galaxies will be minimal.

The ATCA non-detections have a $3\sigma$ detection limit of about
3\,mJy/beam in each velocity channel. Assuming emission to be unresolved
with the 30\arcsec\ beam in each channel, the $3\sigma$ upper limit on
\ion{H}{i} mass for these non-detections is $6\times 10^8 {\rm M}_{\sun}$
(following Vollmer et al. 2001, but with $D=67$\,Mpc rather than
$D=79$\,Mpc, and an assumed linewidth of 200\,km\,s$^{-1}$ rather than 
150\,km\,s$^{-1}$). Taking this limit, the calculated \ion{H}{i} deficiency
for spiral galaxies in the ATCA fields are also added to
Fig.~\ref{hidefatcaplot} (filled triangles). The addition of these lower
limits now reveal a very clear and strong trend of \ion{H}{i} deficiency
for galaxies within $0.4R_{\rm A}$. For the annulus from $0.4R_{\rm A}$ to
$1R_{\rm A}$ and beyond, the \ion{H}{i} deficiency scatters between normal
values to considerably deficient galaxies. This is consistent with what has
been found for other rich clusters, where clear \ion{H}{i} deficiency
manifests itself unambiguously only in the innermost core of the cluster
($R \la 0.5R_{\rm A}$; see e.g., Haynes et al. 1984).

The non-detection rate can also be used to calculate an \ion{H}{i}
deficiency fraction as defined by Giovanelli \& Haynes 1985. In our sample,
we have one detection with an \ion{H}{i} deficiency $>0.3$ within $1R_{\rm
A}$ as well as 13 lower limits close to or above $0.3$. There are an
additional 8 detections and 5 non-detections which may have \ion{H}{i}
deficiencies just under 0.3. The \ion{H}{i} deficiency fraction in this
case is at least 0.58. B\"ohringer et al. (1996) give an X-ray luminosity
for the Norma cluster of $2.2(\pm0.3) \times 10^{44}$\,erg\,s$^{-1}$ in the
ROSAT energy band ($0.1 - 2.4$\,keV). Figure~9 in Giovanelli \& Haynes
shows a relationship between the \ion{H}{i} deficiency fraction and X-ray
luminosity in the ($0.5 - 3.0$\,keV) band. The values for the Norma cluster
fit well onto that relationship.

For completeness, we have checked for \ion{H}{i} deficiency in our other
galaxies. There are only two outliers, WKK\,1294 with $(l,b,v)=(301\fdg7,
10\fdg1, 1919$\,km\,s$^{-1}$) and WKK\,1510 with $(l,b,v)=(303\fdg4,
7\fdg0, 2668$\,km\,s$^{-1}$). Both are located in the vicinity of the two
clusters in Crux area, the CIZA clusters (Clusters in the ZOA, Ebeling et
al. 2002) CIZA\,1324.7--5736 at $(l,b,v)=(307\degr, 5\degr,
5700$\,km\,s$^{-1}$) and less rich Cen-Crux clusters $(l,b,v)=(306\degr,
5\fdg5, 6200$\,km\,s$^{-1}$), but not close enough to expect any \ion{H}{i}
stripping. It is therefore more likely that the morphological types (S3 and
S4, respectively) are wrong, and that these galaxies are of later type.
This is supported for WKK\,1294, which is bright in the $B$-band but barely
visible on the 2MASS $J$-band image. A later type would reduce the
\ion{H}{i} deficiency value.

\section{Large-scale structure in the Crux and GA regions} \label{lss}

In this section, we will discuss the new \ion{H}{i} detections in the
context of the known large-scale structures (LSS) in and across the
ZOA. The new data is presented together with previously known redshifts in
and adjacent to the survey area (extracted from LEDA) in a series of sky
projections and redshift cones (Figs.~10 and~11). Care should be taken in
the interpretation of these plots as they are based on an `uncontrolled'
redshift sample of galaxies.

\subsection{Sky projections}

Figure~10 
displays the galaxy distribution in Galactic coordinates in and around the
survey region sliced in redshift intervals of width $\Delta v =
2000$\,km\,s$^{-1}$ out to 8000\,km\,s$^{-1}$ (the first slice runs from
300 -- 2000\,km\,s$^{-1}$). Higher velocity slices (8000 -- 12000) are not
shown, as the detections for $v > 8000$\,km\,s$^{-1}$ (see
Fig.~\ref{vhistplot}) are too scarce to add useful insight to the LSS at
those redshifts.

\begin{figure}[h!]
\begin{center}
\includegraphics[width=7.5cm]{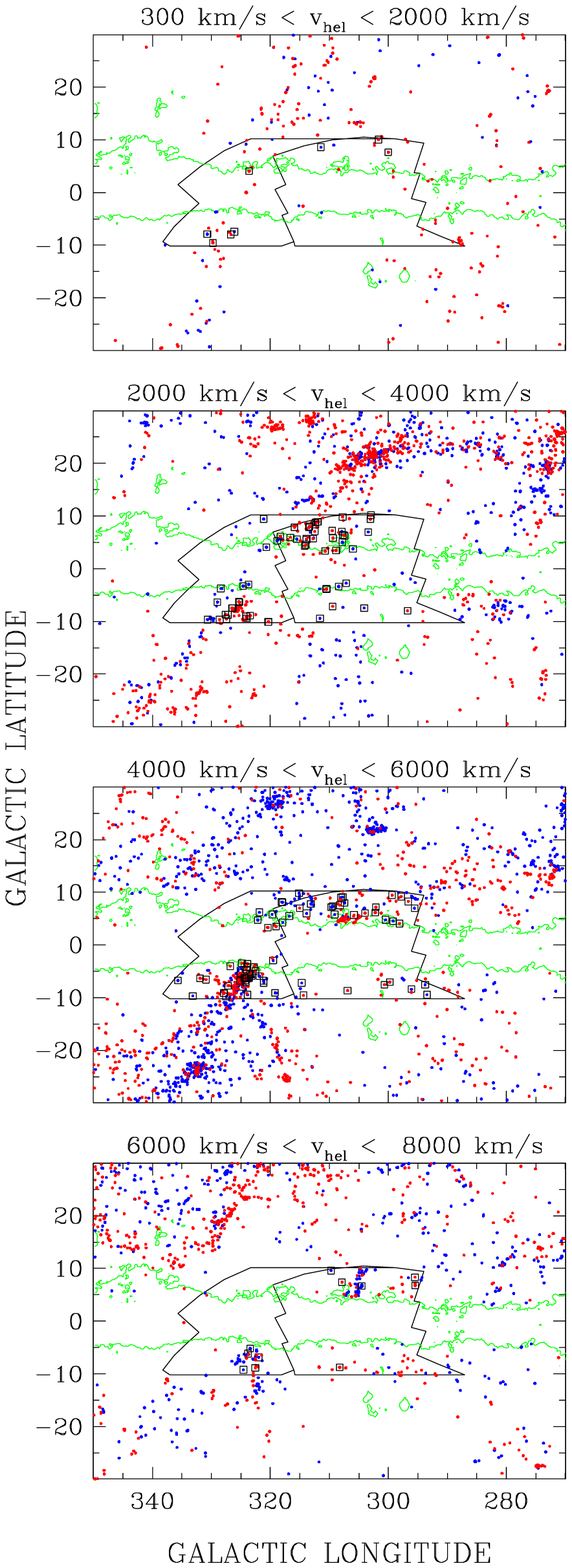} 
\caption[]{Sky projections in Galactic coordinates in and around the
  Crux/GA survey region in redshift intervals of $\Delta
  v=2000$\,km\,s$^{-1}$ for $300<v<8000$\,km\,s$^{-1}$. In each panel, the
  ``nearer'' galaxies ($\Delta v=1000$\,km\,s$^{-1}$) have blue symbols,
  and the more distant red symbols. The squares around the dots identify
  the new \ion{H}{i} detections. The survey area and extinction contour of
  $A_B = 3\fm0$ are outlined.}
\end{center}
\label{lssdet}
\end{figure}

What is immediately conspicuous in all four panels is that the \ion{H}{i}
detections constitute a large fraction of the galaxies with redshifts
within the band limited by $A_B < 3\fm0$ and $b < 10\degr$ (apart from the
\ion{H}{i}-deficient galaxies in the Norma cluster). This emphasises that
\ion{H}{i}-observations, even when dependent on optical (or NIR)
pre-identification, are tremendously powerful in mapping LSS of highly
obscured galaxies in and across the ZOA. Note that optical redshifts for
these survey regions have been obtained earlier with the SAAO 1.9\,m
telescope (Fairall et al. 1998; Woudt et al. 1999), as well as with
multi-object spectroscopy using the 3.6\,m telescope of ESO in Chile (Woudt
et al. 2004).

The first slice is quite sparsely populated. The Supergalactic Plane (SGP;
de Vaucouleurs 1953) is the most prominent feature, visible here from about
($\ell,b) = (335\degr, -30\degr$) to ($315\degr, +30\degr$), crossing the
Galactic Plane (GP) at $\ell = 325\degr$. Most of the detections in this
panel lie along the SGP.

In the second panel (2000\,--\,4000) the Centaurus Wall (CenW) is the
dominant feature (Fairall 1998, Fairall et al. 1998). It enters the panel
at about ($345\degr, -30\degr$) and extends across the ZOA to the Centaurus
clusters ($302\degr, +22\degr$); it is less inclined than the SGP. Apart
from a few detections below the GP in Crux, which do not seem to highlight
any particular structure, most of the \ion{H}{i}-detected galaxies in this
panel follow CenW. However, the detections above the GP spread out over a
much wider area than would be expected from the quite narrow CenW, i.e., it
seems to veer off toward the right ($\sim 310\degr, +5\degr$), away from
CenW. These galaxies probably are the low velocity members of the two
Cen-Crux clusters discussed below which form part of the Norma Great Wall.

Despite the many \ion{H}{i} non-detections at the core of the cluster, the
Norma cluster and the Norma Great Wall -- dubbed so for the first time by
Woudt et al. in 1999 -- are the prominent structures in the third
panel. There are still a few galaxies visible that form part of CenW (blue
dots above the GP around the Crux/GA border). But the majority of the
detections (mostly red dots) follow the Norma Wall (Woudt et al. 1999). The
Norma Wall can be traced from the Pavo II cluster
($\sim\!332\degr,-23\degr$) to the Norma cluster
($\sim\!325\degr,-7\degr$), see also Fig.~11. It crosses the Galactic Plane
at $\sim\!320$\degr, and continues with a much shallower slope with respect
to the Galactic Plane towards two neighbouring galaxy clusters at slightly
higher redshift, the Cen-Crux cluster at $(\ell,b,v) \sim (305\fdg4,
5\degr, 6200$\,km\,s$^{-1}$) (Woudt et al. 1999; Woudt \& Kraan-Korteweg
2001) and the X-ray cluster CIZA J1324.7--736 at $(\ell,b,v) \sim
(307\fdg4, 5\degr, 5700$\,km\,s$^{-1}$) (Ebeling et al. 2002). From there,
the Wall connects with the Vela cluster (Abell S0639;
$(\ell,b,v)=(280\fdg5, -10\fdg9, 6326$\,km\,s$^{-1}$); Stein 1996). The
extent and shape of this wall was suspected in earlier papers (e.g.,
Kraan-Korteweg et al. 1994) and strongly supported in more recent work
(Woudt et al. 2004; Kraan-Korteweg 2005; Ebeling et al. 2005; Radburn-Smith
et al. 2006). A sketch of this Wall, as well as the Centaurus filament, is
given with Fig.~4 in Radburn-Smith et al. (2006).

In the final panel, most of the \ion{H}{i} detections are found at the position
of the Norma cluster and the two Crux clusters. They most certainly are
high velocity dispersion outliers of the massive clusters (finger-of-God
effect) prominent in the previous panel, rather than an indication of
galaxy agglomerations at this higher velocity range.

In conclusion, the new \ion{H}{i} detections mostly help to delineate
filaments and walls deeper into the ZOA. They provide supporting evidence
that the Norma Wall crosses the ZOA, but then turns away from CenW towards
Vela. A much smaller number of galaxies seem to form part of the general
field. The slices also indicate that the \ion{H}{i} observations of
galaxies allow the mapping of LSS to quite high dust column-densities,
higher than can be achieved with optical spectroscopy.

\subsection{Pie diagrams}

The above findings are confirmed with Fig.~11, 
which shows the new detections in a composite of pie diagrams that cover a
wider area than the previous figure to reveal the overall LSS in this part
of the sky. The two top panels show redshift slices out to
12\,000\,km\,s$^{-1}$, the maximum redshift range of our HI
observations. They are 30\degr\ wide in longitude, with the left one
centred on the GA survey region ($340\degr > \ell > 310\degr$) and the
right one centred on the Crux survey region ($310\degr > \ell >
280\degr$). Note that there is a longitude overlap of 10\degr\ to
facilitate the visualisation of the structures running from one wedge
diagram to the next. The pie diagram for these two longitude strips run
from $-$45\degr across the ZOA to +45\degr. The bottom panel displays a pie
diagram of the ZOA ($|b| \le 10\degr$) for the longitude range $360\degr -
270\degr$ as in the sky projection plot in the middle panel. The middle
panel displays the projected LSS distribution of galaxies out to
12\,000\,km\,s$^{-1}$, with blue dots marking the distance range of Norma
and the Norma Wall (3000 -- 6500\,km\,s$^{-1}$), magenta the nearer
galaxies (300 -- 3000\,km\,s$^{-1}$), and cyan the distant galaxies (6500
-- 12000\,km\,s$^{-1}$). This panel is meant for orientation and
interpretation of the pie diagrams. The plot is collapsed in latitude.

\begin{figure*}[tb]
\vspace{-1.cm}
\begin{center}
\includegraphics[height=19cm]{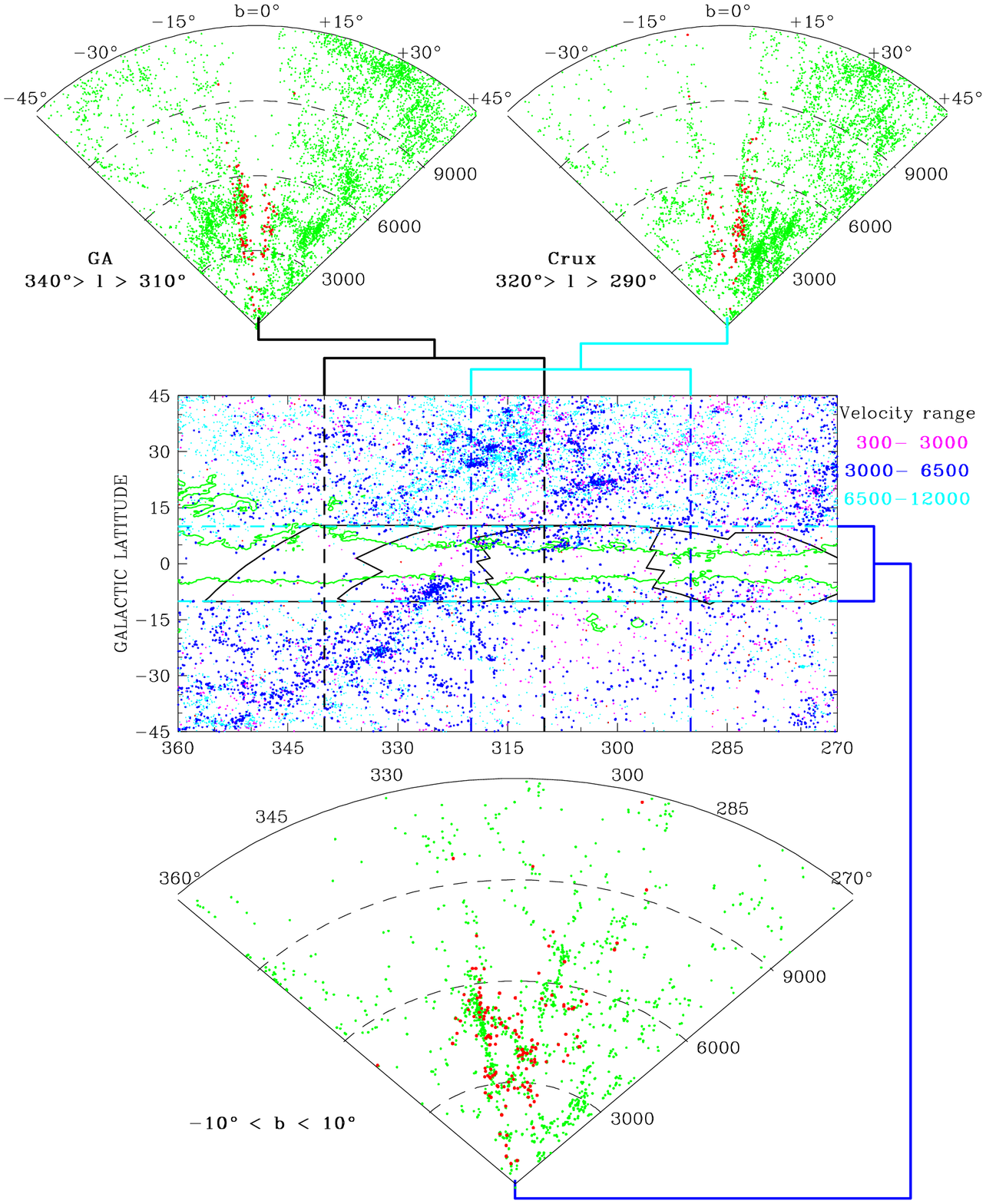} 
\caption[]{Large scale structures in and around the Crux/GA region. The top
  two redshift cones are centred on the GA and Crux regions, respectively,
  and cover the latitude range $b \le \pm45\degr$ out to
  $v=12000$\,km\,s$^{-1}$. The bottom panel displays a redshift cone of the
  ZOA ($|b|\le 10\degr$). Red and green dots represent the Parkes
  \ion{H}{i} detections and previously determined redshifts from LEDA,
  respectively. The middle panel shows the sky distribution of galaxies out
  to the same redshift. It is meant as a guide in the interpretation of the
  redshift cones.
}
\end{center}
\label{lssplot}
\end{figure*}

The pie diagrams confirm that \ion{H}{i}-observations of optically selected
galaxies are quite successful in reducing the ZOA down to $|b| = 5\degr$,
at least out to redshifts of about 6000\,km\,s$^{-1}$ (see also
Fig.~\ref{vhistplot}). For higher velocities the detection efficiency drops
substantially and the ZOA remains increasingly empty. Such a drop in the
detection rate at similar velocities can also seen in the systematic, blind
deep Parkes \ion{H}{i} ZOA survey (HIZOA; $|b| \le 5\degr$; see, e.g.,
Fig.~13 in Kraan-Korteweg 2005). This is not surprising given that the rms
and velocity range of both surveys are similar. In that sense, the pointed
Parkes \ion{H}{i} observations presented here, which are quite successful
in the range $5\degr \la |b| \la 10\degr$, are in fact complementary to the
deep Parkes HIZOA survey of the inner optically opaque ZOA (Henning et al.,
in prep.; for preliminary results see Henning et al. 2005; Kraan-Korteweg
et al. 2005).

The most prominent feature in the GA cone diagram (top left panel) is the
Norma cluster with its finger-of-God ranging between $2500 -
6500$\,km\,s$^{-1}$ evident at ($b= -7\degr$). The cluster stands out even
more prominently in the bottom panel which displays the ZOA slice. The
\ion{H}{i} detections (red dots) mostly hover around Norma. The cluster
itself seems embedded in a wall-like structure centred at a mean velocity
of $4500-5000$\,km\,s$^{-1}$ which can be traced from about $b\sim
-30\degr$ to the Norma cluster ($b=-7\degr$) where we lose it in the
ZOA. This is also evident in the middle panel where we see part of the wall
between the Pavo II cluster (Abell S0805 at 332\degr, --24\degr,
4200\,km\,s$^{-1}$) and Norma. A weak continuation of the wall is visible
on the other side of the GP, together with an overdensity at slightly lower
velocities. The latter is probably related to CenW (see also panel~2 in
Fig.~10). 

We find a clearer continuation of these structures in the Crux cone (top
right panel). Most of the detections lie above the GP. Hardly any
detections or possible extension of the Norma cluster/wall exists below the
GP. The galaxies either form part of CenW which connects with the Centaurus
clusters visible in this cone at lower velocities at $b \simeq 20-30\degr$,
or with the higher velocity Crux clusters (Cen-Crux and CIZA
J1324.7--736). The latter two also stand out in the ZOA-longitude slice
displayed in the bottom panel. The link with the galaxy cluster in Vela
(Abell S0639) is not visible here as the cluster with
$(\ell,b,v)=(280\fdg5, -10\fdg9, 6326$\,km\,s$^{-1}$) (Stein 1996) lies
just beyond the Crux slice (and just beyond the latitude limit of the ZOA
slice).

The bottom panel shows the galaxy distribution in the traditional
``optical'' ZOA, i.e., for $|b| \la 10\degr$ for a major fraction of the
southern ZOA, where near-infrared surveys are also highly incomplete (e.g.,
Fig.~9 in Kraan-Korteweg 2005). It is satisfying to see that real LSS
including voids are now emerging within the ZOA. The \ion{H}{i} data follow
again the previously identified structures, which on average lie at
slightly higher latitudes. Distinct features are the Norma cluster with its
enormous velocity dispersion, as well as the Cen-Crux and CIZA cluster,
clearly identifiable as clusters. As a clump (rather than a finger of God)
we find an overdensity in red and green dots at about 4000\,km\,s$^{-1}$
that most certainly is a signature of the crossing of the Centaurus Wall
across the ZOA.

The wall-like structure of Norma is not obvious in this ZOA pie diagram, as
the larger structures -- from Pavo to Norma and from the Crux clusters to
Vela -- lie just beyond the ZOA, whereas the inner ZOA ($|b| < 5\degr$)
remains unprobed by this project due to the optical pre-selection of spiral
galaxy candidates. However, the pie diagram will soon be filled by the deep
Parkes HIZOA survey, where indeed evidence is found for the continuation of
the Centaurus Wall and the Norma (or GA) Wall suggested here (see
Kraan-Korteweg 2005; Henning et al. 2005 for preliminary results).

\section{Summary} \label{summary}

This paper presents pointed 21\,cm spectral line (\ion{H}{i}) observations
obtained with the 64\,m Parkes radio telescope of partially to
heavily-obscured spiral galaxies uncovered in the deep optical search for
galaxies in the Crux and GA regions of the ZOA (Woudt \& Kraan-Korteweg
2001). Out of a total of 314 observed galaxies we have detected 162. Due to
the high density of galaxies in some areas (in particular near and in the
Norma cluster), a number of pointings contain more than one detection. On
the other hand, some pointings of a non-detection show a chance detection
of a different galaxy (with a distance to the beam centre of up to
20\arcmin).

The average rms for the detections is 4.2\,mJy; 85\% of all detections have
an rms within the range $2 - 6$\,mJy. Non-detections have a slightly higher
rms of 4.7\,mJy. The mean peak flux density is 46\,mJy (with a median of
39\,mJy), which is significantly lower than the average detection limit of
the HIPASS detections ($\sim\!70$\,mJy) but similar to HIZOA (Henning et
al. 2005; Kraan-Korteweg et al. 2005). The \ion{H}{i} parameters compare
well with those found with HIPASS, and the velocities compare well with
optical velocities found in the literature.

The detection rate in the survey area is 52\%, slightly higher than in the
Hydra/Antlia survey region (45\%) obtained in the same way (Paper~I). The
exception is the core of the Norma cluster: within $1R_{\rm A}$, the
detection rate drops to 41\% as expected in case of \ion{H}{i}
deficiency. To explore this further, we have derived \ion{H}{i} deficiency
parameters for all galaxies in and around the Norma cluster. Although only
a few galaxies have actually been detected within $1 R_{\rm A}$, the lower
limits of non-detections confirm that the galaxies in the Norma cluster are
on average strongly \ion{H}{i} deficient within $0.4 R_{\rm A}$. Including
three fields observed with ATCA in the Norma cluster at an earlier stage
(Vollmer et al. 2001) and calculating the upper limits for non-detected
spiral galaxies strengthens this conclusion. It shows furthermore that
non-detected spiral galaxies between $0.4 R_{\rm A}$ and $1.0 R_{\rm A}$
also exhibit \ion{H}{i} deficiencies. The observed trend is similar to that
seen in the Coma cluster within $0.4 R_{\rm A}$ (Bravo-Alfaro et al. 2000)
and to other massive, X-ray bright galaxy clusters within $1.0 R_{\rm A}$
(Giovanelli \& Haynes 1985).

The \ion{H}{i} detections delineate large-scale structures, such as
filaments, walls and voids, deeper into the ZOA than any of the other
previous optical redshift follow-ups of the optically identified
galaxies. Most low-velocity detections lie along the SGP, while
intermediate-velocity galaxies follow the CenW. The higher-velocity
galaxies ($5000-6500$\,km\,s$^{-1}$) support evidence that the Norma Wall
crosses the ZOA, but then turns from the CenW towards Vela. Only a small
number of the detected galaxies seem to form part of the general field.

We have shown that \ion{H}{i} observations, even when dependent on optical
(or NIR) pre-identification, are quite successful in reducing the
(redshift) ZOA down to $|b| \ga 5\degr$ and in mapping LSS of highly
obscured galaxies in and across the ZOA, at least out to redshifts of about
6000\,km\,s$^{-1}$. For higher velocities the detection efficiency drops
substantially and the ZOA remains increasingly empty. Our observations are
complementary to the deep Parkes HIZOA survey of the inner optically opaque
ZOA at $|b| \le 5\degr$ (Kraan-Korteweg et al. 2005; Henning et al. 2005;
Henning et al. in prep).

With the future Square Kilometer Array (SKA) pathfinders, the Australian
ASKAP and the South African MeerKAT, the situation will be greatly
improved: we will be able to close the redshift ZOA to even lower latitudes
and to map the LSSs to higher redshifts and lower \ion{H}{i} masses. For
example, MeerKAT will not only be able to fully map the GA but also the
Shapley cluster concentration and thus possibly solve the GA/Shapley
controversy on what is the dominant contributor to the dipole motion of the
universe (Kraan-Korteweg et al. 2009 and references therein).

\begin{acknowledgements}

We first of all would like to thank P.A. Woudt for many valuable
discussions and the referee for many valuable suggestions. We have made use
of the Lyon-Meudon Extragalactic Database (LEDA), supplied by the LEDA team
at the Centre de Recherche Astronomique de Lyon, Observatoire de Lyon, and
of the NASA/IPAC Extragalactic Database (NED), which is operated by the Jet
Propulsion Laboratory, Caltech, under contract with the National
Aeronautics and Space Administration. Furthermore, this research has made
use of the Digitized Sky Surveys (produced at the Space Telescope Science
Institute under U.S. Government grant NAG W-2166) and of the HIPASS
database provided by the ATNF under the auspices of the Multibeam Survey
Working Group. RKK thanks the South African National Research Foundation
for support.

\end{acknowledgements}

\appendix

\section{Notes to individual galaxies} \label{notes}

In the following we discuss cases where the cross-identification is not
straightforward or where the signal is a combination of the signals from
two or more galaxies. In many cases we quote data from Tables~\ref{cxgadet}
--~\ref{cxgandet} (including the optical galaxy catalogue by Woudt \&
Kraan-Korteweg, 2001), notably morphological types, sizes, and optical
velocities. We have also made use of DSS\footnote{The STScI Digitized Sky
Survey} images and of images from the DENIS (Epchtein et al. 1997) and
2MASS (Skrutskie et al. 2006) surveys.

{\bf WKK\,0491/WKK\,0512:} There is a second detection in the beam of
WKK\,0491. The small lopsided signal is probably due to WKK\,0512
($d=12\farcm4$, $28\arcsec \times 8\arcsec$, SL) which is the largest
late-type spiral in the vicinity. WKK\,0493 at $d=10\farcm2$ ($28\arcsec
\times 6\arcsec$, L) is an early type galaxy and less likely to be the
candidate.

{\bf WKK\,0969/WKK\,1117:} There is a detection in the OFF observation of
WKK\,1117, close to but separated from its signal. The nearest candidate is
WKK\,0969 at $d=8\farcm5$ ($16\arcsec \times 12\arcsec$, unknown type,
$A_B=3\fm4$), which is also the only galaxy found by 2MASS in this area.

{\bf WKK\,1696} has also been detected in the beam of WKK\,1694 (not
detected) at $d=8\farcm7$ with $v=6680$\,km\,s$^{-1}$, $\Delta
V_{50}=300$\,km\,s$^{-1}$, $\Delta V_{20}=318$\,km\,s$^{-1}$,
$I=2.47$\,Jy\,km\,s$^{-1}$, rms = 2.4\,mJy.

{\bf WKK\,2163:} The ID is ambiguous. The observed velocity is
$v=3533$\,km\,s$^{-1}$ , and WKK\,2160 at $d=3\farcm3$ ($27\arcsec \times
24\arcsec$, S) has an optical velocity of $3512\pm58$\,km\,s$^{-1}$
(FW98). It is possible that we have observed WKK\,2160, but if the two
galaxies are genuine companions the signal probably comes from WKK\,2163
($74\arcsec \times 56\arcsec$, S6) which is the larger of the two. HIPASS
can barely resolve the positions, but seems to favour WKK\,2163.

{\bf WKK\,2245:} It is possible that WKK\,2240 (ESO173-G015, $d=2\farcm6$,
S, $85\arcsec \times 12\arcsec$) with an optical velocity of
$v=3006\pm36$\,km\,s$^{-1}$ (SH92) contributes to the signal.

{\bf WKK\,2372/WKK\,2402:} The distance between the galaxies is
$d=16\farcm1$, both have been detected at similar velocities, but the
profiles are not confused.

{\bf WKK\,2377/WKK\,2388/WKK\,2406:} This is a group of four galaxies:
WKK\,2377 (S7:, $70\arcsec \times 23\arcsec$), WKK\,2375 (S5, $62\arcsec
\times 56\arcsec$, $4290\pm37$\,km\,s$^{-1}$ FH95), WKK\,2388 (S5,
$36\arcsec \times 16\arcsec$, $3976\pm40$\,km\,s$^{-1}$, FH95), and
WKK\,2406 (SL, $55\arcsec \times 38\arcsec$). WKK\,2388 lies between
WKK\,2377 and WKK\,2406 and the pointing of WKK\,2388 shows a confusion of
the two latter profiles at $v\simeq4100 - 4400$\,km\,s$^{-1}$. The signal
at $v\simeq3800 - 4050$\,km\,s$^{-1}$ belongs to WKK\,2388 proper and the
line width is assumed to be unaffected. However, due to difficulties with
fitting a good baseline the flux density is somewhat uncertain. Using
HIPASS we have found that WKK\,2375 does not add (significantly) to the
combined signal. The observations of WKK\,2377 and WKK\,2406 at their
respective pointings are not confused at the noise level.

{\bf WKK\,2390/WKK\,2392:} The close galaxy pair (at $d=1\farcm2$ and
$d=3\farcm2$, respectively, from the pointing position) is unresolved, and
the parameters given in Table~\ref{cxgadet} refer to the full profile. The
quoted velocity by VY96 refers to the IRAS detection which is also
unresolved.

{\bf WKK\,2503:} Also observed with HIZSS, the spectrum shows a profile
reminiscent of the blended signal of two objects. WKK\,2503 has a bright
bulge and large faint halo with a bright star superimposed close to the
bulge. No other galaxy in this area is visible either in the optical or in
the NIR. At an extinction of $A_B=3\fm7$ a late-type/LSB galaxy,
WKK\,2503B, could be invisible even in the NIR. Table~\ref{cxgadet} gives
the parameters of the full profile. The two velocities found in the
literature refer to \ion{H}{i} observations and are equally unresolved.

{\bf WKK\,2576/WKK\,2595/WKK\,2597:} The observed profile is a combination
of the signals from three galaxies: WKK\,2595 (S6, $102\arcsec \times
40\arcsec$) and WKK\,2597 (S5, $59\arcsec \times 47\arcsec$,
$3973\pm46$\,km\,s$^{-1}$, SH92) are a close galaxy pair ($d=1\farcm2$);
according to HIPASS, WKK\,2576 at $d=8\farcm4$ (S5, $86\arcsec \times
75\arcsec$, $3948\pm70$\,km\,s$^{-1}$, DN97) has a strong narrow \ion{H}{i}
profile. Table~\ref{cxgadet} gives the measurement of the full profile for
the unresolved pair WKK\,2595/WKK\,2597, while, through comparison with
HIPASS, the width and velocity of the peak at
$v\simeq3800-3950$\,km\,s$^{-1}$ is given for WKK\,2576. By removing this
peak we have re-measured the width and velocity of the underlying
low-intensity profile for the close galaxy pair and found
$v=3889$\,km\,s$^{-1}$, $\Delta V_{50}=266$\,km\,s$^{-1}$, $\Delta
V_{20}=313$\,km\,s$^{-1}$.

{\bf WKK\,2640/WKK\,2644:} There are two detections in the beam of
WKK\,2640: the narrow spike at $v=3705$\,km\,s$^{-1}$ is WKK\,2640 (I,
$51\arcsec \times 42\arcsec$), while the detection at
$v=9404$\,km\,s$^{-1}$ is WKK\,2644 (SM, $26\arcsec \times 9\arcsec$) with
an optical velocity of $9406\pm100$\,km\,s$^{-1}$ (WK04) at
$d=4\farcm3$. Due to the very lopsided profile of WKK\,2644 the
high-velocity end is uncertain.

{\bf WKK\,2844/WKK\,2863:} Two galaxies contribute to the detected
signal. HIPASS shows that WKK\,2863 (S5, $98\arcsec \times 83\arcsec$,
$3778\pm30$\,km\,s$^{-1}$, SE95) at $d=8\farcm7$ has a strong profile with
$v\simeq3600-3850$\,km\,s$^{-1}$. At the position of WKK\,2844 the observed
profile is smaller but extends to $v\simeq3950$\,km\,s$^{-1}$; it is
therefore assumed that WKK\,2844 has been detected but it remains
unresolved, with a velocity slightly larger than the one for
WKK\,2863. Table~\ref{cxgadet} gives the parameters of the full profile for
WKK\,2863, which is considered to be the main contributor to the signal.

{\bf WKK\,2924/WKK\,2938:} There are two detections in the beam of
WKK\,2924: WKK\,2938 (L, $34\arcsec \times 22\arcsec$) at $d=7\farcm3$ has
an optical velocity of $3024\pm157$\,km\,s$^{-1}$ (FW98) which agrees with
the narrow peak at $v=2864$\,km\,s$^{-1}$. WKK\,2924 (S8, $58\arcsec \times
22\arcsec$) is a more likely candidate for the signal at
$v=3410$\,km\,s$^{-1}$. HIPASS also shows the latter detection at the
position of WKK\,2924, while nothing can be seen at the position of
WKK\,2938.

{\bf WKK\,2993} has a close companion (Woudt \& Kraan-Korteweg 2001) which
might contribute to the signal.

{\bf WKK\,3002/WKK\,3006:} The two detections in the beam of WKK\,3002 can
not unambiguously identified: WKK\,3002 (SL?, $56\arcsec \times 20\arcsec$)
is more likely to be the stronger signal at $v=3436$\,km\,s$^{-1}$
(cf.\ HIPASS), while the galaxy at $v=2820$\,km\,s$^{-1}$ is more likely
WKK\,3006 ($13\arcsec \times 8\arcsec$, no type).

{\bf WKK\,3023:} With $A_B=22^{\rm m}$ the galaxy is unlikely to be real,
and nothing is visible on DENIS or 2MASS images.

{\bf WKK\,4016/WKK\,4022:} The profile is due to the blending of two
signals. HIPASS shows that the high narrow peak at
$v\simeq4680$\,km\,s$^{-1}$ is WKK\,4016 (SL, $67\arcsec \times 48\arcsec$)
at $d=12\farcm1$, while the broader profile is probably due to WKK\,4022
proper (S5, $91\arcsec \times 34\arcsec$). Table~\ref{cxgadet} gives the
parameters for the full profile for WKK\,4022 and the measurements of the
narrow peak alone for WKK\,4016; all parameters are uncertain.

{\bf WKK\,5240:} The detection in the beam of WKK\,5267 (not detected) at
$d=11\farcm7$ is WKK\,5240 (S, $157\arcsec \times 13\arcsec$;
cf.\ HIPASS). The profile shape is very noisy and the parameters are
uncertain.

{\bf WKK\,5285:} This galaxy has been detected in the beams of three other
galaxies: in the OFF observations of WKK\,5534 ($d=6\farcm5$) and of
WKK\,5556 ($d=2\farcm7$), as well as in the beam of WKK\,5297 (not
detected) at $d=6\farcm1$. The detection with the smallest distance to the
beam centre is listed in Table~\ref{cxgadet} and shown in
Fig.~\ref{hiprofile}, while the detection in the OFF observation of
WKK\,5524 is least affected by an RFI at $v=5900$\,km\,s$^{-1}$ next to the
signal. The other measurements are: $v=5631$\,km\,s$^{-1}$, $\Delta
V_{50}=353$\,km\,s$^{-1}$, $\Delta V_{20}=383$\,km\,s$^{-1}$,
$I=15.65$\,Jy\,km\,s$^{-1}$, rms $= 5.3$\,mJy (in the beam of WKK\,5297);
and $v=5635$\,km\,s$^{-1}$, $\Delta V_{50}=357$\,km\,s$^{-1}$, $\Delta
V_{20}=396$\,km\,s$^{-1}$, $I=15.17$\,Jy\,km\,s$^{-1}$, rms $=3.3$\,mJy (in
the OFF observation of WKK\,5534).

{\bf WKK\,5366:} The identification is uncertain: the optical velocity is
$4822\pm82$\,km\,s$^{-1}$ (WK04), but both HIPASS and JS00 confirm the
\ion{H}{i} signal to be strongest at the position of WKK\,5366. Since the
extinction here is $A_B=3\fm8$, an obscured galaxy close by cannot be
excluded.

{\bf WKK\,5443OFF:} The detection found in the OFF observation of WKK\,5443
was subsequently searched for the `best' position. It has also been
detected by HIZSS and JS00. No galaxy could be found in the optical or NIR
(DENIS, 2MASS).

{\bf WKK\,5562/WKK\,5616/WKK\,5642:} In the observations of both WKK\,5562
and WKK\,5642 a narrow single peak appears at
$v\simeq4160$\,km\,s$^{-1}$. Using HIPASS we determined that this signal
most likely comes from WKK\,5616, a late-type galaxy ($19\arcsec \times
5\arcsec$) at $d=5\farcm3$ from WKK\,5642 (listed in Table~\ref{cxgadet})
and $d=12\farcm2$ from WKK\,5562 with the following parameter:
$v=4157$\,km\,s$^{-1}$, $\Delta V_{50}=42$\,km\,s$^{-1}$, $\Delta
V_{20}=75$\,km\,s$^{-1}$, $I=1.95$\,Jy\,km\,s$^{-1}$, rms $= 4.7$\,mJy.

{\bf WKK\,5595} is included in the catalogue since it is very close to the
observed but undetected WKK\,5597 ($28\arcsec \times 11\arcsec$, L?) at
$d=0\farcm7$ and is of comparable size ($30\arcsec \times 19\arcsec$, type
unknown), that is, the rms can be considered an upper limit for both
galaxies.

{\bf WKK\,5642/WKK\,5659:} There are three detections in the beam of
WKK\,5642 ($48\arcsec \times 17\arcsec$, SM): the spike at
$v=4167$\,km\,s$^{-1}$ is WKK\,5616 and has been discussed above. The
signal at $v=6446$\,km\,s$^{-1}$ is assumed to belong to WKK\,5642 since it
also has two optical velocities of $6045\pm42$\,km\,s$^{-1}$ (SH92) and
$v=6118\pm100$\,km\,s$^{-1}$ (WK04), though this is only in moderate
agreement. WKK\,5670 ($24\arcsec \times 8\arcsec$, SE:) at $d=6\farcm6$
with an optical velocity of $v=6329\pm44$\,km\,s$^{-1}$ (WK04) can be
excluded as a candidate since it is only $d=5\farcm6$ from WKK\,5694, which
has also been observed and shows no signal at
$v\simeq6400$\,km\,s$^{-1}$. A third signal has been found at
$v=4418$\,km\,s$^{-1}$ which is likely to be WKK\,5659 ($44\arcsec \times
15\arcsec$, S6) at $d=3\farcm4$.

{\bf IC\,4584/IC\,4585:} This is a detection in the beam of WKK\,5581 (not
detected). IC\,4584 and IC\,4585 are two large spiral galaxies at
$d\simeq9\arcmin$ from the pointing. HIPASS confirms their identity but
cannot resolve the pair, though clearly both galaxies contribute to the
signal. The measurement of the full profile is given in
Table~\ref{cxgadet}.

{\bf WKK\,5729:} The three observations of WKK\,5733
($v_{opt}=6215\pm92$\,km\,s$^{-1}$, WK99), WKK\,5694
($v_{opt}=3412\pm36$\,km\,s$^{-1}$, FH95), and WKK\,5709 show a very
similar profile both in line width and peak flux at $v=5729$\,km\,s$^{-1}$,
$v=5730$\,km\,s$^{-1}$, and $v=5723$\,km\,s$^{-1}$, respectively. The left
horn appears to come from WKK\,5768 (also detected in the beam of
WKK\,5780), which lies at a distance of $d=13\farcm5$, $d=17\farcm9$, and
$d=11\farcm7$, respectively, from the three pointings; the peak flux of
this horn varies according to the distance. Due to the similar peak flux
and high velocity end of the rest of the profile we conclude that the
detected galaxy must lie at a similar distance from these three
pointings. WKK\,5729 ($48\arcsec \times 16\arcsec$, SL) lies at
$d=6\farcm9$, $d=9\farcm1$, and $d=5\farcm9$ from WKK\,5733, WKK\,5694, and
WKK\,5709, respectively. It is a late-type spiral and therefore not visible
with 2MASS and DENIS. The signal is too weak to be detectable with HIPASS.

{\bf WKK\,5768} is detected in the beam of WKK\,5780 (not detected) at
$d=9\farcm5$. The low-velocity horn is also visible in the observations of
WKK\,5709 ($d=11\farcm7$; see plot of WKK\,5729), WKK\,5733 ($d=13\farcm5$)
and WKK\,5694 ($d=17\farcm9$).

{\bf WKK\,5993/WKK\,5999:} The detection in the observation of WKK\,5993 is
the blending of two signals. The low-velocity double-horn comes from
WKK\,5999 (also observed, see Fig.~\ref{hiprofile}), while the
high-velocity part is due to WKK\,5993. The parameters for WKK\,5993 in
Table~\ref{cxgadet} have been measured by cutting off the profile of
WKK\,5999 at $v\simeq3350$\,km\,s$^{-1}$. All the parameters are uncertain
since the low-velocity end of WKK\,5993 is undetermined. For the WKK\,5999
profile we have measured $v=3261$\,km\,s$^{-1}$, $\Delta
V_{50}=180$\,km\,s$^{-1}$, $\Delta V_{20}=226$\,km\,s$^{-1}$,
$I=8.83$\,Jy\,km\,s$^{-1}$, rms $= 2.3$\,mJy.

{\bf WKK\,6187} is included in the catalogue since it is very close to the
observed but undetected WKK\,6189 ($13\arcsec \times 8\arcsec$, E) at
$d=0\farcm9$ and is slightly larger ($22\arcsec \times 22\arcsec$, type
unknown), that is, the rms can be considered an upper limit for both
galaxies.

{\bf WKK\,6535/WKK\,6570:} There are two detections in the beam of
WKK\,6570. WKK\,6535 ($39\arcsec \times 9\arcsec$, S5) lies at
$d=6\farcm5$ and is likely to be the detection at
$v\simeq4150$\,km\,s$^{-1}$. Since WKK\,6570 ($60\arcsec \times 27\arcsec$,
S3) is the larger and brighter of the two we have assumed it to be the
closer galaxy at $v=2938$\,km\,s$^{-1}$, but the identities remain
ambiguous.

{\bf WKK\,6594}: The \ion{H}{i} galaxy is probably identical with the IRAS
galaxy at $d=2\farcm2$ with $v=642\pm35$\,km\,s$^{-1}$ (SH92).

{\bf WKK\,6689/WKK\,6732:} The two galaxies with similar velocities lie
$9\farcm6$ apart. The observation of WKK\,6732 shows no significant
confusion with the signal of WKK\,6689 (though the flux density may be
uncertain), while the profile for WKK\,6689 is more uncertain.

{\bf WKK\,7287/WKK\,7289:} The small signal at $v=5740$\,km\,s$^{-1}$
detected in the beam of WKK\,7289 is probably WKK\,7287 at $d=3\farcm3$
($30\arcsec \times 20\arcsec$, I).

{\bf WKK\,7460/WKK\,7463} is an interacting system with a separation of
$1\farcm4$: WKK\,7460 ($198\arcsec \times 105\arcsec$, SL) is the larger
component with an optical velocity of $775\pm36$\,km\,s$^{-1}$ (SH92),
while the profile gives $v=842$\,km\,s$^{-1}$. Table~\ref{cxgadet} gives
the full parameters for WKK\,7460 only, since the contribution by WKK\,7463
($82\arcsec \times 67\arcsec$, S) is uncertain. However, considering the
types and sizes of the two galaxies as well as the \ion{H}{i} velocity as
compared to the optical of WKK\,7460, we can assume that WKK\,7463
contributes to the profile.

{\bf WKK\,7465/WKK\,7198:} WKK\,7198 has been detected in the OFF
observation of WKK\,7465 at $d=7\farcm2$, and the profiles overlap. The
observation of WKK\,7198 (see Fig.~\ref{hiprofile}) shows that the profile
extends from $v\simeq3270$\,km\,s$^{-1}$ to $\sim3540$\,km\,s$^{-1}$. The
profile of WKK\,7465 is therefore truncated and no line widths and flux
could be derived. The systemic velocity is likely to be higher than the one
given.

{\bf WKK\,7652/WKK\,7689:} WKK7652 has been detected in the beam of
WKK\,7689 (not detected) at $d=11\farcm2$. Optical velocities for this
galaxy are $v=1350\pm31$\,km\,s$^{-1}$ (RC3) and
$v=1478\pm38$\,km\,s$^{-1}$ (WK04), while other \ion{H}{i} measurements
find $v=1482\pm6$\,km\,s$^{-1}$ (RC3). While we find
$v=1519$\,km\,s$^{-1}$, HK01 has detected WKK\,7689 at
$v=1559\pm3$\,km\,s$^{-1}$ in \ion{H}{i} with the radio telescope at
Effelsberg, which has a smaller beam size ($9\arcmin$ as compared to
$15\arcmin$ for Parkes). We can therefore not exclude that part of the
signal in our observation comes from WKK\,7689.

{\bf WKK\,7776} has also been detected in the beam of WKK\,7794 at
$d=10\farcm7$ with $v=2790$\,km\,s$^{-1}$, $\Delta
V_{50}=45$\,km\,s$^{-1}$, $\Delta V_{20}=55$\,km\,s$^{-1}$,
$I=4.11$\,Jy\,km\,s$^{-1}$, rms $= 5.6$\,mJy.

\section{Galaxies in the Vela region }

A number of galaxies outside of the Crux and GA region were also observed
within this observing programme, i.e., galaxies in the Vela region
($245\degr \ga \ell \ga 275\degr$). They were taken from the ZOA deep
optical galaxy catalogue (Salem \& Kraan-Korteweg, in prep.; see
Kraan-Korteweg \& Lahav 2000 for preliminary results) that covers the
region between Puppis (Saito et al. 1991) and the Hydra/Antlia region
(Kraan-Korteweg 2000); see Figs.\~2 and~3 in Kraan-Korteweg \& Lahav 2000
for an outline of the surveyed area and the distribution of the uncovered
galaxies.

\begin{figure*}[ht]
\vspace{-11cm}
\resizebox{\hsize}{!}{\includegraphics{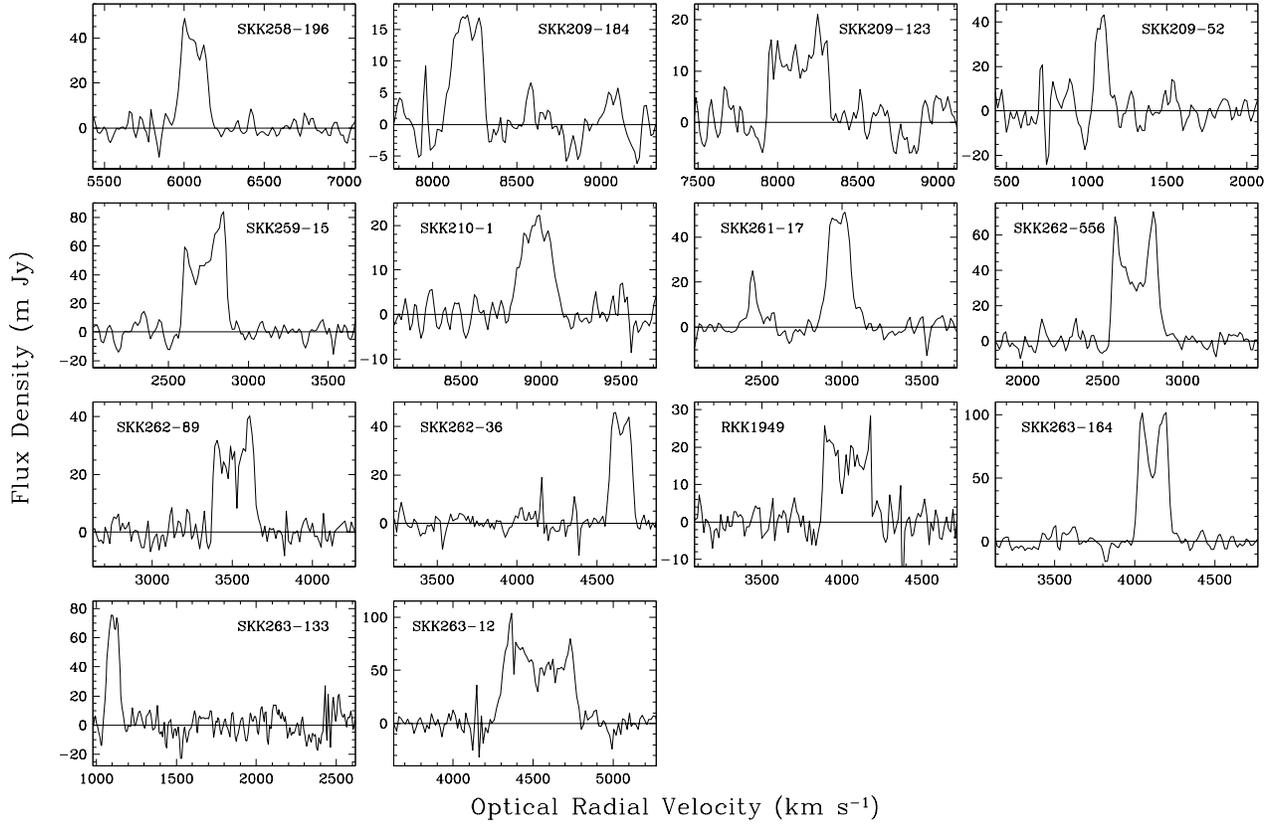}}
\caption[]{\ion{H}{i} profiles of the 14 \ion{H}{i} detections in the Vela
  region. The vertical axis gives the flux density in mJy, the horizontal
  axis the velocity range (radio convention), generally centred on the
  radio velocity of the galaxy displaying a width of 1600\,km\,s$^{-1}$.
  All spectra are baseline-subtracted and generally Hanning-smoothed. The
  respective identifications are given within the panels.
}
\label{velaprofile}
\vspace{-1cm}
\end{figure*}

The data of the detected (N=14) and non-detected galaxies (N=15) are
presented in Tables~\ref{veladet} and~\ref{velandet}, which are equivalent
to Tables~\ref{cxgadet} and~\ref{cxgandet} (see Sects~\ref{det}
and~\ref{ndet} for the column descriptions). The profiles of the detected
galaxies are shown in Fig.~\ref{velaprofile}, and Table~\ref{vel2tab} gives
the independent velocity measurements found in the literature for the
detected galaxies (see the description of Table~\ref{veltab} in
Sec~\ref{det}). The reference for KF95 is Kraan-Korteweg et al. (1995).

\begin{table}[ht]
\normalsize
\caption{Comparison of velocities for Vela galaxies. }
\label{vel2tab}
\scriptsize  
\begin{tabular}{l@{\extracolsep{1mm}}r@{\extracolsep{0mm}}cll@{\extracolsep{3mm}}l}
\noalign{\smallskip}
\hline
\noalign{\smallskip}
 Ident. & {$V_{hel}$} & & \multicolumn{1}{c}{$V_{other}$} & origin & Reference \\
& km/s & & \multicolumn{1}{c}{km/s} & & \\
\vspace{-1mm} \\
\ \ \ (1) & (2) & & \multicolumn{1}{c}{(3)} & (4) & (5) \ \ \\
\noalign{\smallskip}
\hline
\noalign{\smallskip}
SKK259- 15L & 2730& &            $2723\pm\phantom{00}7$  & \ion{H}{i}  & HIPASS   \\
SKK261- 17L & 2979& &            $2974\pm\phantom{0}10$  & \ion{H}{i}  & HIPASS   \\
SKK262- 89L & 3513& &            $3451\pm\phantom{0}50$  & opt  & FW98 \\
SKK262- 36L & 4650& &            $4651\pm\phantom{00}9$  & \ion{H}{i}  & HIPASS   \\
RKK1949     & 4030& &            $4002\pm100$            & opt  & KF95 \\
            &     & &            $4037\pm\phantom{0}36$  & opt  & SH92 \\
SKK263-164L & 4118& &            $4116\pm\phantom{00}1$  & \ion{H}{i}  & DN96 \\
SKK263-133L & 1106& &  $\phantom{0}969\pm\phantom{0}54$  & opt  & FH95 \\
            &     & &            $1106\pm\phantom{00}6$  & \ion{H}{i}  & HIPASS   \\
SKK263- 12L & 4540& &            $4600\pm\phantom{0}40$  & opt  & RC3  \\
            &     & &            $4545\pm\phantom{00}6$  & \ion{H}{i}  & HIPASS   \\
\noalign{\smallskip}
\hline
\end{tabular}
\normalsize
\end{table}

\begin{landscape}  
\begin{table}[h]
 \normalsize
 \renewcommand{\baselinestretch}{0.65}
\caption{\ion{H}{i}-detections in the Vela region}
\label{veladet}
\scriptsize  
\begin{tabular*}{23.9cm}{
  l  @{\extracolsep{1mm}} l @{\extracolsep{3mm}} l@{\extracolsep{-1mm}} l@{\extracolsep{3mm}}      
  l@{\extracolsep{3mm}} l @{\extracolsep{3mm}}  r @{\extracolsep{2mm}} r @{\extracolsep{3mm}} 
  r @{\extracolsep{0mm}}  l @{\extracolsep{0mm}}  c @{\extracolsep{0mm}}   l@{\extracolsep{1mm}}
  r @{\extracolsep{0mm}} l @{\extracolsep{2mm}}
  l @{\extracolsep{0mm}} c @{\extracolsep{0mm}}   
 r @{\extracolsep{0.mm}} c @{\extracolsep{2mm}} r @{\extracolsep{0.mm}} c @{\extracolsep{2mm}}
 r @{\extracolsep{0.mm}} c @{\extracolsep{1mm}} r @{\extracolsep{0.mm}} c @{\extracolsep{1mm}}
 r @{\extracolsep{1mm}} c @{\extracolsep{1mm}} 
 c @{\extracolsep{0mm}} r @{\extracolsep{1mm}} r @{\extracolsep{3mm}} l @{\extracolsep{0mm}}
}
\noalign{\smallskip}
\hline
\noalign{\smallskip}
 Ident. & Other & IR & & \ \ \ \ R.A. & \ \ \ Dec.& gal $\ell$ \ & gal $b$ \ &
 \multicolumn{4}{l}{Type} & 
 \multicolumn{2}{c}{$D$ x $d$} & 
 $B_{J}$ &  $E_{(B-V)}$ &  
 {$V_{hel}$} & & {$\Delta V_{50}$} & & {$\Delta V_{20}$} & & {$I \ \ $} & &
 {rms} & hann & 
 N & dist & {$I_c \ $} &  excised RFI \\
& & & &
(h\,\, m\,\, s) & ($\deg$\,\, $\arcmin$\,\,$\arcsec$) & ($\deg$) \ &($\deg$) \ &
& & & &
\multicolumn{2}{c}{($\arcsec$)} & ($^{\rm m}$) & ($^{\rm m}$) & 
km/s & & km/s & & km/s & & {Jy\,km/s} & & m\,Jy & & & {($\arcmin$)}& {Jy\,km/s} & km/s \\
\vspace{-1mm} \\
\ \ \ (1) & \ \ \ (2) & (3) & & \ \ \ \ (4) & \ \ \ \ (5) & (6) \ & (7) \
 & \multicolumn{4}{c}{(8)} \ & \multicolumn{2}{c}{(9)} \ & \ (10) &
 {(11)} & (12) & & (13) & & (14) & & (15) & & (16)  & (17) & (18) & (19) & (20) & (21) \ \ \\
\noalign{\smallskip}
\hline
\noalign{\smallskip}
SKK258-196L & ESO258-G002 & & & 07 58 13.9 & -47 01 09 & 261.52 & -9.16 & S& & &     & 60x&\phantom{0}40 & 14.9 & 0.32 & 6174& & 177& & 220& &  7.25& &  3.5 & & &&&  \phantom{0}5800 \\
SKK209-184L & ESO209-G016 &M&I& 08 05 29.0 & -48 50 56 & 263.74 & -9.06 & S&B&R&3    & 54x&\phantom{0}54 & 14.6 & 0.29 & 8196& & 198& & 240& &  2.98& &  3.0 & & &&&                  \\
SKK209-123L & ESO209-G017 &M& & 08 08 33.2 & -49 03 50 & 264.20 & -8.74 & S& & &     & 54x&\phantom{0}12 & 16.2 & 0.26 & 8130& & 378& & 395& &  4.71& &  3.3 & & &&&                  \\
SKK209- 52  &             & & & 08 11 34.2 & -47 08 41 & 262.83 & -7.29 & I& & &     & 34x&\phantom{0}13 & 16.9 & 0.61 & 1091&:&  79&:&  99&:&  3.49&:& 10.2 & &*&&&  \phantom{0}1000 \\
SKK259- 15L & ESO259-G001 & & & 08 17 31.4 & -45 37 11 & 262.11 & -5.60 & S& & &L?   & 34x&\phantom{0}19 & 16.1 & 0.81 & 2730& & 278& & 299& & 15.44& &  5.7 & & &&&                  \\[0.1cm]
SKK210-  1  &             & & & 08 46 54.4 & -47 47 53 & 266.85 & -2.82 &? & & &     & 47x&\phantom{0}13 & 16.8 & 1.00 & 8969& & 210& & 285& &  4.20& &  2.8 & & &&&                  \\
SKK261- 17L & ESO261-G006 &M& & 09 23 34.0 & -42 45 39 & 267.42 &  5.34 & S& & &5    & 40x&\phantom{0}40 & 15.5 & 0.65 & 2979& & 156& & 207& &  7.79& &  3.5 & & &&&                  \\
SKK262-556L & ESO262-G004 & & & 09 46 22.5 & -46 38 38 & 273.06 &  5.21 & S& & &     &108x&\phantom{0}11 & 15.8 & 0.46 & 2709& & 288& & 319& & 13.73& &  4.2 & & &&&                  \\                                      
SKK262- 89L & ESO262-G015 & &I& 10 02 39.2 & -45 29 58 & 274.56 &  7.87 & S& & &1    & 60x&\phantom{0}24 & 15.1 & 0.19 & 3513& & 264& & 282& &  7.18&:&  3.5 & & &&&  \phantom{0}4150 \\
SKK262- 36L & ESO262-G016 &M& & 10 03 11.6 & -42 49 08 & 272.99 & 10.06 & S& & &3    & 71x&\phantom{0}44 & 14.5 & 0.16 & 4650& & 144& & 169& &  5.92& &  3.0 & & &&&                  \\[0.1cm]
RKK1949     & ESO263-G004 &M&I& 10 06 46.9 & -47 41 48 & 276.45 &  6.53 & S&B& &1    & 54x&\phantom{0}47 & 14.7 & 0.20 & 4030& & 302& & 316& &  5.26&:&  3.3 & & &&&  3350/4150       \\
SKK263-164L & ESO263-G016 &M& & 10 12 32.4 & -45 14 13 & 275.82 &  9.11 & S& & &3    &108x&\phantom{0}81 & 13.4 & 0.19 & 4118& & 197& & 221& & 16.08& &  5.3 & & &&&                  \\
SKK263-133L & ESO263-G021 & &I& 10 14 42.0 & -44 50 59 & 275.91 &  9.64 &  & & &     & 74x&\phantom{0}60 & 14.1 & 0.21 & 1106& &  83& & 106& &  5.92& &  7.8 & & &&&                  \\
SKK263- 12L & ESO263-G035 & &I& 10 26 14.4 & -45 44 13 & 278.12 & 10.02 & S& & &1    &121x&\phantom{0}47 & 13.6 & 0.18 & 4540& & 450&:& 491&:& 27.61&:&  7.1 & & &&&  \phantom{0}4350 \\
\noalign{\smallskip}
\hline
\noalign{\smallskip}
\end{tabular*}
\newline
{\bf Notes}: SKK209- 52: the RFI is next to the profile and possibly
affects the line width. 
\normalsize
\end{table}

\begin{table}[h]
 \normalsize
 \renewcommand{\baselinestretch}{0.65}
\caption{\ion{H}{i} non-detections in the Vela region}
\label{velandet}
\scriptsize  
\begin{tabular*}{23.5cm}{
 l  @{\extracolsep{2mm}} l @{\extracolsep{3mm}} l@{\extracolsep{-1mm}} l@{\extracolsep{3mm}} 
  l@{\extracolsep{3mm}} l @{\extracolsep{3mm}}  r @{\extracolsep{2mm}} r @{\extracolsep{3mm}} 
  r @{\extracolsep{0mm}}  l @{\extracolsep{0mm}}  c @{\extracolsep{0mm}}   l@{\extracolsep{1mm}}
  r @{\extracolsep{0mm}} l @{\extracolsep{2mm}}
  l @{\extracolsep{0mm}} c @{\extracolsep{1mm}}   
 r @{\extracolsep{2mm}} r @{\extracolsep{3mm}} r @{\extracolsep{2mm}} 
 c @{\extracolsep{1mm}} r @{\extracolsep{3mm}} l @{\extracolsep{2mm}} l @{\extracolsep{3mm}} l                     
}
\noalign{\smallskip}
\hline
\noalign{\smallskip}
 \multicolumn{1}{c}{Ident.} & \multicolumn{1}{c}{Other} & IR & & \multicolumn{1}{c}{R.A.} & \multicolumn{1}{c}{Dec.}
& gal $\ell$ \ & gal $b$ & \multicolumn{4}{l}{Type} & \multicolumn{2}{c}{$D$ x $d$} & 
 $B_{J}$ &  $E_{(B-V)}$ & \multicolumn{1}{c}{$V_{{\rm range}}^{{\rm obs}}$} & rms &
 \multicolumn{1}{c}{$V_{{\rm range}}^{{\rm pert}}$} & N & dist & \multicolumn{1}{c}{$V_{other}$} & Ref & origin\\
& &  &  & (h\,\, m\,\, s) & \ ($\deg$\,\, $\arcmin$\,\, $\arcsec$) & ($\deg$) \ \ & ($\deg$) \ &
& & & & \multicolumn{2}{c}{($\arcsec$)} & ($^{\rm m}$) & ($^{\rm m}$) & 
\multicolumn{1}{c}{km/s} & m\,Jy & \multicolumn{1}{c}{km/s} & & {($\arcmin$)} \ &\multicolumn{1}{c}{km/s} & & \\
\vspace{-1mm} \\
\multicolumn{1}{c}{(1)} & \multicolumn{1}{c}{(2)} & (3) && \multicolumn{1}{c}{(4)} & \multicolumn{1}{c}{(5)} 
& \multicolumn{1}{c}{(6)} & (7) \ & \multicolumn{4}{c}{(8)} \ & \multicolumn{2}{c}{(9)} & (10) &
\multicolumn{1}{c}{(11)} & \multicolumn{1}{c}{(12)} & (13) & \multicolumn{1}{c}{(14)} 
& (15) & (16) & \multicolumn{1}{c}{(17)} & (18) & (19) \\
\noalign{\smallskip}
\hline
\noalign{\smallskip}
SKK258-137  &               &M& & 07 42 49.8 & -44 12 27 & 257.70 &-10.14 & S& & & ? & 47x&\phantom{0}27 & 15.9 & 0.31 &$  300 -          10250 $& 4.0    &                          & &     & & & \\
SKK258- 89  &               &M& & 07 55 07.4 & -45 00 26 & 259.49 & -8.62 & S& & &M  & 54x&\phantom{00}8 & 17.0 & 0.23 &$  300 -          10250 $& 3.2    &   $700 -\phantom{0}1100$ &*&     & & & \\
SKK209-209  &LEDA 3098730   &M& & 08 04 53.1 & -49 08 50 & 263.95 & -9.29 & S& & &   & 54x&\phantom{00}8 & 16.8 & 0.29 &$  500 -          10250 $& 3.5    &                          & &     & & & \\
SKK258- 14  &               &M& & 08 06 10.6 & -43 52 28 & 259.53 & -6.34 & S& & &M  & 40x&\phantom{0}24 & 15.8 & 0.43 &$  400 -          10250 $& 4.5    &                          & &     & & & \\
SKK312-  4  &               &M& & 08 09 18.5 & -38 38 46 & 255.42 & -3.04 &? & & &   & 20x&\phantom{0}20 & 16.6 & 2.45 &$  400 -          10350 $& 3.8    &   $900 -\phantom{0}1100$ & &     & & & \\[0.1cm]
SKK259- 18  &               &M& & 08 14 20.1 & -46 38 07 & 262.65 & -6.62 &?S& & &L  & 67x&\phantom{00}7 & 16.9 & 0.69 &$  500 -          10250 $& 7.3    &                          & &     & & & \\
SKK262-425L &ESO262-G006    &M& & 09 48 40.3 & -45 29 46 & 272.62 &  6.35 & S& & &   & 74x&\phantom{00}5 & 17.0 & 0.19 &$  300 -          10350 $& 4.4    &                          & &     & & & \\
SKK262-361  &ESO262-G009    & & & 09 51 31.9 & -44 01 14 & 272.07 &  7.81 & S& & &   & 67x&\phantom{00}7 & 16.6 & 0.25 &$  400 -          10250 $& 3.3    &   $300 -\phantom{00}400$ & &     & & & \\
            &               & & &            &           &        &       &  & & &   &    &\phantom{0}   &      &      &                         &        &   $850 -\phantom{0}1050$ & &     & & & \\
SKK262-303L &ESO262-G010    &M& & 09 53 47.3 & -45 58 02 & 273.62 &  6.55 & S& & &   & 54x&\phantom{0}13 & 16.0 & 0.22 &$  300 -          10250 $& 4.1    &   $820 -\phantom{0}1020$ & &     & & & \\[0.1cm]
SKK262-127  &ESO316-G015    &M& & 10 00 53.3 & -42 41 03 & 272.57 &  9.91 & S& & &   & 74x&\phantom{0}11 & 16.0 & 0.23 &$  300 -          10250 $& 4.6    &   $800 -\phantom{0}1070$ &*&     &$\phantom{0}9796\pm\phantom{0}10$ & MF96 & opt \\
SKK262- 59  &ESO316-G019    & & & 10 02 47.0 & -42 34 42 & 272.78 & 10.21 & S& & &   & 67x&\phantom{00}9 & 16.3 & 0.19 &$  300 -          10250 $& 4.1    &   $800 -\phantom{0}1030$ &*&     & & & \\
SKK263- 65L &ESO263-G008    & & & 10 07 46.1 & -43 29 40 & 274.08 & 10.02 & S& & &   & 83x&\phantom{0}16 & 15.4 & 0.15 &$ 5100 -          10250 $& 4.0    &                          & &     & & & \\
SKK263-157L &ESO263-G014    &M&I& 10 11 00.2 & -45 09 06 & 275.54 &  9.02 & S& & &1? & 67x&\phantom{0}50 & 14.4 & 0.19 &$ 1600 -\phantom{0}6700 $& 4.5    &                          & &     &$\phantom{0}5200\pm100$ & F83 & opt \\
SKK263- 53L &ESO263-G024    &M&I& 10 15 33.7 & -45 07 58 & 276.20 &  9.50 & S& & &E? & 87x&\phantom{0}40:& 14.3 & 0.20 &$ -600 -\phantom{0}4650 $& 6.3    &  $-100 -\phantom{00}300$ & &     &$\phantom{0}1050\pm100$ & F83 & opt \\[0.1cm]
            &               & & &            &           &        &       &  & & &   &    &\phantom{0}   &      &      &                         &        &   $750 -\phantom{0}1100$ & &     & & & \\
SKK263- 20L &ESO263-G029    & & & 10 21 09.8 & -46 00 53 & 277.52 &  9.31 & S& & &E  &108x&\phantom{0}30 & 14.2 & 0.25 &$  300 -\phantom{0}5500 $& 5.4    &   $900 -\phantom{0}1030$ & &     &$\phantom{0}2960\pm\phantom{0}60$ & RC3 & opt \\    
\noalign{\smallskip}
\hline
\noalign{\smallskip}
\end{tabular*}
\newline
{\bf Notes}: SKK258- 89: detected HVC947 at $v=262$\,km\,s$^{-1}$; SKK262-127: detected CHVC1098 at $v=261$\,km\,s$^{-1}$;
SKK262- 59: detected CHVC1098 at $v=263$\,km\,s$^{-1}$.
\normalsize
\end{table}
\end{landscape}

\end{document}